# Electrical Resistance: an atomistic view

**Supriyo Datta**

**School of Electrical & Computer Engineering**

**Purdue University, W. Lafayette, IN 47907**

*(http://dynamo.ecn.purdue.edu/~datta)*

**Abstract**

This tutorial article presents a "bottom-up" view of electrical resistance starting from something really small, like a molecule, and then discussing the issues that arise as we move to bigger conductors. Remarkably enough, no serious quantum mechanics is needed to understand electrical conduction through something really small, except for unusual things like the Kondo effect that are seen only for a special range of parameters. This article starts with energy level diagrams (Section 2), shows that the broadening that accompanies coupling limits the conductance to a maximum of $q^2/h$ per level (Sections 3, 4), describes how a change in the shape of the self-consistent potential profile can turn a symmetric current-voltage characteristic into a rectifying one (Sections 5, 6), shows that many interesting effects in molecular electronics can be understood in terms of a simple model (Section 7), introduces the non-equilibrium Green's function (NEGF) formalism as a sophisticated version of this simple model with ordinary numbers replaced by appropriate matrices (Section 8) and ends with a personal view of unsolved problems in the field of nanoscale electron transport (Section 9). Appendix A discusses the Coulomb blockade regime of transport, while Appendix B presents a formal derivation of the NEGF equations. MATLAB codes for numerical examples are listed in Appendix C and can be downloaded from www.nanohub.org, where they can also be run without installation.

*Supriyo Datta, Purdue University*

## 1. Introduction

It is common to differentiate between two ways of building a small device: a ***top-down*** approach where we start from something big and chisel out what we want and a ***bottom-up*** approach where we start from something small like atoms or molecules and assemble what we want. When it comes to describing electrical resistance, the standard approach could be called a "top-down" one. We start in college by learning that the conductance, G (inverse of the resistance) of a large macroscopic conductor is directly proportional to its cross-sectional area (A) and inversely proportional to its length (L):

$$G = \sigma \, A/L \qquad \qquad \text{(Ohm's Law)}$$

where the conductivity $\sigma$ is a material property of the conductor. Years later in graduate school we learn about the factors that determine the conductivity and if we stick around long enough we eventually talk about what happens when the conductor is so small that one cannot define its conductivity. In this article I will try to turn this approach around and present a different view of electrical conduction, one that could be called a bottom-up viewpoint [1].

I will try to describe the conductance of something really small, like a molecule, and then explain the issues that arise as we move to bigger conductors. This is not the way the subject is commonly taught, but I believe the reason is that till recently, no one was sure how to describe the conductance of a really small object, or if it even made sense to talk about the conductance of something really small. To measure the conductance of anything we need to attach two large contact pads to it, across which voltage can be applied. No one knew how to attach contact pads to a small molecule till the late twentieth century, and so no one knew what the conductance of a really small object was. But now that we are able to do so, the answers look fairly clear and in this article I will try to convey all the essential principles. Remarkably enough, no serious quantum mechanics is needed to understand electrical conduction through something really small, except for unusual things like the Kondo effect that are seen only for a special range of parameters. Of course, it is quite likely that new effects will be discovered as we experiment more on small conductors and the description presented here is certainly not intended to be the last word. But I think it should be the "first word" since the traditional top-down approach tends to obscure the simple physics of very small conductors.

*datta@purdue.edu*          

*Outline:* To model the flow of current, the first step is to draw an equilibrium energy level diagram and locate the electrochemical potential $\mu$ (also called the Fermi level or Fermi energy) set by the source and drain contacts (**Section 2**). Current flows when an external device like a battery maintains the two contacts at different electrochemical potentials $\mu_1$ and $\mu_2$, driving the channel into a non-equilibrium state (**Section 3**). The current through a really small device with only one energy level in the range of interest, is easily calculated and as we might expect, it depends on the quality of the contacts. But what is not obvious (and was not appreciated before the late 1980's) is that there is a maximum conductance for a one-level device which is a fundamental constant related to the charge on an electron, -q, and the Planck's constant, h.

$$G_0 \equiv q^2/h = 38.7\,\mu S = (25.8\,K\Omega)^{-1} \qquad\qquad (1.1)$$

Actually small devices typically have two levels (one for up spin and one for down spin) making the maximum conductance equal to $2G_0$. One can always measure conductances lower than this, if the contacts are bad. But the point is that there is an upper limit to the conductance that can be achieved even with the most perfect of contacts as explained in **Section 4**. We will then discuss how the shape of the current-voltage (I-V) characteristics depends crucially on the electrostatic potential profile which requires a self-consistent solution of the equations for electrostatics with those for quantum transport (**Section 5**). **Section 6** represents a brief detour, where we discuss the concept of quantum capacitance which can be useful in guessing the electrostatic potential profile without a full self-consistent solution.

**Section 7** presents several toy examples to illustrate how the model can be used to understand different current-voltage (I-V) characteristics that are observed for small conductors. This model, despite its simplicity (I use it to introduce an undergraduate course on nanoelectronics), has a rigorous formal foundation. It is really a special case of the non-equilibrium Green's function (NEGF) formalism applied to a conductor so small that its electrical conduction can be described in terms of a single energy level. More generally, one needs a Hamiltonian matrix to describe the energy levels and the full NEGF equations can be viewed as a sophisticated version of the simple model with ordinary numbers replaced by appropriate matrices as described in **Section 8**. Finally in **Section 9** I will conclude by listing what I view as open questions in the field of nanoscale electron transport.

*Supriyo Datta, Purdue University*

Three supplementary appendices are also included. **Appendix A** describes the multielectron viewpoint needed to describe the new physics (single electron charging effects) that can arise if a device is coupled weakly to both contacts. **Appendix B** provides a formal derivation of the NEGF equations for advanced readers using the second quantized formalism, while **Appendix C** provides a listing of MATLAB codes that can be used to reproduce the numerical examples presented in Section 7 and in Appendices A, B.

## 2. Energy level diagram

**Fig.2.1. Sketch of a nanoscale Field Effect Transistor. The insulator should be thick enough to ensure that no current flows into the gate terminal, but thin enough to ensure that the gate voltage can control the electron density in the channel.**

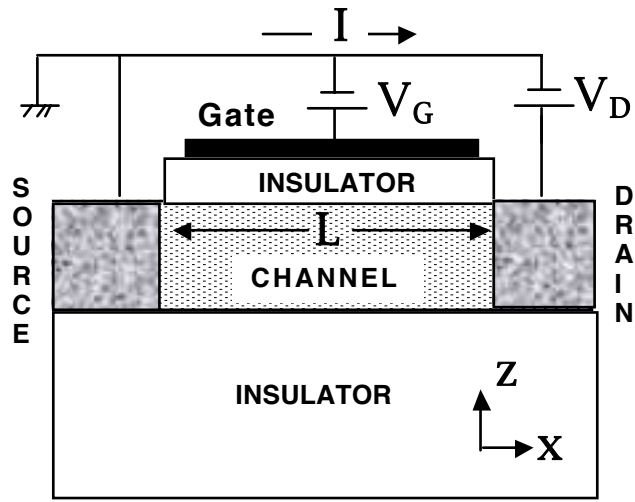

Consider a simple version of a "nanotransistor" consisting of a semiconducting channel separated by an insulator layer (typically silicon dioxide) from the metallic gate surrounding the channel (Fig.2.1). The voltage $V_G$ on the gate is used to control the electron density in the channel and hence its conductance. The regions marked source and drain are the two contact pads which are assumed to be highly conducting. The resistance of the channel determines the current that flows from the source to the drain when a voltage $V_D$ is applied between them. Such a voltage-controlled resistor is the essence of any field effect transistor (FET) although the details differ from one version to another. The channel length, L has been progressively reduced from ~10$\mu$m in 1960 to ~0.1 $\mu$m in 2000, allowing circuit designers to pack $(100)^2 = 10,000$ times more transistors (and hence that much more computing power) into a chip with a given surface area. Laboratory devices have been demonstrated with L = 0.06 $\mu$m which corresponds to approximately 30 atoms! How do we describe current flow through something this small?

The first step in understanding the operation of any inhomogeneous device structure is to draw an *equilibrium* energy level diagram (sometimes called a "band diagram") assuming that there is no voltage applied between the source and the drain. Electrons in a semiconductor occupy a set of energy levels that form bands as sketched in Fig.2.2. Experimentally, one way to measure the occupied energy levels is to find the minimum energy of a photon required to knock an electron out into vacuum (photoemission or PE experiments). We can describe the process symbolically as

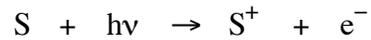

where "S" stands for the semiconductor device (or any material for that matter!).

**Fig. 2.2. Allowed energy levels that can be occupied by electrons in the active region of the device like the channel in Fig.2. A positive gate voltage $V_G$ moves the energy levels down while the electrochemical potential $\mu$ is fixed by the source and drain contacts which are assumed to be in equilibrium with each other ($V_D = 0$).**

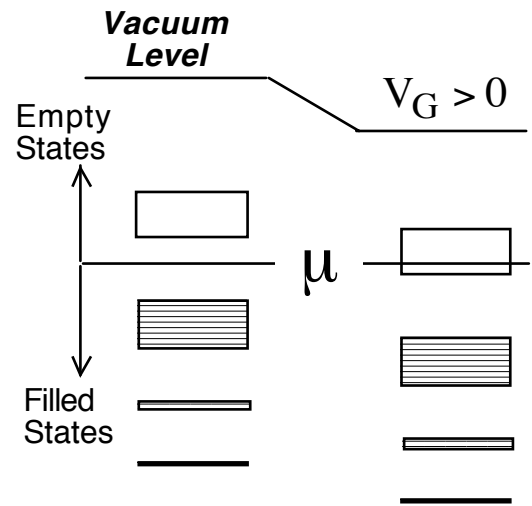

The empty levels, of course cannot be measured the same way since there is no electron to knock out. We need an inverse photoemission (IPE) experiment where an incident electron is absorbed with the emission of photons.

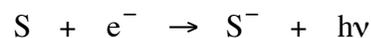

Other experiments like optical absorption also provide information regarding energy levels. All these experiments would be equivalent if electrons did not interact with each other and we could knock one electron around without affecting everything else around it. In the real world this is not true and subtle considerations are needed to relate the measured energies to those we use, but we will not get into this question [2].

We will assume that the large contact regions (labeled source and drain in Fig.2.1) have a continuous distribution of states. This is true if the contacts are metallic, but not exactly true of semiconducting contacts and interesting effects like a decrease in the current with an increase in the voltage (sometimes referred to as negative differential resistance, NDR) can arise as we will see in Section 7 (see also the article by Hersam et.al. [6]). But for the moment let us ignore this possibility and assume the distribution of states to be continuous. They are occupied upto some energy $\mu$ (called the electrochemical potential) which too can be located using photoemission measurements. The work function is defined as the minimum energy of a photon needed to knock a photoelectron out of the metal and it tells us how far below the vacuum level $\mu$ is located.

***Fermi function:*** If the source and drain regions are coupled to the channel (with $V_D$ held at zero), then electrons will flow in and out of the device bringing them all in equilibrium with a common electrochemical potential, $\mu$ just as two materials in equilibrium acquire a common temperature, T. In this equilibrium state, the average (over time) number of electrons in any energy level is typically not an integer, but is given by the Fermi function:

$$f_0(E - \mu) \quad = \quad \frac{1}{1 + \exp\left((E - \mu)/k_B T\right)} \qquad (2.1)$$

which is 1 for energies far below $\mu$ and 0 for energies far above $\mu$.

***n-type operation:*** A positive gate voltage $V_G$ applied to the gate lowers the energy levels in the channel. However, the energy levels in the source and drain contacts are unchanged and hence the electrochemical potential $\mu$ (which must be the same everywhere) remain unaffected. As a result the energy levels move with respect to $\mu$ driving $\mu$ into the empty band as shown in Fig. 2.2. This makes the channel more conductive and turns the transistor ON, since, as we will see in the next Section, the current flow under bias depends on the number of energy levels available around E $=\mu$. The threshold gate voltage $V_T$ needed to turn the transistor ON is thus determined by the energy difference between the equilibrium electrochemical potential $\mu$ and the lowest available empty state (Fig.2.2) or what is called the conduction band edge.

***p-type operation:*** Note that the number of electrons in the channel is not what determines the current flow. A negative gate voltage ($V_G < 0$), for example, reduces the number of electrons in the channel. Nevertheless the channel will become more conductive once the electrochemical potential is driven into the filled band as shown in Fig.2.3, due to the availability of states (filled or otherwise) around E $= \mu$.

**Fig. 2.3. Example of p-type or hole conduction. A negative gate voltage ($V_G < 0$), reduces the number of electrons in the channel. Nevertheless the channel will become more conductive once the electrochemical potential $\mu$ is driven into the filled band since conduction depends on the availability of states around E = $\mu$ and not on the total number of electrons.**

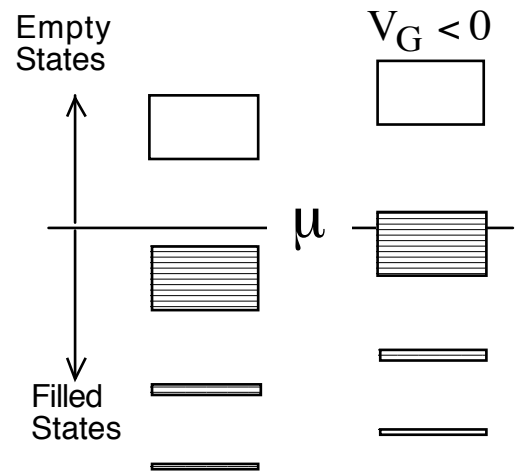

This is an example of p-type or "hole" conduction as opposed to the example of n-type or electron conduction shown in Fig.2.2. The point is that for current flow to occur states are needed near E = $\mu$, but they need not be empty states. Filled states, are just as good and it is not possible to tell from this experiment whether conduction is n-type (Fig.2.2) or p-type (Fig.2.3). This point should get clearer in the next section when we discuss why current flows in response to a voltage applied across the source and drain contacts.

Figs.2.2 and 2.3 suggest that the same device can be operated as a n-type or a p-type device simply by reversing the polarity of the gate voltage. This is true for short devices if the contacts have a continuous distribution of states as we have assumed. But in general this need not be so: for example, long devices can build up "depletion layers" near the contacts whose shape can be different for n- and p-type devices.

### 3. What makes electrons flow ?

We have stated that conduction depends on the availability of states around $E = \mu$; it does not matter if they are empty or filled. To understand why, let us ask what makes electrons flow from the source to the drain? The battery lowers the energy levels in the drain contact with respect to the source contact (assuming $V_D$ to be positive) and maintains them at distinct electrochemical potentials separated by $qV_D$

$$\mu_1 - \mu_2 \;=\; qV_D \tag{3.1}$$

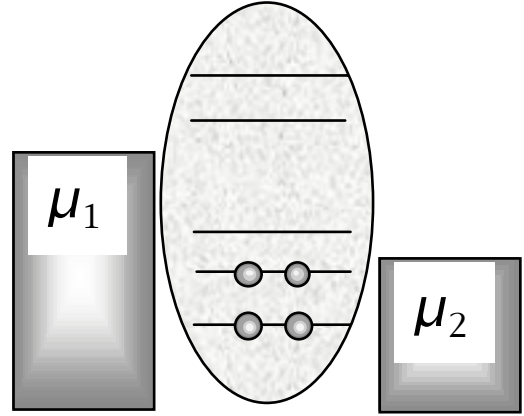

**Fig.3.1. A positive voltage** $V_D$ **applied to the drain with respect to the source lowers the electrochemical potential at the drain:** $\mu_2 = \mu_1 - qV_D$. **Source and drain contacts now attempt to impose different Fermi distributions as shown and the device goes into a state intermediate between the two.**

giving rise to two different Fermi functions:

$$f_1(E) \;=\; \frac{1}{1 + \exp\left((E - \mu_1)/k_B T\right)} \;=\; f_0(E - \mu_1) \tag{3.2a}$$

$$f_2(E) \;=\; \frac{1}{1 + \exp\left((E - \mu_2)/k_B T\right)} \;=\; f_0(E - \mu_2) \tag{3.2b}$$

Each contact seeks to bring the active device into equilibrium with itself. The source keeps pumping electrons into it hoping to establish equilibrium. But equilibrium is never achieved as the drain keeps pulling electrons out in its bid to establish equilibrium with itself. The device is thus forced into a balancing act between two reservoirs with different agendas which sends it into a non-equilibrium state intermediate between what the source would like to see and what the drain would like to see.

***Rate equations for a one-level model:*** This balancing act is easy to see if we consider a simple one level system, biased such that its energy $\varepsilon$ lies between the electrochemical potentials in the two contacts (Fig.3.2). Contact 1 would like to see $f_1(\varepsilon)$ electrons, while contact 2 would like to see $f_2(\varepsilon)$ electrons occupying the state where $f_1$ and $f_2$ are the source and drain Fermi functions defined in Eq.(3.1). The average number of electrons N at steady state will be something intermediate between $f_1$ and $f_2$. There is a net flux $I_1$ across the left junction that is proportional to ($f_1$ − N), dropping the argument $\varepsilon$ for clarity:

$$I_1 \;=\; (-q)\frac{\gamma_1}{h}\left(f_1 - N\right) \qquad\qquad (3.3a)$$

where − q is the charge per electron. Similarly the net flux $I_2$ across the right junction is proportional to ($f_2$ - N) and can be written as

$$I_2 \;=\; (-q)\frac{\gamma_2}{h}\left(f_2 - N\right) \qquad\qquad (3.3b)$$

**Fig.3.2. Flux of electrons into and out of a one-level device at the source and drain ends: Simple rate equation picture.**

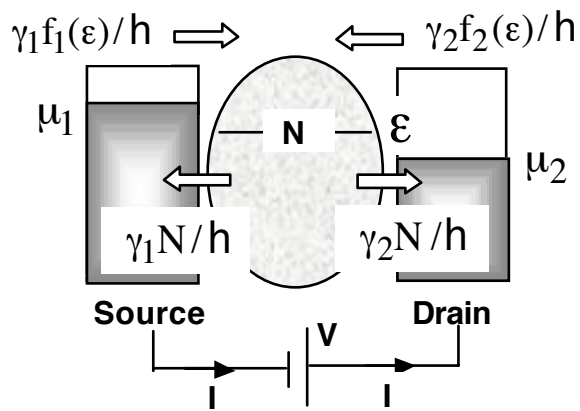

*Supriyo Datta, Purdue University*

We can interpret the rate constants $\gamma_1/\hbar$ and $\gamma_2/\hbar$ as the rates at which an electron placed initially in the level $\varepsilon$ will escape into the source and drain contacts respectively. In principle, we could experimentally measure these quantities which have the dimension per second ($\gamma_1$ and $\gamma_2$ have the dimension of energy). At the end of this Section I will say a few more words about the physics behind these equations. But for the moment, let us work out the consequences.

***Current in a one-level model:*** At steady state there is no net flux into or out of the device: $I_1 + I_2 = 0$, so that from Eqs.(3.2a,b) we obtain the reasonable result

$$N \;\; = \;\; \frac{\gamma_1 f_1 + \gamma_2 f_2}{\gamma_1 + \gamma_2} \tag{3.3}$$

that the occupation $N$ is a weighted average of what contacts 1 and 2 would like to see. Substituting this result into Eq.(3.2a) or (3.2b) we obtain an expression for the steady-state current:

$$I \;\; = \;\; I_1 = -I_2 \;\; = \;\; \frac{q}{\hbar}\,\frac{\gamma_1\gamma_2}{\gamma_1 + \gamma_2}\,\bigl[f_1(\varepsilon) - f_2(\varepsilon)\bigr] \tag{3.4}$$

This is the current per spin. We should multiply it by 2, if there are two spin states with the same energy.

This simple result serves to illustrate certain basic facts about the process of current flow. Firstly, no current will flow if $f_1(\varepsilon) = f_2(\varepsilon)$. A level that is way below both electrochemical potentials $\mu_1$ and $\mu_2$ will have $f_1(\varepsilon) = f_2(\varepsilon) = 1$ and will not contribute to the current, just like a level that is way above both potentials $\mu_1$ and $\mu_2$ and has $f_1(\varepsilon) = f_2(\varepsilon) = 0$. It is only when the level lies within a few $k_B T$ of the potentials $\mu_1$ and $\mu_2$ that we have $f_1(\varepsilon) \neq f_2(\varepsilon)$ and a current flows as a result of the ***"difference in agenda"*** between the contacts. Contact 1 keeps pumping in electrons striving to bring the number up from $N$ to $f_1$ while contact 2 keeps pulling them out striving to bring it down to $f_2$. The net effect is a continuous transfer of electrons from contact 1 to 2 corresponding to a current $I$ in the external circuit (Fig.3.2). Note that the current is in a direction opposite to that of the flux of electrons, since electrons have negative charge.

It should now be clear why the process of conduction requires the presence of states around $E = \mu$. It does not matter if the states are empty (n-type, Fig.2.2) or filled (p-type, Fig.2.4) in equilibrium, before a drain voltage is applied. With empty states, electrons are first injected by the negative contact and subsequently collected by the positive contact. With filled states, electrons are first collected by the positive contact and subsequently refilled by the negative contact. Either way, we have current flowing in the external circuit in the same direction.

***Inflow / Outflow:*** Eqs.(3.3a,b) look elementary and I seldom hear anyone question it. But it hides many subtle issues that could bother more advanced readers and so I feel obliged to mention these issues briefly at the risk of confusing satisfied readers. The right hand sides of Eqs.(3.2a,b) can be interpreted as the difference between the influx and the outflux from the source and drain respectively (see Fig.3.2). For example, consider the source. The outflux of $\gamma_1 N / \hbar$ is easy to justify since $\gamma_1 / \hbar$ represents the rate at which an electron placed initially in the level $\varepsilon$ will escape into the source contact. But the influx $\gamma_1 f_1 / \hbar$ is harder to justify since there are many electrons in many states in the contacts, all seeking to fill up one state inside the channel and it is not obvious how to sum up the inflow from all these states. A convenient approach is to use a thermodynamic argument as follows: If the channel were in equilibrium with the source, there would be no net flux, so that the influx would equal the outflux. But the outflux under equilibrium conditions would equal $\gamma_1 f_1 / \hbar$ since N would equal $f_1$. Under non-equilibrium conditions, N differs from $f_1$ but the influx remains unchanged since it depends only on the condition in the contacts which remains unchanged (note that the outflux does change giving a net current that we have calculated above).

***"Pauli blocking"?*** Advanced readers may disagree with the statement I just made, namely that the influx "depends only on the condition in the contacts". Shouldn't the influx be reduced by the presence of electrons in the channel due to the exclusion principle ("Pauli blocking")? Specifically one could argue that the inflow and outflow (at the source contact) be identified respectively as

$$\gamma_1 f_1 (1 - N) \quad \textit{and} \quad \gamma_1 N (1 - f_1)$$

instead of      $\gamma_1 f_1 \quad \textit{and} \quad \gamma_1 N$

as we have indicated in Fig.3.2. It is easy to see that the net current given by the difference between inflow and outflow is the same in either case, so that the argument might appear "academic". What is not academic, however, is the level broadening that accompanies the process of coupling to the contacts, something we need to include in order to get quantitatively correct results (as we will see in the next section). I have chosen to define inflow and outflow in such a way that the outflow per electron ($\gamma_1 = \gamma_1 N / N$) is equal to the broadening (in addition to their difference being equal to the net current). Whether this broadening (due to the source) is $\gamma_1$ or $\gamma_1(1-f_1)$ or something else is not an academic question. It can be shown that as long as energy relaxing or inelastic interactions are not involved in the inflow / outflow process, the broadening is $\gamma_1$ independent of the occupation factor $f_1$ in the contact.

## 4. The quantum of conductance

Consider a device with a small voltage applied across it causing a splitting of the source and drain electrochemical potentials (Fig.4.1a). We can write the current through this device from Eq.(3.4) and simplify it by assuming $\mu_1 > \varepsilon > \mu_2$ and the temperature is low enough that $f_1(\varepsilon) \equiv f_0(\varepsilon - \mu_1) \approx 1$ and $f_2(\varepsilon) \equiv f_0(\varepsilon - \mu_2) \approx 0$ (see eqs. (3.2)):

$$I \;=\; \frac{q}{h}\,\frac{\gamma_1\gamma_2}{\gamma_1+\gamma_2} \;=\; \frac{q\gamma_1}{2h} \qquad \text{if} \qquad \gamma_2 = \gamma_1 \qquad\qquad (4.1a)$$

This suggests that we could pump unlimited current through this one-level device by increasing $\gamma_1$ $(=\gamma_2)$, that is by coupling it more and more strongly to the contacts. However, one of the seminal results of mesoscopic physics is that the maximum conductance of a one-level device is equal to $G_0$ (see Eq.(I.1)). What have we missed?

What we have missed is the broadening of the level that inevitably accompanies any process of coupling to it. This causes part of the energy level to spread outside the energy range between $\mu_1$ and $\mu_2$ where current flows. The actual current is then reduced below what we expect from Eq.(4.1) by a factor $(\mu_1 - \mu_2)/C\gamma_1$ representing the fraction of the level that lies in the window between $\mu_1$ and $\mu_2$, where $C\gamma_1$ is the effective width of the level, C being a numerical constant. Since $\mu_1 - \mu_2 = qV_D$, we see from Eq.(4.1)

$$I = \frac{q\gamma_1}{2h}\frac{qV_D}{C\gamma_1} \rightarrow G = \frac{I}{V_D} = \frac{q^2}{2Ch} \qquad (4.1b)$$

that the conductance indeed approaches a constant value independent of the strength of the coupling ($\gamma_1 = \gamma_2$) to the contacts. We will now carry out this calculation a little more quantitatively so as to obtain a better estimate for 'C'.

**Fig.4.1. (a) A device with a small voltage applied across it causing a splitting of the source and drain electrochemical potentials $\mu_1 > \varepsilon > \mu_2$ .**

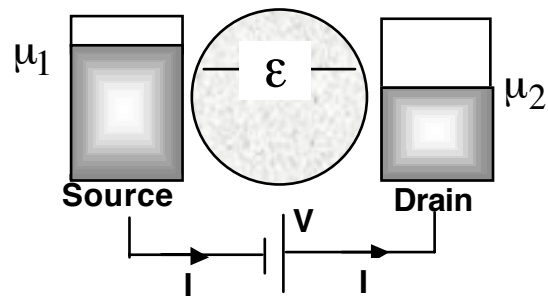

**Fig.4.1. (b) The process of coupling to the device inevitably broadens it thereby spreading part of the energy level outside the energy range between $\mu_1$ and $\mu_2$ where current flows.**

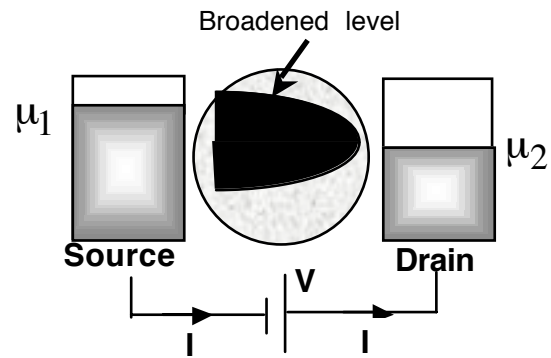

One way to understand this broadening is to note that, ***before*** we couple the channel to the source and the drain, the density of states (DOS), D(E) looks something like this (dark indicates a high DOS)

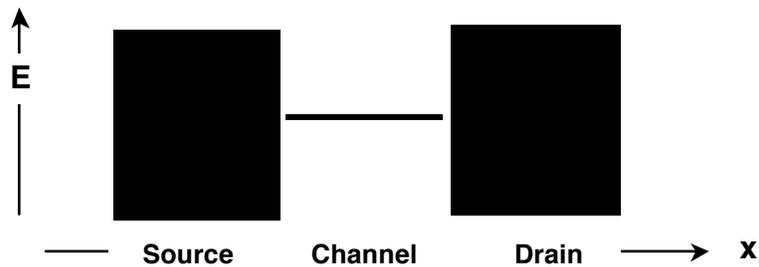

We have one sharp level in the channel and a continuous distribution of states in the source and drain contacts. On coupling, these states "spill over": The channel "loses" part of its state as it spreads into the contacts, but it also "gains" part of the contact states that spread into the channel. Since the loss occurs at a fixed energy while the gain is spread out over a range of energies, the overall effect is to broaden the channel DOS from its initial sharp structure into a more diffuse structure.

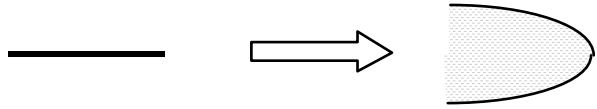

There is a "sum rule" that requires the loss to be exactly offset by the gain, so that integrated over all energy, the level can still hold only one electron. It is common to represent the broadened DOS by a Lorentzian function centered around $E = \varepsilon$ (whose integral over all energy is equal to one):

$$D_\varepsilon(E) \quad = \quad \frac{\gamma/2\pi}{(E - \varepsilon)^2 + (\gamma/2)^2} \qquad\qquad (4.2)$$

The initial delta function can be represented as the limiting case of $D_\varepsilon(E)$ as the broadening tends to zero: $\gamma \rightarrow 0$. The broadening $\gamma$ is proportional to the strength of the coupling as we might expect. Indeed it turns out that $\gamma = \gamma_1 + \gamma_2$, where $\gamma_1/\hbar$ and $\gamma_2/\hbar$ are the escape rates introduced in the last Section. This comes out of a full quantum mechanical treatment, but we could rationalize it as a consequence of the "uncertainty principle" that requires the product of the lifetime ( $= \hbar/\gamma$) of a state and its spread in energy ($\gamma$) to equal $\hbar$ [3].

Another way to justify the broadening that accompanies the coupling is to note that the coupling to the surroundings makes energy levels acquire a finite lifetime, since an electron inserted into a state with energy $E = \varepsilon$ at time t = 0 will gradually escape from that state making its wavefunction look like

$$\exp(-i\varepsilon t/\hbar) \exp(-|t|/2\tau) \quad \textit{instead of} \quad \text{just} \qquad \exp(-i\varepsilon t/\hbar)$$

This broadens its Fourier transform from a delta function at $E = \varepsilon$ to the Lorentzian function of width $\gamma = \hbar/\tau$ centered around $E = \varepsilon$ given in Eq.(4.2). There is thus a simple relationship between the lifetime of a state and its broadening: a lifetime of one picosecond (ps) corresponds to approximately 1.06e-22 joules or 0.7 meV. In general the escape of electrons from a level need not follow a simple exponential and the corresponding lineshape need not be Lorentzian. This is usually reflected in an energy-dependent broadening $\gamma(E)$.

The coupling to the contacts thus broadens a single discrete energy level into a continuous density of states given by Eq.(4.2) and we can include this effect by modifying our expression for the current

$$I \;=\; \frac{q}{h} \frac{\gamma_1 \gamma_2}{\gamma_1 + \gamma_2} \big[ f_1(\varepsilon) - f_2(\varepsilon) \big] \qquad \text{(same as Eq.(3.4))}$$

to account for it:    $I \;=\; \dfrac{q}{h} \displaystyle\int_{-\infty}^{+\infty} dE \; D_\varepsilon(E) \; \dfrac{\gamma_1 \gamma_2}{\gamma_1 + \gamma_2} \big[ f_1(E) - f_2(E) \big]$    (4.3)

Eq.(4.3) for the current extends our earlier result in Eq.(3.4) to include the effect of broadening. We could write it in the form

$$I \;=\; \frac{q}{h} \int_{-\infty}^{+\infty} dE \; \overline{T}(E) \; \big[ f_1(E) - f_2(E) \big] \qquad (4.4)$$

where the ***transmission*** $\overline{T}(E)$ is defined as (making use of Eq.(4.2))

$$\overline{T}(E) \;\equiv\; 2\pi D_\varepsilon(E) \frac{\gamma_1 \gamma_2}{\gamma_1 + \gamma_2} \;=\; \frac{\gamma_1 \, \gamma_2}{(E - \varepsilon)^2 + (\gamma/2)^2} \qquad (4.5)$$

At low temperatures, we can write    $f_1(E) - f_2(E) \;=\; 1 \quad \text{if } \mu_1 > E > \mu_2$
$$= 0, \;\; \text{otherwise}$$

so that the current is given by    $I \;=\; \dfrac{q}{h} \displaystyle\int_{\mu_2}^{\mu_1} dE \; \overline{T}(E)$

If the bias is small enough that we can assume the density of states and hence the transmission to be constant over the range $\mu_1 > E > \mu_2$, so that using Eq.(4.5) we can write

$$I \;=\; \frac{q}{h}\,\big[\mu_1 - \mu_2\big]\,\frac{\gamma_1\gamma_2}{(\mu - \varepsilon)^2 + \big((\gamma_1 + \gamma_2)/2\big)^2}$$

The maximum current is obtained if the energy level $\varepsilon$ coincides with $\mu$, the average of $\mu_1$ and $\mu_2$. Noting that $\mu_1 - \mu_2 = q\,V_D$, we can write the maximum conductance as

$$G \;\equiv\; \frac{I}{V_D} \;=\; \frac{q^2}{h}\,\frac{4\,\gamma_1\gamma_2}{(\gamma_1 + \gamma_2)^2} \;=\; \frac{q^2}{h} \quad \text{if} \qquad \gamma_1 = \gamma_2$$

We can also extend the expression for the number of electrons N (see Eqs.(3.3)) to account for the broadened density of states:

$$N \;=\; \int_{-\infty}^{+\infty} dE \; n(E) \quad \text{where} \; n(E) \;\equiv\; D_\varepsilon(E)\,\frac{\gamma_1 f_1(E) + \gamma_2 f_2(E)}{\gamma_1 + \gamma_2} \quad (4.6)$$

## 5. Potential profile

Now that we have included the effect of level broadening, there is one other factor that we should include in order to complete our model for a one-level conductor. This has to do with the fact that the voltages applied to the external electrodes (source, drain and gate) change the electrostatic potential in the channel and hence the energy levels. It is easy to see that this can play an important role in determining the shape of the current-voltage characteristics [4]. Consider a one-level device with an equilibrium electrochemical potential $\mu$ located slightly above the energy level $\varepsilon$ as shown.

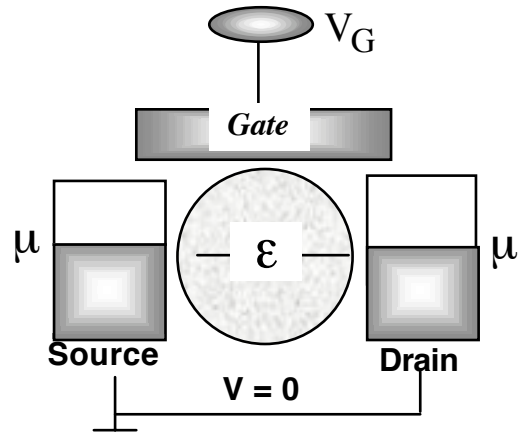

When we apply a voltage between the source and drain, the electrochemical potentials separate by $qV$: $\mu_1 - \mu_2 = qV$. We know that a current flows (at low temperatures) only if the level $\varepsilon$ lies between $\mu_1$ and $\mu_2$. Depending on how the energy level $\varepsilon$ moves we have different possibilities.

If we ignore the gate we might expect the potential in the channel to be lie halfway between the source and the drain: $\varepsilon \rightarrow \varepsilon - (V / 2)$ leading to the picture shown in Fig.5.1 for positive and negative voltages (note that we are assuming the source potential, relative to which the other potentials are changing, to be held constant):

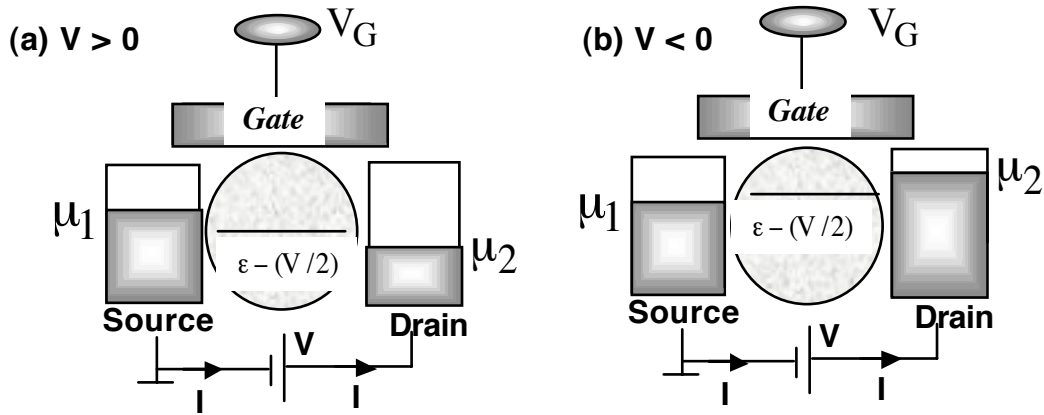

Fig.5.1. If the channel potential lies halfway between the source and drain potentials, significant current will flow for either bias polarity and the current-voltage characteristics will look symmetric.

It is apparent that the energy level lies halfway between $\mu_1$ and $\mu_2$ for either bias polarity ($V > 0$ or $V < 0$), leading to a current-voltage characteristic that is symmetric in V.

A different picture emerges, if we assume that the gate is so closely coupled to the channel that the energy level follows the gate potential and is unaffected by the drain voltage or in other words, $\varepsilon$ remains fixed (Fig.5.2):

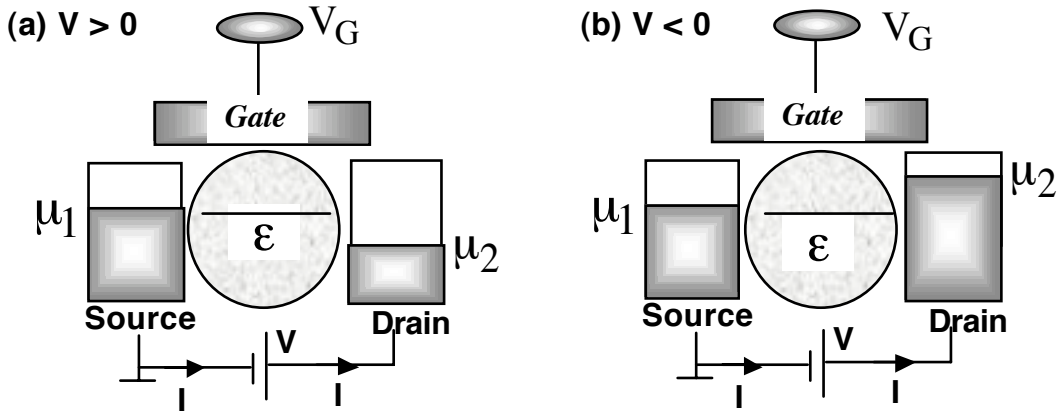

**Fig.5.2. If the channel potential is tied to the source and unaffected by the drain potential, significant current will flow for V>0, but not for V<0, making the current-voltage characteristics look rectifying.**

In this case the energy level lies between $\mu_1$ and $\mu_2$ for positive bias ($V > 0$) but not for negative bias ($V < 0$), leading to a current-voltage characteristic that can be very asymmetric in V.

The point I wish to make is that the shape of the current-voltage characteristic is affected strongly by the potential profile and even the simplest model needs to account for it. One often hears the question: How do we design a molecule that will rectify? The above example shows that the same molecule could rectify or not rectify depending on how close the gate electrode is located!

So how do we calculate the potential inside the channel? If the channel were an insulator, we could solve Laplace's equation ($\varepsilon_r$: relative permittivity which could be spatially varying)

$$\vec{\nabla} \cdot \left( \varepsilon_r \vec{\nabla} V \right) \;=\; 0$$

subject to the boundary conditions that $V = 0$ (source electrode), $V = V_G$ (gate electrode) and $V = V_D$ (drain electrode). We could visualize the solution to this equation in terms of the capacitive circuit model shown in Fig.5.3, if we treat the channel as a single point ignoring any variation in the potential inside it.

The potential energy in the channel is obtained by multiplying the electrostatic potential, V by the electronic charge, - q:

$$U_L = \frac{C_G}{C_E}(-qV_G) \quad + \quad \frac{C_D}{C_E}(-qV_D) \tag{5.1a}$$

Here we have labeled the potential energy with a subscript 'L' as a reminder that it is calculated from the Laplace equation ignoring any change in the electronic charge, which is justified if there are very few electronic states in the energy range around $\mu_1$ and $\mu_2$.

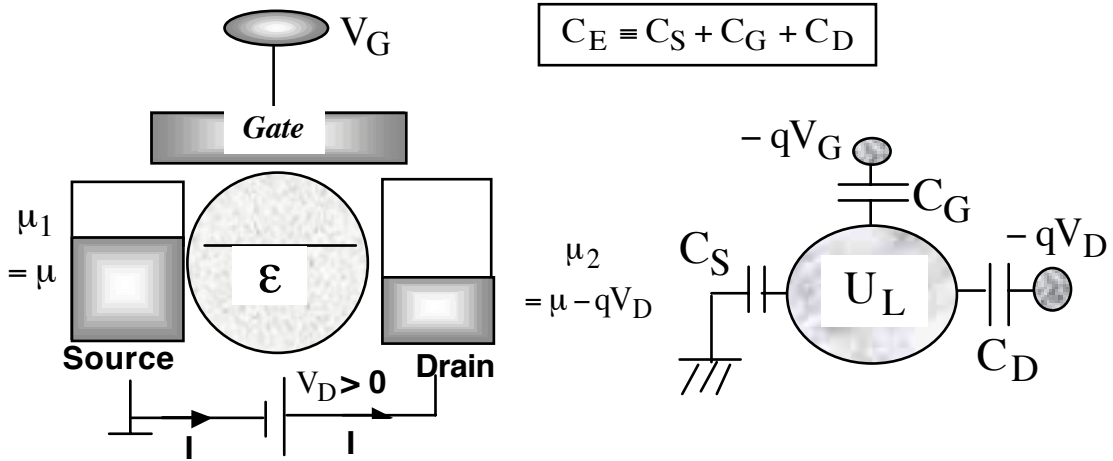

**Fig.5.3. A simple capacitive circuit model for the "Laplace" potential** $U_L$ **of the active region in response to the external gate and drain voltages,** $V_G$ **and** $V_D$. **The actual potential 'U' can be different from** $U_L$ **if there is a significant density of electronic states in the energy range around** $\mu_1$ **and** $\mu_2$. **The total capacitance is denoted** $C_E$, **where 'E' stands for electrostatic.**

Otherwise there is a change $\Delta\rho$ in the electron density in the channel and we need to solve the Poisson equation

$$\vec{\nabla} \cdot \left( \varepsilon_r \vec{\nabla} V \right) \quad = \quad - \Delta\rho/\varepsilon_0$$

*Supriyo Datta, Purdue University*

for the potential. In terms of our capacitive circuit model, we could write the change in the charge as a sum of the charges on the three capacitors:

$$- q\Delta N \;\; = \;\; C_S\, V \;\; + \;\; C_G\left(V - V_G\right) \;\; + \;\; C_D\left(V - V_D\right)$$

so that the potential energy $U = - qV$ is given by the sum of the Laplace potential and an additional term proportional to the change in the number of electrons:

$$U \;\; = \;\; U_L \;\; + \;\; \frac{q^2}{C_E}\,\Delta N \tag{5.1b}$$

The constant $q^2/C_E \equiv U_0$ tells us the change in the potential energy due to **one** extra electron and is called the single electron charging energy, whose significance we will discuss further in the next Section. The **change** $\Delta N$ in the number of electrons is calculated with respect to the reference number of electrons, $N_0$, originally in the channel, corresponding to which its energy level $\varepsilon$ is known.

***Iterative procedure for self-consistent solution:*** For a small device, the effect of the potential U is to raise the density of states in energy and can be included in our expressions for the number of electrons, N (Eq.(4.6)) and the current, I (equation (4.3)) in a straightforward manner:

$$N \;\; = \;\; \int\limits_{-\infty}^{+\infty} dE \;\; D_\varepsilon(E - U)\, \frac{\gamma_1 f_1(E) + \gamma_2 f_2(E)}{\gamma_1 + \gamma_2} \tag{5.2}$$

$$I \;\; = \;\; \frac{q}{h}\int\limits_{-\infty}^{+\infty} dE \; D_\varepsilon(E - U)\;\frac{\gamma_1\gamma_2}{\gamma_1 + \gamma_2}\Big[f_1(E) - f_2(E)\Big] \tag{5.3}$$

Eq.(5.2) has an U appearing on its right hand side which in turn is a function of N through the electrostatic relation (Eq.(5.1)). This requires a simultaneous or "self-consistent" solution of the two equations which is usually carried out using the iterative procedure depicted in Fig.5.4. We start with an initial guess for U, calculate N from Eq.(5.2) with $D_\varepsilon(E)$ given by Eq.(4.2), calculate an appropriate U from Eqs.(5.1b), with $U_L$ given by Eq.(5.1a) and compare with our starting guess for U.

If this new U is not sufficiently close to our original guess, we revise our guess using a suitable algorithm, say something like

$$U_n = U_o + \alpha \left[ U_c - U_o \right] \tag{5.4}$$

**New guess**        **Old guess**        **Calculated**

where $\alpha$ is a positive number (typically < 1 ) that is adjusted to be as large as possible without causing the solution to diverge (which is manifested as an increase in $U_c - U_o$ from one iteration to the next). The iterative process has to be repeated till we find a U that yields an 'N' that leads to a new U which is sufficiently close (say within a fraction of $k_BT$) to the original value. Once a converged U has been found, the current can be calculated from Eq.(5.3).

Fig.5.4. Iterative procedure for calculating N and U self-consistently.

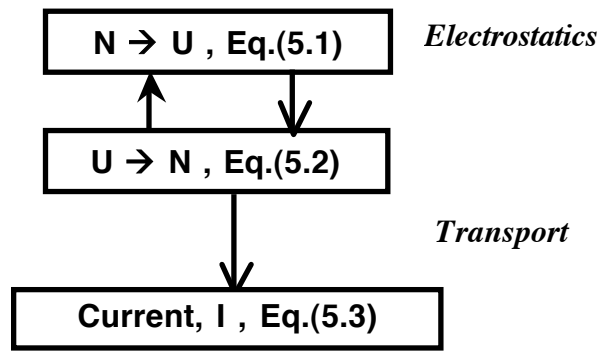

The self-consistent charging model based on the Poisson equation that we have just discussed represents a good zero-order approximation (sometimes called the Hartree approximation) to the problem of electron-electron interactions, but it is generally recognized that it tends to overestimate the effect. Corrections for the so-called exchange and correlation effects are often added, but the description is still within the one-electron picture which assumes that a typical electron feels some average potential, U due to the other electrons. Failure of this one-electron picture is known to give rise to profound effects like magnetism. As we may expect, related effects can manifest themselves in nanoscale transport as well and will continue to be discovered as the field progresses. Such effects are largely outside the scope of this article. However, there is one aspect that is fairly well understood and can affect our picture of current flow even for a simple one-level device putting it in the so-called Coulomb blockade or single-electron charging regime. A proper treatment of this regime requires the multielectron picture described in Appendix A.

### 6. Quantum capacitance

As we have seen, the actual potential U inside the channel plays an important role in determining the shape of the I-V characteristics. Of course, this comes out automatically from the self-consistent calculation described above, but it is important not merely to calculate but to understand the result. Quantum capacitance is a very useful concept that helps in this understanding [5].

We are performing a simultaneous solution of two relations connecting the potential, U to the number of electrons, N: An electrostatic relation (Eq.(5.1)) which is strictly linear and is based on freshman physics, and a transport relation (Eq.(5.2)) which is non-linear and in general could involve advanced quantum statistical mechanics, although we have tried to keep it fairly simple so far. It is this latter equation that is relatively unfamiliar and one could get some insight by linearizing it around an appropriate point. For example, we could define a potential $U = U_N$, which makes $N = N_0$ and keeps the channel exactly neutral:

$$N_0 = \int\limits_{-\infty}^{+\infty} dE \ D_\varepsilon(E - U_N) \frac{\gamma_1 f_1(E) + \gamma_2 f_2(E)}{\gamma_1 + \gamma_2}$$

Any increase in U will raise the energy levels and reduce N, while a decrease in U will lower the levels and increase N. So, for small deviations from the neutral condition, we could write

$$\Delta N \equiv N - N_0 \approx C_Q [U_N - U]/q^2$$

$$\text{where} \ \ C_Q \equiv -q^2 [dN/dU]_{U=U_N} \tag{6.1}$$

is called the quantum capacitance and depends on the density of states around the energy range of interest, as we will show. We can substitute this linearized relation into Eq.(5.1b) to obtain

$$U = U_L + \frac{C_Q}{C_E}[U_N - U] \rightarrow U = \frac{C_E U_L + C_Q U_N}{C_E + C_Q} \tag{6.2}$$

showing that the actual channel potential U is intermediate between the Laplace potential, $U_L$ and the neutral potential, $U_N$. How close it is to one or the other depends on the relative magnitudes of the electrostatic capacitance, $C_E$ and the quantum capacitance, $C_Q$. This is easily visualized in terms of a capacitive network obtained by extending Fig.5.1 to include the quantum capacitance, as shown in Fig.6.1.

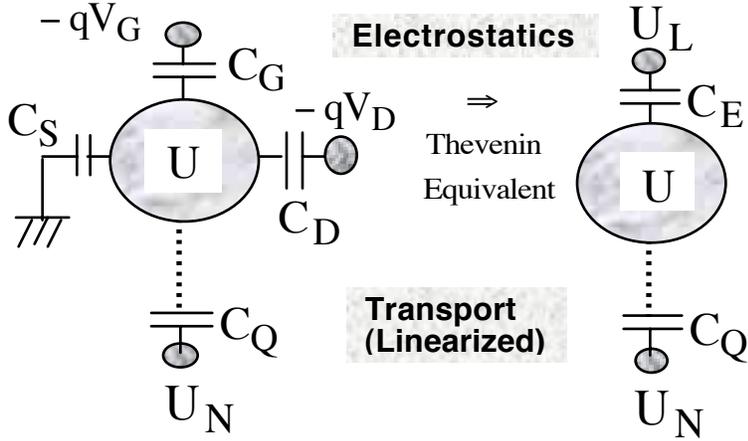

**Fig.6.1. Extension of the capacitive network in Fig.5.1 to include the quantum capacitance.**

We will now show that a channel with a low density of states in the energy range of interest has a low $C_Q$ making $U=U_L$ as we expect for an insulator. A channel with a high density of states in the energy range of interest has a high $C_Q$, making $U=U_N$ as we expect for a metal.

***Relation between*** $C_Q$ ***and the density of states: To*** establish the connection between the quantum capacitance and the density of states, we rewrite Eq.(5.2) in the form

$$N = \int_{-\infty}^{+\infty} dE \; D_\varepsilon(E) \frac{\gamma_1 f_0(E + U - \mu_1) + \gamma_2 f_0(E + U - \mu_2)}{\gamma_1 + \gamma_2}$$

and then make use of Eq.(6.1) for $C_Q$ :

$$C_Q \equiv -q^2 \left[ dN/dU \right]_{U=U_N}$$

$$= q^2 \int_{-\infty}^{+\infty} dE \left[ D_1(E) F_T(E + U_N - \mu_1) + D_2(E) F_T(E + U_N - \mu_2) \right] \quad (6.3)$$

where $\quad D_1(E) = D_\varepsilon(E) \dfrac{\gamma_1}{\gamma_1 + \gamma_2} \quad$ and $\quad D_2(E) = D_\varepsilon(E) \dfrac{\gamma_2}{\gamma_1 + \gamma_2}$

and we have introduced the thermal broadening function $F_T$ defined as

$$F_T(E) \equiv -\frac{df_0}{dE} = \frac{1}{4 k_B T} \operatorname{sech}^2 \left( \frac{E}{2 k_B T} \right) \quad (6.4)$$

Its maximum value is $\left( 1/4 k_B T \right)$ while its width is proportional to $k_B T$. It is straight forward to show that the area obtained by integrating this function is equal to one, independent of $k_B T$. This means that at low temperatures $F_T(E)$ becomes very large but very narrow while maintaining a constant area of one and can be idealized as a delta function: $F_T(E) \rightarrow \delta(E)$, which allows us to simplify the expression for the quantum capacitance:

$$C_Q \approx q^2 \left[ D_1(\mu_1 - U_N) + D_2(\mu_2 - U_N) \right] \quad (6.5)$$

This expression, valid at low temperatures, shows that the quantum capacitance depends on the density of states around the electrochemical potentials $\mu_1$ and $\mu_2$, after shifting by the potential $U_N$.

## 7. Toy examples

In this Section I will first summarize the model that we have developed here and then illustrate it with a few toy examples. We started by calculating the current through a device with a single discrete level ($\varepsilon$) in Section 3, and then extended it to include the broadening of the level into a Lorentzian density of states

$$D_\varepsilon(E) = 2 \,(\text{for spin}) \, x \, \frac{\gamma / 2\pi}{\left( E - \varepsilon \right)^2 + \left( \gamma/2 \right)^2} \, , \, \gamma \equiv \gamma_1 + \gamma_2 \quad (7.1)$$

in Section 4 and the self-consistent potential in Section 5

$$U = U_L + U_0 (N - N_0) \qquad (7.2)$$

$$U_L = \frac{C_G}{C_E}(-qV_G) + \frac{C_D}{C_E}(-qV_D)$$

$$U_0 = q^2/C_E \quad , \quad C_E = C_G + C_S + C_D \qquad (7.3)$$

The function $D_\varepsilon(E)$ in Eq.(7.1) is intended to denote the density of states (DOS) obtained by broadening a single discrete level $\varepsilon$. What about a multi-level conductor with many energy levels looking something like this?

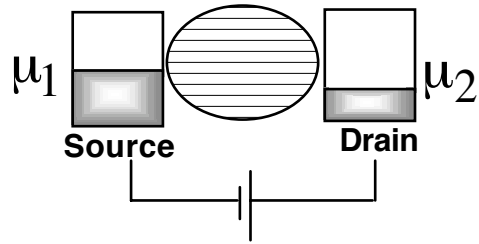

Fig. 7.1

If we make the rather cavalier assumption that all levels conduct independently, then we could use exactly the same equations as for the one-level device, replacing the one-level DOS, $D_\varepsilon(E)$ in Eq.(7.1) with the total DOS, $D(E)$. With this in mind, I will use $D(E)$ instead of $D_\varepsilon(E)$ to denote the density of states and refer to the results summarized below as the ***independent level model*** rather than the single level model.

*Independent level model: summary :* In this model, the number of electrons, N is given by

$$N = \int_{-\infty}^{+\infty} dE \ n(E)$$

$$\text{where} \quad n(E) = D(E-U)\left(\frac{\gamma_1}{\gamma}f_1(E) + \frac{\gamma_2}{\gamma}f_2(E)\right) \qquad (7.4)$$

while the current at the two terminals are given by

$$I_1 = \frac{q}{h} \int_{-\infty}^{+\infty} dE \, \gamma_1 \left[ D(E-U) f_1(E) - n(E) \right] \qquad (7.5a)$$

$$I_2 = \frac{q}{h} \int_{-\infty}^{+\infty} dE \, \gamma_2 \left[ D(E-U) f_2(E) - n(E) \right] \qquad (7.5b)$$

At steady state, the sum of the two currents is equated to zero to eliminate n(E):

$$I = \frac{q}{h} \int_{-\infty}^{+\infty} dE \, \overline{T}(E-U) \left[ f_1(E) - f_2(E) \right]$$

$$\text{where} \quad \overline{T}(E) = D(E) \, 2\pi \, \gamma_1 \gamma_2 / \gamma \qquad (7.6)$$

is called the **transmission** a concept that plays a central role in the transmission formalism widely used in mesoscopic physics [9]. Note that the Fermi functions $f_1$ and $f_2$ are given by

$$f_1(E) = f_0(E - \mu_1) \quad , \quad f_2(E) = f_0(E - \mu_2) \qquad (7.7)$$

$$\text{where} \quad f_0(E) = \left( 1 + \exp\left(E / k_B T\right) \right)^{-1}$$

where the electrochemical potentials in the source and drain contacts are given by

$$\mu_1 = \mu \quad , \quad \mu_2 = \mu - qV_D \, , \qquad (7.8)$$

$\mu$ being the equilibrium electrochemical potential.

### 7.1. Negative Differential Resistance (NDR)

To see how the model works, consider first a one-level device with a broadened DOS given by Eq.(7.1) with parameters as listed in Fig.7.2. As we might expect the current increases once the applied drain voltage is large enough that the energy level comes within the energy window between $\mu_1$ and $\mu_2$. The current then increases towards a maximum value of $(2q / \hbar) \, \gamma_1 \gamma_2 / (\gamma_1 + \gamma_2)$ over a voltage range $\sim (\gamma_1 + \gamma_2 + k_B T) \, C_E / C_D$ as shown in Fig.7.2a. Here we have assumed the broadening due to the two contacts $\gamma_1$ and $\gamma_2$ to be constants equal to 0.005 eV.

 

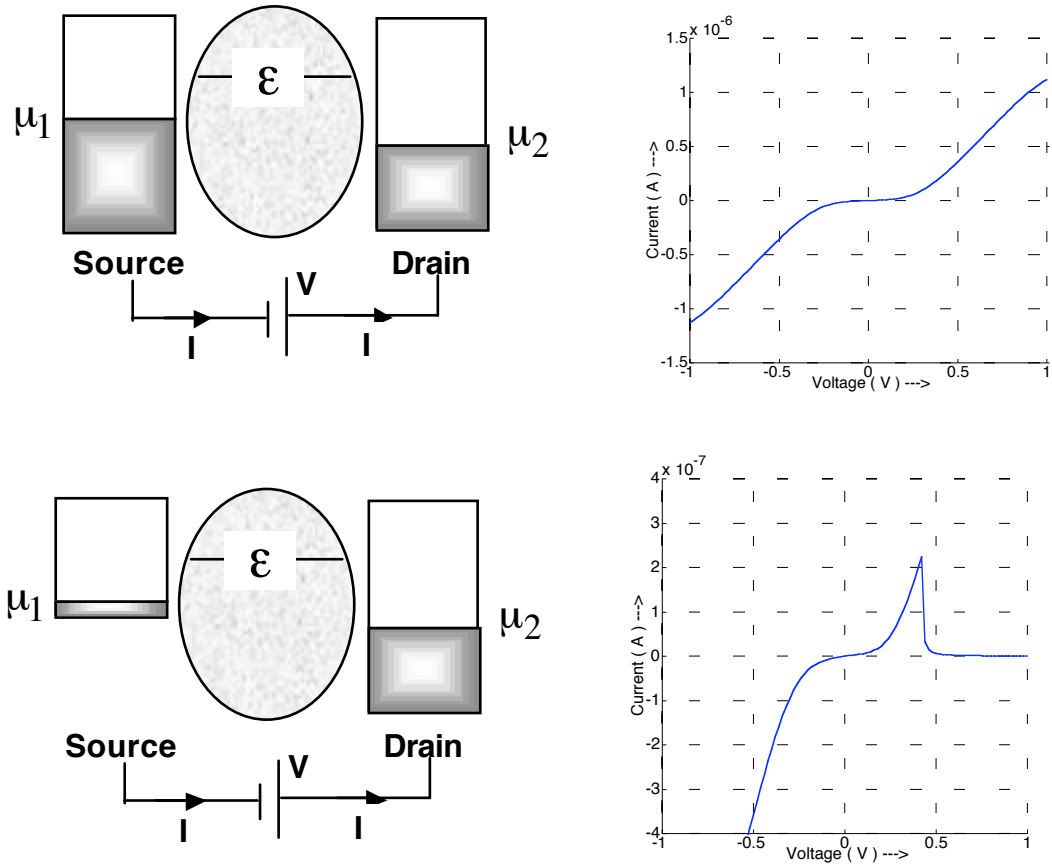

**Fig.7.2. Current vs. voltage calculated using Eqs.(7.1) − (7.8) with**

$\mu = 0, \varepsilon = 0.2 \, \text{eV}, \ V_G = 0, \ k_B T = 0.025 \, \text{eV}, U_0 = 0.25 \, \text{eV}, \ C_D / C_E = 0.5,$

**and** $\gamma_1 = \gamma_2 = 0.005 \, \text{eV}$ **. The only difference between (a) and (b) is that in (a),** $\gamma_1$

**is independent of energy, while in (b)** $\gamma_1$ **is zero for energies less than zero.**

**In either case** $\gamma_2$ **is assumed to be independent of energy.**

Now suppose $\gamma_1$ is equal to 0.005 eV for E > 0, but is ***zero*** for E < 0 ($\gamma_2$ is still independent of energy and equal to 0.005 eV). The current-voltage characteristics now show negative differential resistance (NDR), that is, a drop in the current with an increase in the voltage, in one direction of applied voltage but not the other as shown in Fig.7.2b. This simple model may be relevant to the experiment described in [6] though the nature and location of the molecular energy levels remain to be established quantitatively.

*Supriyo Datta, Purdue University*

### 7.2. Thermoelectric effect

We have discussed the current that flows when a voltage is applied between the two contacts. In this case the current depends on the density of states near the Fermi energy and it does not matter whether the equilibrium Fermi energy $\mu_1$ lies at the (a) lower end (n-type) or at the (b) upper end (p-type) of the density of states:

(a) "n-type"                    (b) "p-type"

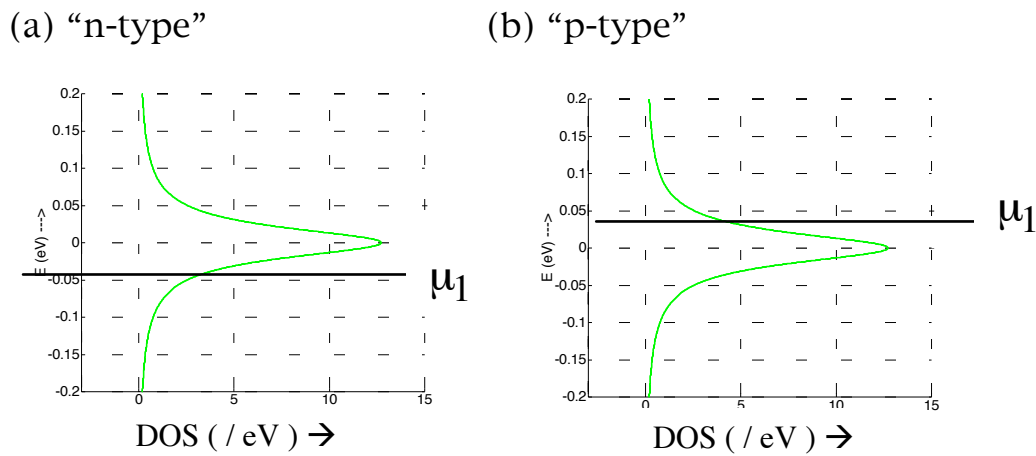

**Fig.7.3: We can define n- and p-type conduction depending on whether the electrochemical potential lies on an up-slope or a down-slope of the DOS.**

However, if we simply heat up one contact relative to the other so that $T_1 > T_2$ (with no applied voltage) a thermoelectric current will flow whose direction will be different in case (a) and in case (b). To see this we could calculate the current from our model with $U = 0$ (there is no need to perform a self-consistent solution), $V_D = 0$ and $V_G = 0$, and with

$$f_1(E) \equiv \frac{1}{1 + \exp\left(\dfrac{E - \mu_1}{k_B T_1}\right)} \quad \text{and} \quad f_2(E) \equiv \frac{1}{1 + \exp\left(\dfrac{E - \mu_1}{k_B T_2}\right)}$$

As shown in Fig.7.4 the direction of the current is different for n- and p-type samples. This is of course a well-known result for bulk solids where hot point probes are routinely used to identify the type of conduction. But the point I am trying to make is that it is true even for ballistic samples and can be described by the elementary model described here [7].

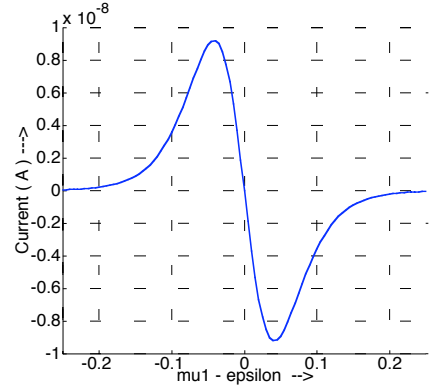

**Fig.7.4. The thermoelectric current reverses direction from n-type ($\mu_1 < 0$) to p-type ($\mu_1 > 0$) samples.** $\gamma_1 = \gamma_2 = 0.005$ eV, $k_B T_1 = 0.026$ eV and $k_B T_2 = 0.025$ eV.

## 7.3. Nanotransistor

As another example of the independent level model, let us model a nanotransistor [8] by writing the DOS as (see Fig.7.5, W: width in the y-direction)

$$D(E) \;=\; m_c W L / \pi \hbar^2 \; \vartheta(E - E_c) \tag{7.9}$$

making use of the well-known result that the DOS per unit area in a large 2D conductor described by an electron effective mass $m_c$ is equal to $m_c / \pi \hbar^2$, for energies greater than the energy $E_c$ of the conduction band edge. The escape rates can be written down assuming that electrons are removed by the contact with a velocity $v_R$ :

$$\gamma_1 \;=\; \gamma_2 \;=\; \hbar v_R / L \tag{7.10}$$

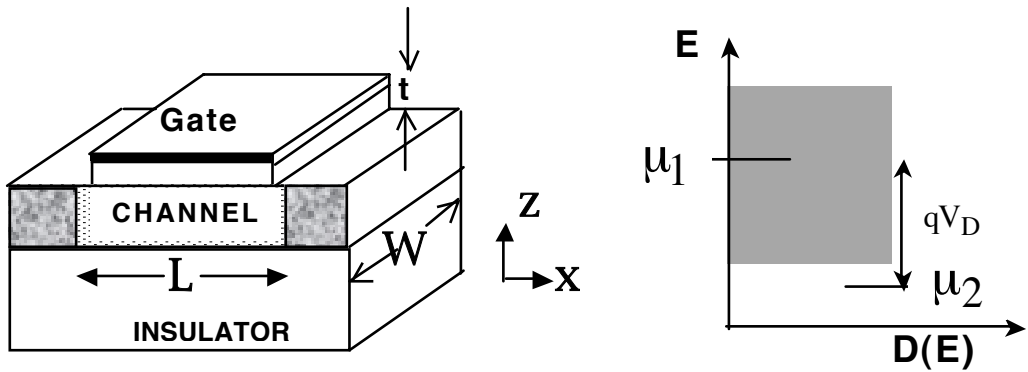

Fig.7.5. A nanotransistor: Physical structure and assumed
        density of states (DOS) in the channel region.

*Supriyo Datta, Purdue University*

The current-voltage relations shown in Fig.7.6 were obtained using these model parameters: Ec=0, $\mu_1$ = -0.2 eV, $m_c$ = 0.25 m, $C_G = 2\varepsilon_r\varepsilon_0 WL/t$ , $C_S = C_D$ = 0.05 $C_G$, W = 1 $\mu$m, L = 10 nm, insulator thickness, t = 1.5 nm, $v_R = 10^7$ cm / sec. At high drain voltages ($V_D$) the current saturates when $\mu_2$ drops below $E_c$ since there are no additional states to contribute to the current. Note that the gate capacitance $C_G$ is much larger than the other capacitances, which helps to hold the channel potential fixed relative to the source as the drain voltage is increased (see Eq.(7.3)). Otherwise, the bottom of the channel density of states, $E_c$ will "slip down" with respect to $\mu_1$ when the drain voltage is applied, so that the current will not saturate. The essential feature of a well-designed transistor is that the gate is much closer to the channel than 'L' allowing it to hold the channel potential constant despite the voltage $V_D$ on the drain.

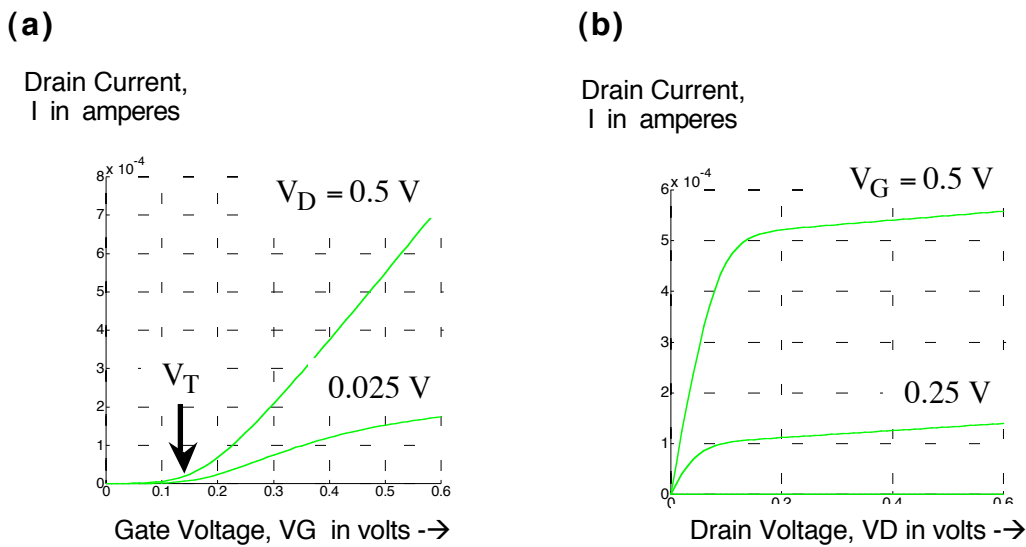

**(a)**

Drain Current,
I in amperes

**(b)**

Drain Current,
I in amperes

Gate Voltage, VG in volts -→

Drain Voltage, VD in volts -→

**Fig. 7.6. (a) Drain current (I) as a function of the gate voltage ($V_G$) for different values of the drain voltage ($V_D$); (b) Drain current as a function of the drain voltage for different values of the gate voltage.**

I should mention that our present model ignores the profile of the potential along the length of the channel, treating it as a little box with a single potential U given by Eq.(7.2). Nonetheless the results (Fig.7.6) are surprisingly close to experiments / realistic models, because the current in well-designed nanotransistors is controlled by a small region in the channel near the source whose length can be a small fraction of the actual length L. Luckily we do not need to pin down the precise value of this fraction, since the present model gives the same current independent of L [8].

***Ohm's Law***: It is natural to ask whether the independent level model would lead to Ohm's law if we were to calculate the low bias conductance of a large conductor of length L and cross-sectional area S? Since the current is proportional to the DOS, D(E) (see Eq.(7.5)) which is proportional to the volume SL of the conductor, it might seem that the conductance G ~ S L. However, the coupling to the contacts decreases inversely with the length L of the conductor, since the longer a conductor is, the smaller is its coupling to the contact (see eq.(7.10)). While the DOS goes up as the volume, the coupling to the contact goes down as 1/L, so that the conductance

$$G \sim S \, L \, / \, L \; = S$$

But Ohm's law tells us that the conductance should scale as S/L; we are predicting that it should scale as "S". The reason is that we are really modeling a ***ballistic*** conductor, where electrons propagate freely, the only resistance arising from the contacts. The conductance of such a conductor is indeed independent of its length. The length dependence of the conductance comes from scattering processes within the conductor that are not yet included in our thinking [9].

For example, in a uniform channel the electronic wavefunction is spread out uniformly. But a scatterer in the middle of the channel could split up the wavefunctions into two pieces, one on the left and one on the right with different energies. One has a small $\gamma_2$ while the other has a small $\gamma_1$, and so neither conducts very well. This ***localization*** of wavefunctions would seem to explain why the presence of a scatterer contributes to the resistance, but to get the story quantitatively correct it is in general necessary to go beyond the independent-level model to account for interference between multiple paths. This requires a model that treats $\gamma$ as a ***matrix*** rather than as simple numbers.

Such "coherent" scatterers, lead to many interesting phenamona but not to Ohm's law: R ~ 1/L (Ohm's law). The full story requires us to include phase-breaking scattering processes that cause a change in the state of an external object. For example, if an electron gets deflected by a rigid (that is unchangeable) defect in the lattice, the scattering is said to be coherent. But, if the electron transfers some energy to the atomic lattice causing it to start vibrating that would constitute a phase-breaking or incoherent process. Purely coherent scatterers can give rise to a measurable resistance R, but cannot give rise to any dissipation, since no energy is removed from the electrons. Indeed there is experimental evidence that the associated Joule heating ($I^2R$) occurs in the contacts outside the channel, allowing experimentalists to pump a lot more current through a small conductor without burning it up.

Much of the work on small conductors is usually in the coherent limit, but it is clear that including phase-breaking scattering will be important in developing quantitative models. In Section 7.4 I will show how this can be done within our simple one-level model. This will lead naturally to the full-fledged non-equilibrium Green's function (NEGF) formalism described in Section 8.

### 7.4. Inelastic Spectroscopy

For the purpose of including phase-breaking it is useful to re-cast the equations listed at the beginning of this Section in a slightly different form by defining a Green's function G

$$G \;=\; \frac{1}{E - \varepsilon - U + (i\gamma/2)} \qquad \text{where} \quad \gamma \;=\; \gamma_1 + \gamma_2 \qquad (7.11)$$

such that

$$2\pi D(E) \;=\; G(E)\,\gamma(E)\,G^*(E) \;=\; i\,[G - G^*] \qquad (7.12)$$

The electron density can then be written as (cf. Eq.(7.4))

$$2\pi n(E) \;=\; G(E)\,\gamma^{in}(E)\,G^*(E) \qquad (7.13)$$

in terms of the inscattering function defined as $\gamma^{in} \;=\; \gamma_1^{in} + \gamma_2^{in}$, where

$$\gamma_1^{in} \;=\; \gamma_1\,f_1 \qquad \text{and} \qquad \gamma_2^{in} \;=\; \gamma_2\,f_2 \qquad (7.14a)$$

It is also useful to define an outscattering function $\gamma^{out} \;=\; \gamma_1^{out} + \gamma_2^{out}$, where

$$\gamma_1^{out} \;=\; \gamma_1\,(1 - f_1) \qquad \text{and} \qquad \gamma_2^{out} \;=\; \gamma_2\,(1 - f_2) \qquad (7.14b)$$

       

Noting that    $\gamma_i = \gamma_i^{out} + \gamma_i^{in}$                                    (7.15)

Subtracting Eq.(7.13) from (7.12) we obtain

$$2\pi p(E) = G(E)\,\gamma^{out}(E)\,G^*(E)$$                                    (7.16)

for the ***hole density***    $p(E) = D(E) - n(E)$                                    (7.17)

obtained by subtracting the electron density from the density of states.

Phase-breaking scattering processes can be visualized as a fictitious terminal 's' with its own inscattering and outscattering functions, so that

$$\gamma^{in} = \gamma_1^{in} + \gamma_2^{in} + \gamma_s^{in}$$                                    (7.18a)

$$\gamma^{out} = \gamma_1^{out} + \gamma_2^{out} + \gamma_s^{out}$$                                    (7.18b)

The current (per spin) at any terminal 'i' can be calculated from

$$I_i = (q/\hbar)\int_{-\infty}^{+\infty} dE\;\tilde{I}_i(E)$$                                    (7.19)

with    $\tilde{I}_i = [\gamma_i^{in}D] - [\gamma_i n]$                                    (7.20)

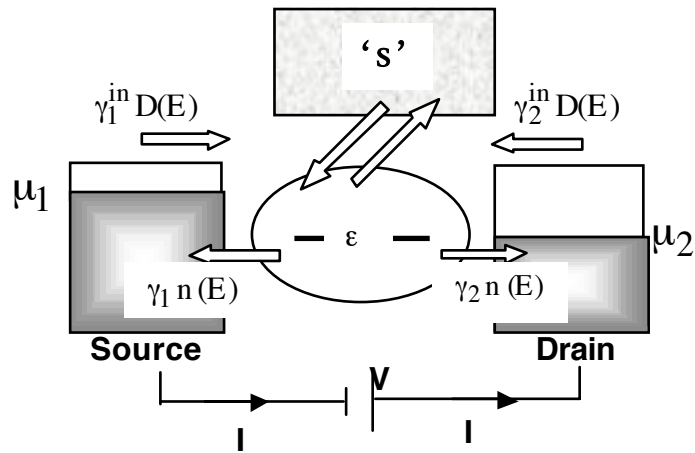

**Fig.7.7. Phase-breaking scattering processes can be visualized as a fictitious terminal 's' with its own inscattering and outscattering functions.**

To find $\gamma_s^{in}$ and $\gamma_s^{out}$, one approach is to view the scattering terminal 's' like a real terminal whose electrochemical potential $\mu_s$ is adjusted to make the current $I_s = 0$, following the phenomenological approach widely used in mesoscopic physics [9e]. The scattering terminal, however, cannot in general be described by a Fermi function that we can use in Eqs.(7.14a,b), The NEGF formalism allows us to evaluate $\gamma_s^{in}$ and $\gamma_s^{out}$ to any desired approximation from a microscopic theory. In the self-consistent Born approximation,

$$\gamma_s^{in}(E) \;\; = \;\; \int d(\hbar\omega)\, D^{ph}(\hbar\omega)\, n(E + \hbar\omega) \tag{7.21a}$$

$$\text{and} \quad \gamma_s^{out}(E) \;\; = \;\; \int d(\hbar\omega)\, D^{ph}(\hbar\omega)\, p(E - \hbar\omega) \tag{7.21b}$$

where the "phonon" spectral function can be written as the sum of an emission term (positive frequencies) and an absorption term (negative frequencies)

$$D^{ph}(\hbar\omega) \;\; = \;\; \sum_i D_i\, [(N_i + 1)\, \delta(\hbar\omega - \hbar\omega_i) + N_i\, \delta(\hbar\omega + \hbar\omega_i)] \tag{7.22}$$

with $N_i$ representing the number of phonons of frequency $\hbar\omega_i$, and $D_i$ its coupling. We assume $N_i$ to be given by the Bose Einstein factor, but it is conceivable that the phonons could be driven off equilibrium requiring $N_i$ to be evaluated from a transport equation for the phonons. Low frequency phonons with $\hbar\omega_i$ much smaller than other relevant energy scales can be treated as elastic scatterers with $\hbar\omega_i \sim 0$, $D_i(N_i + 1) \approx D_i N_i \equiv D_0^{ph}$. Eqs.(7.21) then simplify to

$$\gamma_s^{in} \;\; = \;\; D_0^{ph}\, n(E) \quad \text{and} \quad \gamma_s^{out} \;\; = \;\; D_0^{ph}\, p(E)$$

$$\text{so that} \;\; \gamma_s \;\; = \;\; \gamma_s^{in} + \gamma_s^{out} \;\; = \;\; D_0^{ph}\, D(E) \tag{7.23}$$

Fig.7.8 shows a simple example where the energy level $\varepsilon = 5$ eV lies much above the equilibrium electrochemical potential $\mu = 0$, so that current flows by tunneling. The current calculated without any phonon scattering (all $D_i = 0$) and with phonon scattering ($D_1$=0.5, $\hbar\omega_1$=0.075 eV and $D_2$=0.7, $\hbar\omega_2$=0.275 eV) shows no discernible difference. The difference, however, shows up in the conductance dI/dV where there is a discontinuity proportional to $D_i$ when the applied voltage equals the phonon frequency $\hbar\omega_i$. This discontinuity shows up as peaks in $d^2I/dV^2$ whose location along the voltage axis corresponds to molecular vibration quanta, and this is the basis of the field of inelastic electron tunneling spectroscopy (IETS) [10].

<div align="center">

(a) I vs. V          (b) dI/dV vs. V          (c) $d^2I/dV^2$ vs. V

</div>

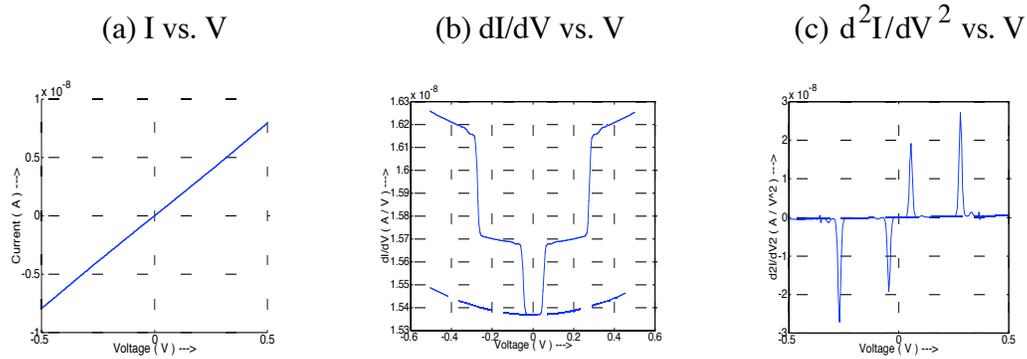

**Fig.7.8. (a) Current (I), (b) conductance (dI/dV) and (c) $d^2I/dV^2$ as a function of voltage calculated without phonon scattering (dashed line) and with scattering by phonons (solid line) with two distinct frequencies having slightly different coupling strengths** ($D_1$=0.5, $\hbar\omega_1$=0.075 eV and $D_2$=0.7, $\hbar\omega_2$=0.275 eV).

Note that the above prescription for including inelastic scattering (Eqs.(7.21), (7.22)) is based on the NEGF formalism. This is different from many common theories where exclusion principle factors (1-f) appropriate to the contacts are inserted somewhat intuitively and as such cannot be applied to long devices, By contrast the NEGF prescription can be extended to long devices by replacing numbers with matrices as we will describe in the next Section. Indeed as we mentioned in the introduction, what we have described so far can be viewed as a special case of the NEGF formalism applied to a device so small that it is described by a single energy level or a "(1x1) Hamiltonian matrix". Let us now look at the general formalism.

## 8.  From numbers to matrices: NEGF formalism

The one level model serves to identify the important concepts underlying the flow of current through a conductor, such as the location of the equilibrium *electrochemical potential μ* relative to the *density of states* D(E), the *broadening* of the level $\gamma_{1,2}$ due to the coupling to contacts 1 and 2 etc.. In the general model for a multilevel conductor with 'n' energy levels, all the quantities we have introduced are replaced by a corresponding matrix of size *(n x n):*

$\varepsilon \;\rightarrow\; [H]$                                             *Hamiltonian* matrix

$\gamma_i \;\rightarrow\; [\Gamma_i(E)]$                                       *Broadening* matrix

$2\pi D(E) \;\rightarrow\; [A(E)]$                                             *Spectral function*

$2\pi n(E) \;\rightarrow\; [G^n(E)]$                                           *Correlation function*

$2\pi p(E) \;\rightarrow\; [G^p(E)]$                                           *Hole correlation function*

$U \;\rightarrow\; [U]$                                                        *Self-consistent potential* matrix

$N \;\rightarrow\; [\rho] \;=\; \int (dE/2\pi)\,[G^n(E)]$   *Density matrix*

$\gamma_i^{in} \;\rightarrow\; [\Sigma_i^{in}(E)]$                             *Inscattering* matrix

$\gamma_i^{out} \;\rightarrow\; [\Sigma_i^{out}(E)]$                           *Outscattering* matrix

Actually, the effect of the contacts is described by a *"self-energy" matrix,* $[\Sigma_{1,2}(E)]$, whose anti-Hermitian part is the broadening matrix: $\Gamma_{1,2} = i\,[\,\Sigma_{1,2} - \Sigma_{1,2}^+\,]$. The Hermitian part effectively adds to [H] thereby shifting the energy levels – an effect we ignored in the simple model. The Hermitian and anti-Hermitian parts are Hilbert transform pairs . Also, I should mention that I have used

$G^n(E)\,,\,G^p(E)\,,\,\Sigma^{in}(E)\,,\,\Sigma^{out}(E)$

to denote what is usually written in the literature [11-12] as

$-iG^<(E)\,,\,+iG^>(E)\,,\,-i\Sigma^<(E)\,,\,+i\Sigma^>(E)\,,$

in order to emphasize their physical significance.

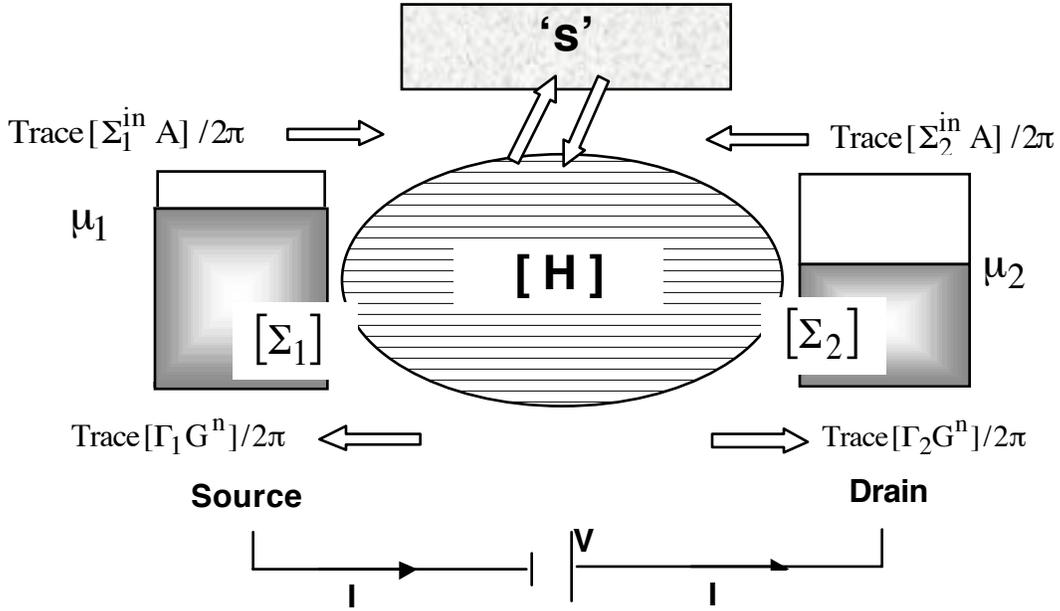

**Fig.8.1. (cf. Fig.7.7) From numbers to matrices: General matrix model, based on the NEGF formalism. Without the "s-contact" this model is equivalent to Eq.(6) of [12b]. The "s-contact" distributed throughout the channel, describes incoherent scattering processes [12c]. In general this "contact" cannot be described by a Fermi function, unlike the real contacts.**

The **NEGF equations** for dissipative quantum transport look much like those discussed in Section 7.4, but with num bers replaced by matrices:

$$G^n \;=\; G \, \Sigma^{in} \, G^+ \tag{8.1}$$

$$G \;=\; [EI - H_0 - U - \Sigma]^{-1} \tag{8.2}$$

$$A \;=\; i[G - G^+] \;\;,\;\; \Gamma \;=\; i[\Sigma - \Sigma^+] \tag{8.3}$$

where $\quad \Sigma^{in} \;=\; \Sigma_1^{in} + \Sigma_2^{in} + \Sigma_s^{in}$

$$\Sigma \;=\; \Sigma_1 + \Sigma_2 + \Sigma_s \tag{8.4}$$

These equations can be used to calculate the correlation function $G^n$ and hence the density matrix $\rho$ whose diagonal elements give us the electron density.

$$\rho \quad = \quad \int dE \; G^n(E)\,/\,2\pi \tag{8.5}$$

The current (per spin) at any terminal 'i' can be calculated from

$$I_i \quad = \quad (q/\hbar) \int_{-\infty}^{+\infty} dE \; \tilde{I}_i(E)\,/\,2\pi \tag{8.6}$$

$$\text{with} \quad \tilde{I}_i \quad = \quad \text{Trace}\,[\Sigma_i^{in} A] - \text{Trace}\,[\Gamma_i G^n] \tag{8.7}$$

which is shown in Fig.8.1 in terms of an inflow ($\Sigma_i^{in} A$) and an outflow ($\Gamma_i G^n$). The full time-dependent versions of these equations are derived in Section B.2, B.3 and B.4 from which the steady-state versions stated above are obtained.

***Input parameters:*** To use these equations, we need a channel Hamiltonian $[H_0]$ and the inscattering $[\Sigma^{in}]$ and broadening $[\Gamma]$ functions. For the two contacts, these are related:

$$\Sigma_1^{in} \quad = \quad \Gamma_1 f_1 \qquad \text{and} \qquad \Sigma_2^{in} \quad = \quad \Gamma_2 f_2 \tag{8.8}$$

and the broadening / self-energy for each contact can be determined from a knowledge of the surface spectral function (a) / surface Green's function (g) of the contact and the matrices $[\tau]$ describing the channel contact coupling:

$$\Gamma \quad = \quad \tau a \tau^+ \; \text{ and } \quad \Sigma \quad = \quad \tau g \tau^+ \tag{8.9}$$

Finally one needs a model (Hartree-Fock, density functional theory etc) for relating the self-consistent potential U to the density matrix. This aspect of the problem needs further work, since not much of the work in quantum chemistry has been geared towards transport problems.

***"Scattering contact":*** The NEGF equations without the 's' contact is often used to analyze small devices and in this form it is identical to the result obtained by Meir and Wingreen (see Eq.(6) of Ref.[12b]). The third "contact" labeled 's' represents scattering processes, without which we cannot make the transition to Ohm's law. Indeed it is only with the advent of mesoscopic physics in the 1980's that the importance of the contacts ($\Gamma_1$ and $\Gamma_2$) in interpreting experiments became widely recognized.

Prior to that, it was common to ignore the contacts as minor experimental distractions and try to understand the physics of conduction in terms of the 's' contact, though no one (to my knowledge) thought of scattering as a "contact" till Buttiker introduced the idea phenomenologically in the mid-80's (see Ref.[9e]). Subsequently, it was shown [12c] from a microscopic model that incoherent scattering processes in the NEGF method act like a fictitious "contact" distributed throughout the channel that extracts and reinjects electrons. Like the real contacts, coupling to this "contact" too can be described by a broadening matrix $\Gamma_s$. However, unlike the real contacts, the scattering contact in general cannot be described by a Fermi function so that although the outflow is given by Trace $[\Gamma_s G^n / 2\pi]$, the inflow is more complicated. For the scattering "terminal", unlike the contacts, there is no simple connection between $\Sigma_s^{in}$ and $\Sigma_s$ (or $\Gamma_s$). Moreover, these quantities are related to $G^n$ and have to be computed self-consistently. The relevant equations are derived in Section B.4 can be viewed as the matrix version of eqs. (7.21 a, b) [17].

***Derivation of NEGF equations:*** The full set of equations are usually derived using the non-equilibrium Green's function (NEGF) formalism, also called the Keldysh or the Kadanoff-Baym formalism initiated by the works of Schwinger, Baym, Kadanoff and Keldysh in the 1960s. However, their work was motivated largely by the problem of providing a systematic perturbative treatment of electron-electron interactions, a problem that demands the full power of this formalism. By contrast, we are discussing a much simpler problem, with interactions treated only to lowest order.

Indeed it is quite common to ignore interactions completely (except for the self-consistent potential) assuming "coherent transport". The NEGF equations for coherent transport can be derived from a one-electron Schrödinger equation without the advanced formal machinery [12d]. We start by partitioning the Schrödinger equation into three parts, the channel and the source and drain contacts (fig. 8.2)

$$i\hbar \frac{d}{dt} \begin{Bmatrix} \Phi_S \\ \psi \\ \Phi_D \end{Bmatrix} = \begin{bmatrix} H_S - i\eta & \tau_S^+ & 0 \\ \tau_S & H & \tau_D \\ 0 & \tau_D & H_D - i\eta \end{bmatrix} \begin{Bmatrix} \Phi_S \\ \psi \\ \Phi_D \end{Bmatrix} \qquad (8.10)$$

***Supriyo Datta, Purdue University***

with an infinitesimal $i\eta$ added to represent the extraction and injection of electrons from each of the contacts.

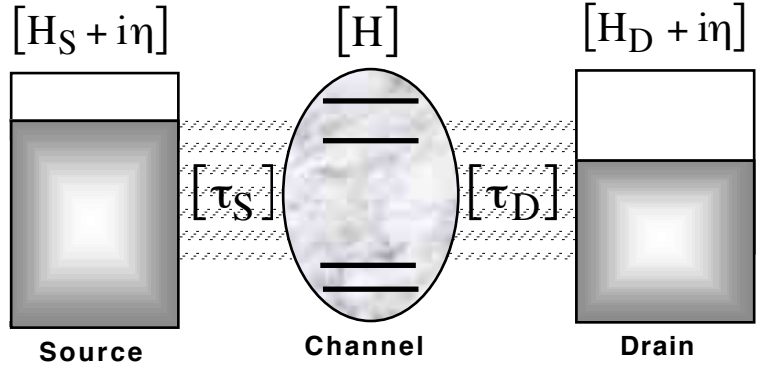

**Fig.8.2. A channel connected to two contacts.**

It is possible to eliminate the contacts, to write a Schrodinger-like equation for the channel alone

$$i\hbar\frac{d\psi}{dt} \;-\; H\psi \;-\; \Sigma\psi \;=\; S$$

$$\qquad\qquad\quad \textit{Outflow}\quad\;\textit{Inflow} \qquad\qquad (8.11)$$

with an additional self-energy term '$\Sigma\psi$' and a source term 'S' that give rise to outflow and inflow respectively. Note that unlike [H], the self-energy [$\Sigma$] is non-Hermitian and gives rise to an outflow of electrons. The additional terms in Eq.(8.2) are reminiscent of the frictional term and the noise term added to Newton's law to obtain the Langevin equation

$$m\frac{dv}{dt} \;+\; \gamma v \;=\; F \;+\; N(t)$$

$$\qquad\;\; \textit{Friction}\quad\;\; \textit{Force}\qquad \textit{Noise}$$

describing a Brownian particle [13]. Equivalently, one can move to a collective picture and balance inflow with outflow to obtain the Boltzmann equation. With quantum dynamics too we can express the inflow and outflow in terms of the correlation functions: $G^n \sim \psi\psi^+$, $\Sigma^{in} \sim SS^+$ and relate them to obtain the NEGF equations (sometimes called the quantum Boltzmann equation).

***Beyond the one-electron picture:*** A proper derivation of the NEGF equations, however, requires us to go beyond this one-electron picture, especially if non-coherent processes are involved. For example, the self-energy term '$\Sigma\psi$' in Eq.(8.2) represents the outflow of the electrons and it is natural to ask if $\Sigma$ (whose imaginary part gives the broadening or the inverse lifetime) should depend on whether the final state (to which outflow occurs) is empty or full. Such exclusion principle factors do not appear as long as purely coherent processes are involved. But they do arise for non-coherent interactions in a non-obvious way that is hard to rationalize from the one-electron picture.

In the one-electron picture, individual electrons are described by a one-electron wavefunction $\psi$ and the electron density is obtained by summing $\psi^*\psi$ from different electrons. A more comprehensive viewpoint describes the electrons in terms of field operators 'c' such that '$c^+c$' is the number operator which can take on one of two values '0' or '1' indicating whether a state is empty or full. These "second quantized" operators obey differential equations

$$ i\hbar\frac{d}{dt}c \;-\; Hc \;-\; \Sigma c \;=\; S \qquad\qquad (8.12) $$

that look much like the ones describing one-electron wavefunctions (see Eq.(8.11)). But unlike $\psi^*\psi$ which can take on any value, operators like $c^+c$ can only take on one of two values '0' or '1', thereby reflecting a particulate aspect that is missing from the Schrödinger equation. This advanced formalism is needed to progress beyond coherent quantum transport to inelastic interactions and onto more subtle many-electron phenomena like the Kondo effect.

A derivation of Eq.(8.12) leading to the NEGF equations is provided in Appendix B using second quantization for the benefit of advanced readers. However, in this derivation I have not used advanced concepts like the "Keldysh contour" which are needed for a systematic treatment of higher order processes. While future works in the field will undoubtedly require us to go beyond the lowest order treatment discussed here, it is not clear whether a higher order perturbative treatment will be useful or whether non-perturbative treatments will be required that describe the transport of composite or dressed particles obtained by appropriate unitary transformations of the bare electron operator 'c'.

## 9. Open questions

Let me end by listing what I see as the open questions in the field of nanoscale electronic transport.

***Model Hamiltonian:*** Once the matrices [H] and [Σ] are known the NEGF equations provide a well-defined prescription for calculating the current-voltage characteristics. For concrete calculations one needs to adopt a suitable basis like tight-binding / Huckel / extended Huckel / Gaussian described in the literature [14] in order to write down the matrices [H] and [Σ].

We could visualize the Hamiltonian [H] as a network of unit cells described by matrices [$H_{nn}$] whose size (bxb) is determined by the number of basis functions (b) per unit cell. Different unit cells are coupled through the "bond matrices" [$H_{nm}$].

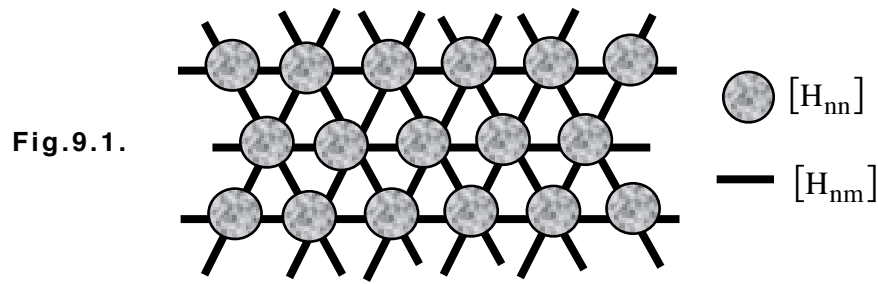

**Fig.9.1.**

$[H_{nn}]$

$[H_{nm}]$

The overall size of [H] is (Nb x Nb), N being the number of unit cells. The self-energy matrix [Σ] is also of the same size as [H], although it represents the effect of the infinite contacts. It can be evaluated from knowledge of the coupling matrices [$\tau_S$] and [$\tau_D$] (see Fig.8.2) and the surface properties of the contacts, as expressed through its surface Green's function (see eq. (8.9)). The matrices [H] and [Σ] thus provide a kind of intellectual partitioning: [H] expresses the properties of the channel while [Σ] depends on the interface with the contacts. In specific problems it may be desirable to borrow [H] and [Σ] from two different communities (like quantum chemists and surface physicists), but the process is made difficult by the fact that they often use different basis functions and self-consistent fields (see below). Much work remains to be done along these lines. Indeed, sometimes it may not even be clear where the channel ends and the contact starts!

*Transient transport:* Most of the current work to date has been limited to steady-state transport, but it is likely that future experiments will reveal transient effects whose time constants are controlled by the quantum dynamics, rather than circuit or RC effects [18]. The time-dependent NEGF equations [19] should be useful in modeling such phenomena.

*Self-consistent field:* An important conceptual issue in need of clarification is the treatment of electron-electron interactions. Discovering an appropriate self-consistent field U(N) to replace our simple ansatz (cf. Eq.(5.1b))

$$U(N) \quad = \quad q^2 \left[ N - N_0 \right] / C_E$$

is arguably one of the central topics in many-electron physics. Quantum chemists have developed sophisticated models for the self-consistent field like Hartree-Fock (HF) and Density Functional Theory (DFT) in additon to a host of semi-empirical approaches which can all give very different energy level structures. A lot of work has gone into optimizing these models but largely with respect to ground-state calculations and it is not clear what the best choice is for electron transport problems.

One could argue that electron transport involves adding and removing electrons and as such one should be looking at difference between the energies of the (N±1) electron system relative to the ground state of the N-electron system. However, for large broadening, the wavefunctions are significantly delocalized from the channel into the contacts, so that the number of electrons in the channel can change by fractional amounts. The best choice of a self-consistent field for transport problems a careful consideration of the degree of delocalization as measured by the relative magnitudes of the broadening and the charging.

*Transport regimes:* In this context it is useful to distinguish broadly between three different transport regimes for small conductors depending on the degree of delocalization:

**Self-consistent field (SCF) regime :** If the thermal energy $k_BT$ and / or the broadening $\gamma$ are comparable to the single-electron charging energy $U_0$, we can use the scf method described in this article. However, the optimum choice of the self-consistent potential needs to be clarified.

**Coulomb blockade (CB) regime:** If $U_0$ is well in excess of both $k_BT$ and $\gamma$, the scf method is not adequate, at least not the restricted one. More correctly, one could use (if practicable) the multielectron master equation described in appendix A [15].

**Intermediate Regime:** If $U_0$ is comparable to the larger of $k_BT$, $\gamma$, there is no simple approach: The scf method does not do justice to the charging, while the master equation does not do justice to the broadening and a different approach is needed to capture the observed physics[16].

With large conductors too we can envision three regimes of transport that evolve out of these three regimes. We could view a large conductor as an array of unit cells as shown in Fig.9.2.

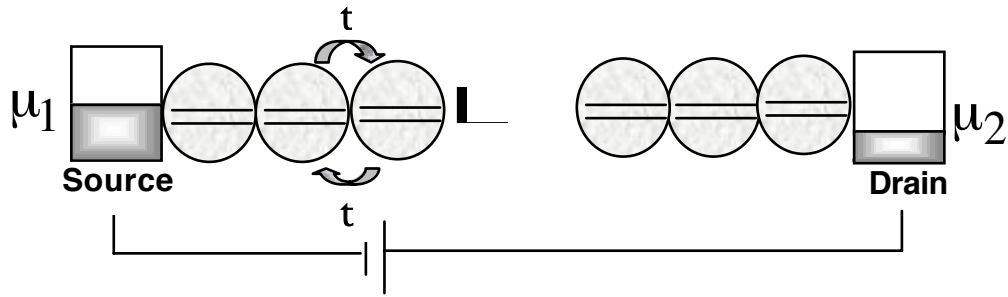

**Fig.9.2. A large conductor can be viewed as an array of unit cells. If the conductor is extended in the transverse plane, we should view each unit cell as representing an array of unit cells in the transverse direction.**

The inter-unit coupling energy 't' has an effect somewhat (but not exactly) similar to the broadening '$\gamma$' that we have associated with the contacts. If t $\geq$ $U_0$, the overall conduction will be in the **SCF** regime and can be treated using the method described here. If t << $U_0$, it will be in the **CB** regime and can in principle be treated using the multielectron master equation under certain conditions (specifically if " t " is much less than the level broadening $\gamma_s$). On the other hand, large conductors with $\gamma_s \ll t \leq U_0$ belong to an intermediate regime that presents major theoretical challenges [20], giving rise to intriguing possibilities. Indeed many believe that the high-$T_c$ superconductors (whose microscopic theory is yet to be discovered) consist of unit cells whose coupling is delicately balanced at the borderline of the SCF and the CB regimes.

I believe the field of nanoelectronics is currently at a very exciting stage where important advances can be expected from both applied and basic points of view. On the one hand we will continue to acquire a better quantitative understanding of nanoscale devices based on nanowires, nanotubes, molecules and other nanostructured materials. Although many of the observations appear to be described well within the basic self-consistent field model discussed here, much remains to be done in terms of discovering better basis functions for representing the Hamiltonian [H] and self-energy [$\Sigma$] matrices (see Fig.8.1), including inelastic scattering processes and implementing more efficient algorithms for the solution of the quantum transport equations. At the same time we can hope to discover new quantum transport phenomena (both steady-state and time dependent) involving strong electron-phonon and electron-electron interactions, which are largely unexplored. A notable exception is Coulomb blockade arising from strong electron-electron interactions which is fairly well understood. In Appendix A, I have tried to provide a brief introduction to this transport regime and relate it to the self-consistent field regime that forms the core of this tutorial. But all this I believe represents the "tip of the iceberg". The progress of molecular electronics should lead to greater control of the degree of hybridization between the localized strongly interacting molecular states and the delocalized contact states, thereby allowing a systematic study of different transport regimes. Such a study should reveal many more subtle phenomena involving electrons "dressed" by a variety of strong interactions that require non-perturbative treatments far beyond those described here.

***Acknowledgements:*** It is a pleasure to thank my colleagues Ron Reifenberger, Magnus Paulsson, Mark Lundstrom and Avik Ghosh for looking at preliminary versions of this manuscript and providing me with useful feedback.

# References

In keeping with the tutorial spirit of this article, I have only listed a few related references that the reader may find helpful in clarifying the concepts and viewpoint presented here. This is not intended to provide a comprehensive (or even representative) list of the extensive literature in the field of quantum transport.

   d.  L.P. Kouwenhoven and P.L. McEuen,"Single Electron Transport through
       a Quantum Dot", Chapter 13 in Nano- Science and Technology, ed. G.
       Timp, AIP Press (1997).

   e.  Bonet E, Deshmark M and Ralph DC, "Solving rate equations for electron
       tunneling via discrete quantum states", Phys. Rev. B vol. 65, 045317
       (1992).

16. For  further reading about the Kondo resonance observed in this transport regime,
see for example

       a. L. Kouwenhoven and L. Glazman, "Revival of the Kondo Effect", Physics
       World, p.33, January 2001.

       b. P. Fulde," Electron Correlations in Molecules and Solids", Springer-
       Verlag (1991).

The Green's function equations described for the Kondo resonance in Appendix B.5
are the same as those presented in

       c. Y. Meir, N.S. Wingreen and P.A. Lee, "Transport through a Strongly
       Interacting Electron Syatem: Theory of Periodic Conductance Oscillations",
       Phys. Rev. Lett., **66**, 3048 (1991).

17. The lowest order treatment of electron-phonon interaction described in Appendix
B.4 can be compared to that described in Ref.[12a]. If we assume both electron and
phonon distributions to be in equilibrium, our results can be shown to reduce to what
is known as Migdal's "Theorem" in the theory of electron-phonon interactions in
metals. See for example, Section II of

       P.B. Allen and B. Mitrovic,"Theory of Superconducting $T_c$", in Solid State
       Physics, vol.37, p.1, ed. H. Ehrenreich, F. Seitz and D. Turnbull, Academic
       (1982). See Eq.(3.47), p.20.

18. See for example, Fedorets, Gorelik L.Y., Shekhler R. I. and Jonson M,
"Vibrational Instability due to Coherent Tunneling of Electrons", Europhys. Lett.
Vol. 58, p.99 (2002).

19. The time-dependent equations in Appendix B.3 can be compared to those in

       A.P. Jauho, N.S. Wingreen and Y. Meir, "Time-dependent transport in
       interacting and non-interacting resonant tunneling systems", Phys. Rev.
       **B50**, 5528 (1994).

20. See for example, Georges A, "Strongly Correlated Electron Materials: Dynamical
Mean Field Theory and Electronic Structure", cond-mat 0403123.

*Supriyo Datta, Purdue University*

# Electrical Resistance:
# An Atomistic View  <u>Appendices A, B and C</u>

## Supriyo Datta
**School of Electrical & Computer Engineering**
**Purdue University, W. Lafayette, IN  47907**
*(http://dynamo.ecn.purdue.edu/~datta)*

**Abstract**

This tutorial article presents a "bottom-up" view of electrical resistance starting from the conductance of something really small, like a molecule, and then discussing the issues that arise as we move to bigger conductors. Remarkably enough, no serious quantum mechanics is needed to understand electrical conduction through something really small, except for unusual things like the Kondo effect that are seen only for a special range of parameters. This article starts with energy level diagrams (Section 2), shows that the broadening that accompanies coupling limits the conductance to a maximum of $q^2/h$ per level (Sections 3, 4), describes how a change in the shape of the self-consistent potential profile can turn a symmetric current-voltage characteristic into a rectifying one (Sections 5, 6), shows that interesting effects in molecular electronics can be understood in terms of a simple model (Section 7), introduces the non-equilibrium Green's function (NEGF) formalism as a sophisticated version of this simple model with ordinary numbers replaced by appropriate matrices (Section 8) and ends with a personal view of unsolved problems in the field of nanoscale electron transport (Section 9). Appendix A discusses the Coulomb blockade regime of transport, while Appendix B presents a formal derivation of the NEGF equations for advanced readers. MATLAB codes for numerical examples are listed in Appendix C and can be downloaded from www.nanohub.org, where they can also be run without installation.

*Supriyo Datta, Purdue University*

**Appendix A / Coulomb blockade**

Our discussion of charging effects in Sections 5 and 6 is based on the one-electron picture which assumes that a typical electron feels some average field due to the other electrons. Failure of this one-electron picture is known to give rise to observable effects that will continue to be discovered as the field progresses. Such effects are largely outside the scope of this article. However, there is one aspect that is fairly well understood and can affect our picture of current flow even for a simple one-level device putting it in the so-called Coulomb blockade or single-electron charging regime. In this Appendix I will try to explain what this means and how it relates to the self-consistent field model.

Energy levels come in pairs, one up-spin and one down-spin which are degenerate, that is have the same energy. Even the smallest device has two levels rather than one, and its maximum conductance will be twice the conductance quantum $G_0 \equiv q^2/h$ discussed earlier (see Eq.(1.1)). Usually this simply means that all our results have to be multiplied by two. Consider for example, a channel with two spin-degenerate levels (Fig.A.1), containing one electron when neutral ($N_0 = 1$, see Eq.(5.1b)). We expect the broadened density of states to be twice our previous result (see Eq.(4.2))

$$D_\varepsilon(E) \quad = \quad 2\,(\text{for spin})\ \text{x}\ \frac{\gamma/2\pi}{\left(E-\varepsilon\right)^2 + \left(\gamma/2\right)^2} \qquad (A.1)$$

where the total broadening is the sum of those due to each of the two contacts individually: $\gamma = \gamma_1 + \gamma_2$, as before. Since the available states are only half filled for a neutral channel, the electrochemical potential will lie exactly in the middle of the broadened density of states, so that we would expect a lot of current to flow when a bias is applied to split the electrochemical potentials in the source and drain as shown.

However, under certain conditions the density of states looks like one of the two possibilities shown in Fig.A.2. The up-spin and the down-spin density of states splits into two parts separated by the single-electron charging energy

$$U_0 \quad \equiv \quad q^2/C_E \qquad (A.2)$$

Very little current flows when we apply a small bias since there are hardly any states between $\mu_1$ and $\mu_2$ and this "Coulomb blockade" has been experimentally observed for systems where the charging energy $U_0$ exceeds the broadening $\gamma$.

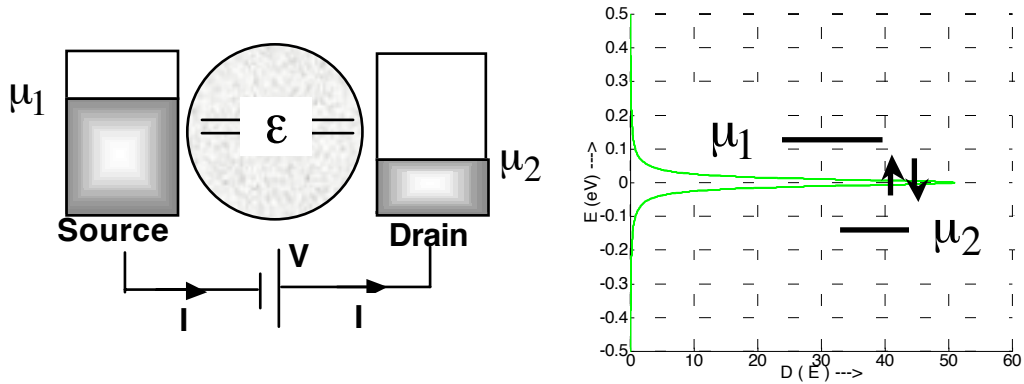

**Fig.A.1. A channel with two spin-degenerate levels containing one electron is expected to have an electrochemical potential that lies in the center of its broadened density of states, so that current should flow easily under bias ($\gamma = 0.05$ eV).**

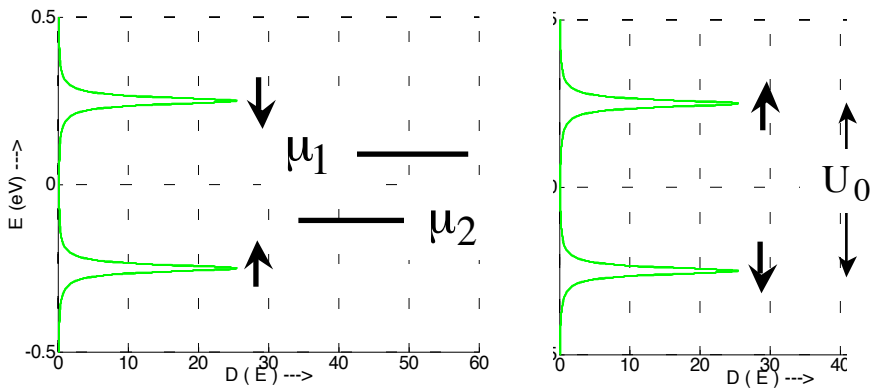

**Fig.A.2. Under certain conditions, the up-spin and down-spin density of states splits into two parts separated by the single-electron charging energy, $U_0$, instead of one single degenerate peak as shown in Fig.A.1 ($\gamma = 0.05$ eV, $U_0 = 0.25$ eV).**

It is hard to understand why the two peaks should separate based on the simple scf picture. Two peaks with the same energy ("degenerate") should always remain degenerate as long as they feel the same self-consistent potential U. One way to understand this is by arguing that ***no electron feels any potential due to itself.*** Suppose the up-spin level gets filled first causing the down-spin level to float up by $U_0$. But the up-spin level does not float up because it does not feel any self-interaction, leading to the picture shown on the left in Fig.A.2. Of course, it is just as likely that the down-spin will fill up first leading to the picture on the right. In either case the density of states near $\mu$ is suppressed relative to the scf picture (Fig.A.1).

A proper description of the flow of current in this Coulomb blockade regime requires a very different point of view that I will try to explain in the rest of this Appendix. But when do we have to worry about Coulomb blockade effects? Answer: Only if $U_0$ exceeds both $k_B T$ and $\gamma$ ( = $\gamma_1 + \gamma_2$). Otherwise, the scf method will give results that look much like those obtained from the correct treatment (see Fig.A.5). So what determines $U_0$? Answer: the extent of the electronic wavefunction. If we smear out one electron over the surface of a sphere of radius R, then we know from freshman physics that the potential of the sphere will be $q / 4\pi\epsilon R$ , so that the energy needed to put another electron on the sphere will be $q^2 / 4\pi\epsilon_r\epsilon_0 R \cong U_0$, which is $\sim 0.025$ eV if R = 5 nm and $\epsilon_r = 10$. Levels with well-delocalized wavefunctions (large R) have a very small $U_0$ and the SCF method provides an acceptable description even at the lowest temperatures of interest. But if R is small, then the charging energy $U_0$ can exceed $k_B T$ and one could be in a regime dominated by single-electron charging effects that is not described well by the self-consistent field method.

The self-consistent field (SCF) method is widely used because the exact method based on a multielectron picture is usually impossible to implement. However, it is possible to solve the multielectron problem exactly if we are dealing with a small channel, like the one-level system discussed in this article and assume that it is coupled very weakly to its surroundings (small $\gamma_{1,2}$). It is instructive to re-do this one-level problem in the multielectron picture and compare with the results obtained from the SCF method.

***One-electron vs. multielectron energy levels:*** If we have one spin degenerate level with energy $\boldsymbol{\varepsilon}$, the one-electron and multielectron energy levels would look as shown in Fig.A.3. Since each one-electron energy level can either be empty ('0') or occupied ('1'), multielectron states can be labeled in the form of binary numbers with a number of digits equal to the number of one-particle states. 'N' one-electron states thus give rise to $2^N$ multielectron states, which quickly diverges as N increases, making a direct treatment impractical. That is why SCF methods are so widely used, even though they are only approximate.

*One-electron energy levels*          *Multi-electron energy levels*

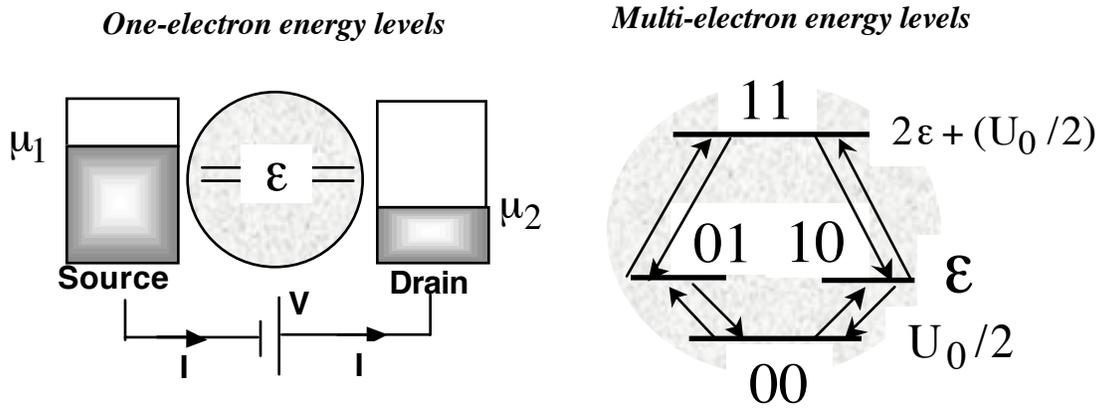

**Fig.A.3. One-electron vs. multielectron energy levels in a channel with one spin degenerate level having energy $\varepsilon$.**

For a small system with two one-electron states, there are only four multielectron states: 00, 01, 10 and 11 whose energy E(N) can be written in terms of the number of electrons as

$$E(N) \;=\; \varepsilon N + (U_0/2)\,(N - N_0)^2 \tag{A.3}$$

where the second term is the electrostatic energy stored in the capacitor formed by the channel and the surrounding contacts and is obtained from the relation $Q^2/2C_E$, by noting that $Q = q(N - N_0)$ is the net charge in this capacitor and $U_0$ is the single electron charging energy defined in Eq.(A.2). Eq.(A.3) can be used to obtain the energies for the different multielectron states 00, 01, 10 and 11 shown in Fig.A.3b if we set the number of electrons $N_0$ in a neutral molecule equal to one.

*Master equation:* In the multielectron picture, the overall system has different probabilities $P_\alpha$ of being in one of the $2^N$ possible states $\alpha$ and all the probabilities must add up to one:

$$\sum_{\alpha} P_{\alpha} \;=\; 1 \;\rightarrow\; P_{00} \;+\; P_{01} \;+\; P_{10} \;+\; P_{11} \;=\; 1 \qquad \text{(A.4a)}$$

We can calculate the individual probabilities by noting that the system is continually shuffled among these states and under steady state conditions there must be no net flow into or out of any state.

$$\sum_{\beta} R(\alpha \rightarrow \beta)\, P_{\alpha} \;=\; \sum_{\beta} R(\beta \rightarrow \alpha)\, P_{\beta} \qquad \text{(A.4b)}$$

Knowing the rate constants, we can calculate the probabilities by solving Eq.(A.4). Equations involving probabilities of different states are called master equations. We could call Eq.(A.4b) a multielectron master equation.

The rate constants $R(\alpha \rightarrow \beta)$ can be written down assuming a specific model for the interaction with the surroundings. For example, if we assume that the interaction only involves the entry and exit of individual electrons from the source and drain contacts then for the '00' and '01' states the rate constants are given by

$$\frac{\gamma_1}{h} f_1' + \frac{\gamma_2}{h} f_2' \qquad\qquad \overline{\quad 01 \quad} \;\; \varepsilon \qquad\qquad \frac{\gamma_1}{h}\left(1 - f_1'\right) + \frac{\gamma_2}{h}\left(1 - f_2'\right)$$

$$\overline{\quad 00 \quad}\;\; U_0/2$$

where $\qquad f_1' \;\equiv\; f_0(\varepsilon_1 - \mu_1) \qquad$ and $\qquad f_2' \;\equiv\; f_0(\varepsilon_1 - \mu_2)$

tell us the availability of electrons with energy $\qquad \varepsilon_1 \;=\; \varepsilon - (U_0/2) \qquad$ (A.5a)

in the source and drain contacts respectively. The entry rate is proportional to the available electrons, while the exit rate is proportional to the available empty states. The same picture applies to the flow between the '00' and the '10' states, assuming that up- and down-spin states are described by the same Fermi function in the contacts, as we would expect if each contact is locally in equilibrium.

Similarly we can write the rate constants for the flow between the '01' and the '11' states

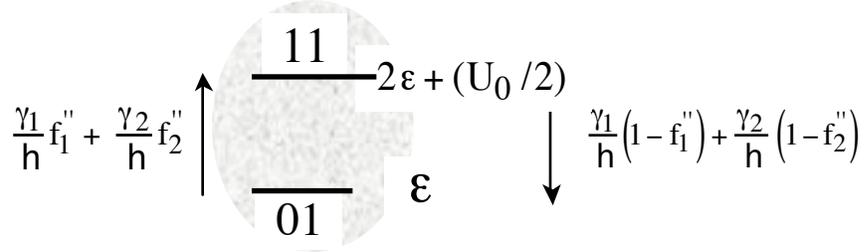

where            $f_1'' \equiv f_0(\varepsilon_2 - \mu_1)$ and $f_2'' \equiv f_0(\varepsilon_2 - \mu_2)$

tell us the availability of electrons with energy    $\varepsilon_2 = \varepsilon + (U_0/2)$           (A.5b)

corresponding to the energy difference between the 01 and 11 states. This is larger than the energy difference $\varepsilon$ between the 00 and 01 states because it takes an extra energy of $U_0$ to add an electron when one electron is already present.

Using these rate constants it is straightforward to show from Eq.(A.4b) that

$$\frac{P_{10}}{P_{00}} = \frac{P_{01}}{P_{00}} = \frac{\gamma_1 f_1' + \gamma_2 f_2'}{\gamma_1\left(1 - f_1'\right) + \gamma_2\left(1 - f_2'\right)} \qquad \text{(A.6a)}$$

and    $$\frac{P_{11}}{P_{10}} = \frac{P_{11}}{P_{01}} = \frac{\gamma_1 f_1'' + \gamma_2 f_2''}{\gamma_1\left(1 - f_1''\right) + \gamma_2\left(1 - f_2''\right)} \qquad \text{(A.6b)}$$

Together with Eq.(A.4a), this gives us all the individual probabilities. Fig.A.4 shows the evolution of these probabilities as the gate voltage $V_G$ is increased holding the drain voltage $V_D$ equal to zero. The gate voltage shifts the one-electron level $\varepsilon \rightarrow \varepsilon + U_L$ (we have assumed $U_L = -qV_G$) and the probabilities are calculated from Eqs.(A.6a,b) and (A.4a). The system starts out in the '00' state ($P_{00}=1$), shifts to the '01' and '10' states ($P_{01}=P_{10}=0.5$) once $\varepsilon_1 + U_L$ drops below $\mu$, and finally goes into the '11' state ($P_{11}=1$) when $\varepsilon_2 + U_L$ drops below $\mu$.

***Relation between the multielectron picture and the one-electron levels:*** One-electron energy levels represent ***differences*** between energy levels in the multielectron picture corresponding to states that differ by ***one electron***.

***Supriyo Datta, Purdue University***

**Fig.A.4. Evolution of the energy levels of a channel with one spin-degenerate level as the gate voltage $V_G$ is made more positive, holding the drain voltage $V_D$ equal to zero.**

$\mu = 0\,,\varepsilon = 0.2\,\text{eV}\,,k_B T = 0.025\,\text{eV}$

$U_0 = 0.25\,\text{eV}\,,U_L = -\,qV_G$

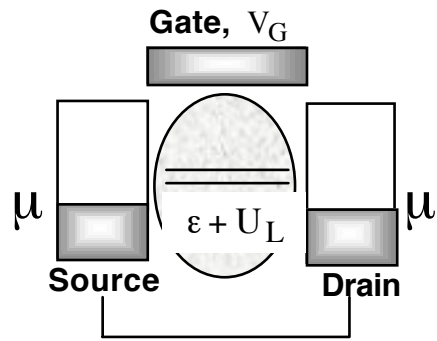

Gate, $V_G$

$\mu$

Source

$\varepsilon + U_L$

$\mu$

Drain

Evolution of one-electron energy levels

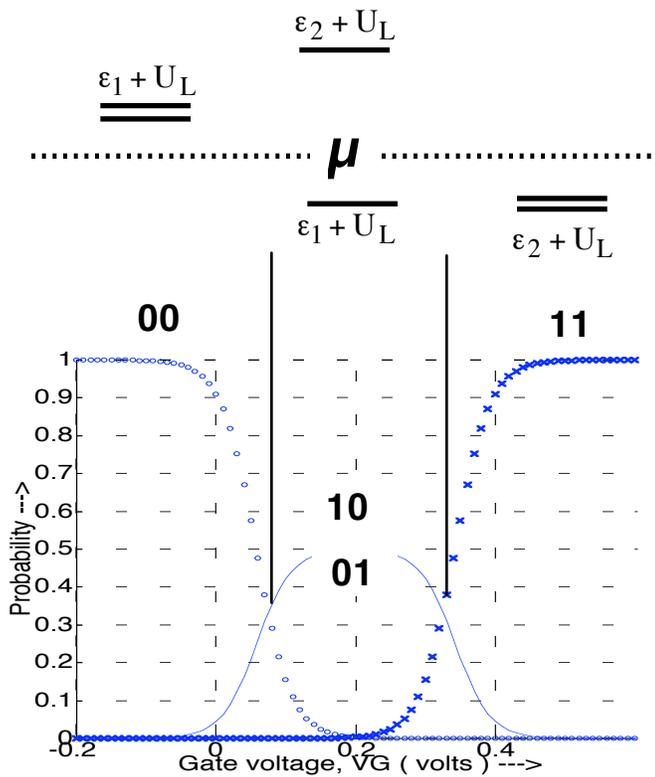

$\varepsilon_2 + U_L$

$\varepsilon_1 + U_L$

$\mu$

$\varepsilon_1 + U_L$

$\varepsilon_2 + U_L$

$\varepsilon_1 \equiv \varepsilon - (U_0\,/2)$

$\varepsilon_2 \equiv \varepsilon + (U_0\,/2)$

Evolution of many-electron state with gate voltage

Transitions involving the addition of one electron are called ***affinity*** levels while those corresponding to the removal of one electron are called ***ionization*** levels. For example (see Fig.A.4), if the system is in the '00' state then there are two degenerate one-electron levels $\varepsilon_1 + U_L$ representing

$$\varepsilon_1 + U_L = E\ (10) - E\ (00) = E(01) - E\ (00) \qquad \textbf{\textit{Affinity levels}}$$

Once it is in the '10' state there are two one-electron levels

$$\varepsilon_1 + U_L \qquad = E\ (10) - E\ (00) \qquad \textbf{\textit{Ionization level}}$$

$$\text{and} \quad \varepsilon_2 + U_L = E\ (11) - E\ (10) \qquad \textbf{\textit{Affinity level}}$$

In the '11' state there are two degenerate one-electron levels

$$\varepsilon_2 + U_L = E\ (11) - E\ (10) = E\ (11) - E\ (01) \qquad \textbf{\textit{Ionization levels}}$$

Affinity levels lie above $\mu$, while ionization levels lie below $\mu$ as shown in Fig.A.4. This is a very important general concept regarding the interpretation of the one-electron energy levels when dealing with complicated interacting objects. The occupied (or ionization) levels tell us the energy levels for removing an electron while the unoccupied (or affinity) levels tell us the energy levels for adding an extra electron. Indeed that is one way these levels are measured experimentally, the occupied levels by photoemission and the unoccupied levels by inverse photoemission as mentioned in Section 2.

***Law of equilibrium:*** Fig.A.4 represents an equilibrium calculation with both source and drain contacts having the same Fermi function: $f_1 = f_2$. Equilibrium problems do not really require the use of a master equation like Eq.(A.4). We can use the general principle of equilibrium statistical mechanics which states that the probability $P_\alpha$ that the system is in a multielectron state $\alpha$ with energy $E_\alpha$ and $N_\alpha$ electrons is given by

$$P_\alpha = \frac{1}{Z} \exp\left(- (E_\alpha - \mu N_\alpha)/k_B T\right) \qquad (A.7)$$

where the constant Z (called the partition function) is determined so as to ensure that the probabilities of all the states add up to one (Eq.(A.7)):

$$Z = \sum_{\alpha} \exp\left(-\left(E_{\alpha} - \mu N_{\alpha}\right)/k_B T\right) \qquad (A.8)$$

This is the central law of equilibrium statistical mechanics that is applicable to any system of particles (electrons, photons, atoms etc), interacting or otherwise [3.4]. The Fermi function is just a special case of this general relation that can be obtained by applying it to a system with just a single one-electron energy level, corresponding to two multielectron states:

| $\alpha$ | $N_{\alpha}$ | $E_{\alpha}$ | $P_{\alpha}$ |
|---|---|---|---|
| **0** | **0** | **0** | **1 / Z** |
| **1** | **1** | $\varepsilon$ | **(1 / Z)** $\exp\left[(\mu - \varepsilon)/k_B T\right]$ |

so that $Z = 1 + \exp\left[(\mu - \varepsilon)/k_B T\right]$ and it is straightforward to show that the average number of electrons is equal to the Fermi function (Eq.(2.1)):

$$N = \sum_{\alpha} N_{\alpha} P_{\alpha} = P_1 = \frac{\exp\left[(\mu - \varepsilon)/k_B T\right]}{1 + \exp\left[(\mu - \varepsilon)/k_B T\right]} = f_0\left[\varepsilon - \mu\right]$$

For multielectron systems, we can use the Fermi function only if the electrons are not interacting. It is then justifiable to single out one level and treat it independently ignoring the occupation of the other levels. The self-consistent field method uses the Fermi function assuming that the energy of each level depends on the occupation of the other levels. But this is only approximate. The exact method is to abandon the Fermi function altogether and use Eq.(A.7) instead to calculate the probabilities of the different multiparticle states.

Eq.(A.7), however, can only be used to treat equilibrium problems. Our primary interest is in calculating the current under non-equilibrium conditions and that is one reason we have emphasized the master equation approach based on Eq.(A.4). For equilibrium problems, it gives the same answer. However, it also helps to bring out an important conceptual point. One often hears concerns that the law of equilibrium is a statistical one that can only be applied to large systems. But it is apparent from the master equation approach that the law of equilibrium (Eq.(A.7)) is not a property of the system. It is a property of the contacts or the "reservoir". The only assumptions we have made relate to the energy distribution of the electrons that come in from the contacts. As long as the "reservoirs" are simple, it does not matter how complicated or how small the "system" is.

*Current calculation:* Getting back to non-equilibrium problems, once we have solved the master equation for the individual probabilities, the source current can be obtained from

$$I_1 \;=\; -q \sum_{\beta} (\pm) \, R_1(\alpha \rightarrow \beta) \, P_\alpha \qquad\qquad (A.10)$$

> **'+' if $\beta$ has one more electron than $\alpha$**
>
> **'-' if $\beta$ has one less electron than $\alpha$**

where $R_1$ represents the part of the total transition rate (R) associated with the source contact. In our present problem this reduces to evaluating the expression

$$I_1 \;=\; 2\gamma_1 f_1' P_{00} \quad - \quad \gamma_1\left(1-f_1'\right)\left(P_{01}+P_{10}\right)$$
$$+ \quad \gamma_1 f_1''\left(P_{01}+P_{10}\right) \quad - \quad 2\gamma_1\left(1-f_1''\right)P_{11}$$

Fig.A.5 shows the current-drain voltage ($I$-$V_D$) characteristics calculated from the approach just described. The result is compared with a calculation based on the restricted SCF method described in the main article. The two approaches agree well for $U_0 = 0.025$ eV, but differ appreciably for $U_0 = 0.25$ eV showing evidence for Coulomb blockade or single-electron charging.

The multielectron master equation provides a suitable framework for the analysis of current flow in the Coulomb blockade regime where the single electron charging energy $U_0$ is well in excess of the level broadening $\gamma_{1,2}$ and/or the thermal energy $k_B T$. We cannot use this method more generally for two reasons. Firstly, the size of the problem goes up exponentially and becomes prohibitive. Secondly, it is not clear how to incorporate broadening into this picture and apply it to the transport regime where the broadening is comparable to the other energy scales. And so it remains a major challenge to provide a proper theoretical description of the intermediate transport regime $U_0 \sim \gamma_{1,2}, k_B T$: the regime where electronic motion is "strongly correlated" making a two-electron probability like P(11) very different from the product of one-electron probabilities like P(01)P(10). A lot of work has gone into trying to discover a suitable self-consistent field within the one-electron picture that will capture the essential physics of correlation. For example, in the simple one-level problem we have used a self-consistent potential (see Eq.(5.1b)) $U_{scf} = U_0 \Delta N$, which is the same for all energy levels or orbitals. One could use an "unrestricted" self-consistent field that is orbital-dependent such that the potential felt by level 'j' excludes any self-interaction due to electrons in that level

$$U_{scf}(j) = U_0(\Delta N - \Delta n_j)$$

Such approaches can lead to better agreement with the results from the multielectron picture but have to be carefully evaluated, especially for non-equilibrium problems.

**(a)** $U_0 = 0.25$ eV                          **(b)** $U_0 = 0.025$ eV

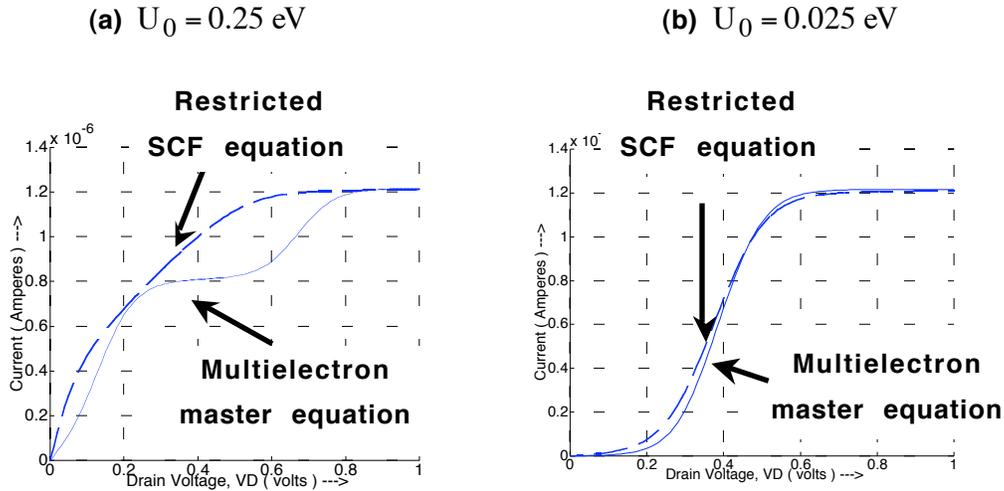

**Fig.A.5. Current vs, drain voltage $V_D$ calculated assuming $V_G = 0$ with $\mu = 0, \varepsilon = 0.2$ eV, $k_B T = 0.025$ eV , $\gamma_1 = \gamma_2 = 0.005$ eV , $U_L = -q V_D / 2$. The two approaches (the self-consistent field and the multielectron master equation) agree well for $U_0 = 0.025$ eV , but differ appreciably for $U_0 = 0.25$ eV showing evidence for Coulomb blockade or single-electron charging.**

## Appendix B / Formal derivation of NEGF equations

B.1. Second quantization

B.2. Non-equilibrium density matrix

B.3. Inflow and outflow

B.4. Incoherent flow

B.5. Coulomb blockade / Kondo resonance

The general "matrix" model listed in Section 8 (see Eqs.(8.1)-(8.9), Fig.8.1) can be derived starting from the one-electron Schrodinger equation

$$i\hbar\frac{d\psi}{dt} - H\psi = 0 \qquad\qquad (B.1.)$$

describing the time evolution of the one-electron wavefunction $\psi$ from which the electron density is obtained by summing $\psi^*\psi$ for different electrons. Although it is a little tricky explaining exactly what one means by "different electrons", this procedure is adequate for dealing with simple problems involving coherent interactions where the background remains unaffected by the flow of electrons. But to go beyond such phenomena onto more complex processes involving phase-breaking interactions or strong electron-electron interactions it is desirable to use a more comprehensive viewpoint that describes the electrons terms of field operators 'c'. For non-interacting electrons these second quantized operators obey differential equations

$$i\hbar\frac{d}{dt}c - Hc = 0 \qquad\qquad (B.2)$$

that look much like the one describing the one-electron wavefunction (see Eq.(B.1)). But unlike $\psi^*\psi$ which can take on any value, the number operator $c^+c$ can only take on one of two values '0' or '1' indicating whether a state is full or empty. At equilibrium, the average value of '$c^+c$' for a one-electron state with energy $\epsilon$ is given by the corresponding Fermi function $<c^+c> = f_0(\epsilon - \mu)$. However, this is true only if our channel consists of non-interacting electrons (perhaps with interactions described by a self-consistent field) in equilibrium. For non-equilibrium

*Supriyo Datta, Purdue University*

problems, a transport equation is needed that allows us to calculate 'c⁺c' based on our knowledge of the source term from the contacts which are assumed to remain locally in equilibrium, and hence described by the equilibrium relations.

In this Appendix, I will (1) introduce the field operators and define the correlation functions in ***Section B.1***, (2) use these advanced concepts to start from the second quantized description of the composite channel-contact system and eliminate the contact variables to obtain an effective equation for the channel having the form

$$i\hbar\frac{d}{dt}c \;-\; Hc \;-\; \underset{\text{Outflow}}{\sum_t c} \;=\; \underset{\text{Inflow}}{S}$$

from which we derive an expression for the non-equilibrium density matrix in ***Section B.2***. We will then (3) derive an expression for the current in ***Section B.3***, (4) derive the inscattering and broadening functions for incoherent processes in ***Section B.4***, and (5) show how the approach is extended to include strong electron-electron interactions leading to Coulomb blockade and the Kondo resonance (***Section B.5***). The approach presented here is different from the standard derivations, since it does not make use of the "Keldysh contour" which I believe is necessary only for a systematic treatment of perturbations to a higher order.

## B.1. Correlation functions

***Creation and annihilation operators:*** Consider a set of one-electron states labeled by 'j'. In the multielectron picture, we have two states $0_j$ and $1_j$ for each such one-electron state 'j' The creation operator $c_j^+$ inserts an electron in state 'j' taking us from $0_j$ to $1_j$ and can be represented in the form

$$
\begin{array}{cc} 0_j & 1_j \end{array}
$$

$$c_j(t) \;=\; \begin{bmatrix} 0 & 1 \\ 0 & 0 \end{bmatrix} \exp\left(-i\varepsilon_j t/\hbar\right) \qquad \text{(B.1.1a)}$$

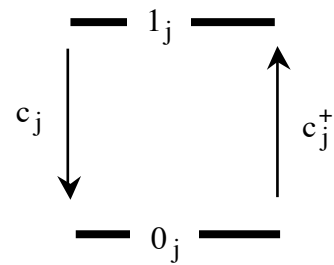

while the annihilation operator $c_j$ represented by

$$c_j^+(t) \quad = \quad \begin{bmatrix} 0 & 0 \\ 1 & 0 \end{bmatrix} \exp(+i\varepsilon_j t/\hbar) \qquad (B.1.1b)$$

takes an electron out of state 'j' taking us from $1_j$ to $0_j$. It is straightforward to show that

$$0_j \quad 1_j \qquad\qquad\qquad 0_j \quad 1_j$$

$$c_j^+(t)\, c_j(t) \quad = \quad \begin{bmatrix} 0 & 0 \\ 0 & 1 \end{bmatrix} \qquad c_j(t)\, c_j^+(t) \quad = \quad \begin{bmatrix} 1 & 0 \\ 0 & 0 \end{bmatrix}$$

independent of the time 't'. The former is called the number operator since its eigenvalues '0' and '1' represent the number of electrons in the corresponding state, while the latter tells us the number of empty states. Their expectation values are interpreted as the number of electrons and the number of holes in state 'j':

$$< c_j^+(t)\, c_j(t) > \quad = \quad f_j \quad \text{and} \quad < c_j(t)\, c_j^+(t) > \quad = \quad 1 - f_j \qquad (B.1.2)$$

It can be checked that

$$c_j(t)\, c_j^+(t) + c_j^+(t)\, c_j(t) \quad = \quad 1,$$

$$c_j(t)\, c_j(t) \quad = \quad 0 \quad \text{and} \quad c_j^+(t)\, c_j^+(t) \quad = \quad 0$$

What is much less intuitive is the relation between the field operators for two different states 'i' and 'j':

$$c_i(t)\, c_j^+(t) + c_j^+(t)\, c_i(t) \quad = \quad \delta_{ij} \qquad\qquad (B.1.3a)$$

$$c_i(t)\, c_j(t) + c_j(t)\, c_i(t) \quad = \quad 0 \qquad\qquad (B.1.3b)$$

$$c_i^+(t)\, c_j^+(t) + c_j^+(t)\, c_i^+(t) \quad = \quad 0 \qquad\qquad (B.1.3c)$$

Ordinarily we would expect the operators for two distinct states to be independent of each other, so that for $i \neq j$, $c_i(t) c_j^+(t) = c_j^+(t) c_i(t)$. Eq.(B.1.3a) would then imply that each is equal to zero. However, due to the exclusion principle, two distinct states are not really independent and $c_i(t) c_j^+(t) = -c_j^+(t) c_i(t)$. We can show that this is ensured if we modify Eqs.(B.1.1) to read

$$c_j(t) \quad = \quad \begin{bmatrix} 0 & (-1)^\nu \\ 0 & 0 \end{bmatrix} \exp(-i\,\varepsilon_j t /\hbar)$$

and

$$c_j^+(t) \quad = \quad \begin{bmatrix} 0 & 0 \\ (-1)^\nu & 0 \end{bmatrix} \exp(+i\,\varepsilon_j t/\hbar)$$

$\nu$ being the number of occupied states to the 'left' of state 'j'. This means that when dealing with a number of one-electron states, we need to agree on a specific order (does not matter what order we choose) and stick to it, so that 'left' has a well-defined meaning throughout the calculation. In practice, however, we do not need to worry about this, since we will be manipulating the operators making use of the algebra described by Eqs.(B.1.3). I just want to point out that this algebra, which is an expression of the exclusion principle, implies that putting an electron in state 'j' is affected by the presence or absence of another electron in state 'i', even before we have included any interactions between electrons.

***Correlation and spectral functions:*** In general we can define a two time electron correlation function (in the literature this is commonly written as $-i\,G_{ij}^<(t,t')$)

$$G_{ij}^n(t,t') \quad \equiv \quad <c_j^+(t')\,c_i(t)> \qquad\qquad \text{(B.1.4a)}$$

and a two time hole correlation function (in the literature this is commonly written as $+i\,G_{ij}^>(t,t')$)

$$G_{ij}^p(t,t') \quad \equiv \quad <c_i(t)\,c_j^+(t')> \qquad\qquad \text{(B.1.4b)}$$

whose values for equal time $t' = t$ give us the number operators in Eq.(B.1.2). Their sum is defined as the spectral function:

$$A_{ij}(t,t') \quad \equiv \quad G_{ij}^p(t,t') + G_{ij}^n(t,t') \qquad\qquad (B.1.4c)$$

*Fourier transformed functions:* Under steady-state conditions the correlation functions depend only on the difference between the two time coordinates and it is convenient to work with the Fourier transformed functions: *(t − t')* ➔ *E* :

$$G_{ij}^n(E) \quad \equiv \quad \int_{-\infty}^{+\infty} (d\tau/\hbar) \exp[\,i\,E\tau/\hbar\,]\, G_{ij}^n(t, t-\tau) \qquad (B.1.5a)$$

The inverse transform is given by

$$\begin{aligned} G_{ij}^n(t, t-\tau) \quad &\equiv \quad < c_j^+(t-\tau)\,c_i(t) > \\ &= \quad \int_{-\infty}^{+\infty} (dE/2\pi)\, G_{ij}^n(E) \exp(-i\,E\,\tau/\hbar) \qquad (B.1.5b) \end{aligned}$$

so that the equal time correlation function can be written as

$$G_{ij}^n(t) \quad \equiv \quad < c_j^+(t)\,c_i(t) > \quad = \quad \int_{-\infty}^{+\infty} (dE/2\pi)\, G_{ij}^n(E) \qquad\qquad (B.1.6)$$

Similar relations hold for the hole correlation function $G^p$ and the spectral function B. Note that the equal time correlation function in Eq.(B.6.6) is the same as the quantity usually called the *density matrix*.

*Equilibrium:* In general the electron and hole correlation functions in can take on any value so long as they add up to give the spectral function as indicated in Eq.(B.1.4c). But at equilibrium, the former is proportional to the Fermi function so that (I am using lower case symbols to indicate equilibrium quantities)

$$g_{ij}^n(E) \;=\; a_{ij}(E)\, f_0(E-\mu) \qquad\qquad (B.1.7a)$$

$$g_{ij}^p(E) \;=\; a_{ij}(E)\, (1-f_0(E-\mu)) \qquad\qquad (B.1.7b)$$

For an isolated system described by a Hamiltonian [h], the spectral function is written down easily in the eigenstate representation

$$a_{rs}(E) \;=\; \delta_{rs}\, \delta(E-\varepsilon_r) \qquad\qquad (B.1.7c)$$

where r,s are the eigenstates of [h], so that

$$g_{rs}^n(E) \;=\; \delta_{rs}\, \delta(E-\varepsilon_r) f_r \;,\; \text{with} \quad f_r \;\equiv\; f_0(\varepsilon_r-\mu) \qquad (B.1.8a)$$

$$g_{rs}^n(t,t') \;=\; \delta_{rs}\, f_r \exp\left(-i\varepsilon_r(t-t')/\hbar\right)\; \exp\left(-\eta|t-t'|/\hbar\right) \quad (B.1.8b)$$

$\eta$ being a positive infinitesimal. Our general approach is to use these relations with the appropriate Fermi functions for the contacts and then calculate the ***resulting*** correlation functions in the region of interest, namely the channel.

***Boson operators:*** In discussing problems that involve phonon or photon emission, we will need to include operators $b_\alpha^+$, $b_\alpha$ describing the phonon / photon fields. These operators obey a somewhat different algebra

$$b_\beta(t)\, b_\alpha^+(t) - b_\alpha^+(t)\, b_\beta(t) \;=\; \delta_{\alpha\beta} \qquad\qquad (B.1.9a)$$

$$b_\alpha(t)\, b_\beta(t) - b_\beta(t)\, b_\alpha(t) \;=\; 0 \qquad\qquad (B.1.9b)$$

$$b_\alpha^+(t)\, b_\beta^+(t) - b_\beta^+(t)\, b_\alpha^+(t) \;=\; 0 \qquad\qquad (B.1.9c)$$

where $\alpha$ , $\beta$ are the different phonon / photon modes. Comparing with Eqs.(B.1.3) for the electron operators, it is easy to see that the difference lies in replacing a positive sign with a negative one. However, this "minor" change makes these operators far more intuitive since distinct modes $\alpha \neq \beta$ now function independently

$$b_\beta(t)\, b_\alpha^+(t) = b_\alpha^+(t)\, b_\beta(t) \qquad\qquad , \qquad\qquad \alpha \quad \beta$$

as 'common sense' would dictate. Indeed, readers who have taken a course in quantum mechanics will recognize that in an operator treatment of the harmonic oscillator problem, one defines creation and annihilation operators that are linear combinations of the position (x) and momentum (p) operators which obey precisely the same algebra as in Eq.(B.1.12). What this means is that, unlike electron operators, we could represent the phonon/photon operators using ordinary differential operators. However, in this chapter we will not really use any representation. We will simply manipulate these operators making use of the algebra described in Eq.(B.1.12). One consequence of this change in the algebra is that instead of Eq.(B.1.2) we have

$$< b_\alpha^+(t) b_\alpha(t) > \; = N_\alpha \quad \text{and} \quad < b_\alpha(t) b_\alpha^+(t) > \; = 1 + N_\alpha \qquad \text{(B.1.10)}$$

where $N_\alpha$ is the number of phonons. As with the electron operators (see Fig.B.1.4), we can define two-time correlation functions

$$G_{\alpha\beta}^{ab}(t,t') \;\; \equiv \;\; < b_\beta^+(t') b_\alpha(t) > \qquad\qquad \text{(B.1.11a)}$$

$$\text{and} \quad G_{\alpha\beta}^{em}(t,t') \;\; \equiv \;\; < b_\alpha(t) b_\beta^+(t') > \qquad\qquad \text{(B.1.11b)}$$

which we have labeled with 'ab' and 'em' (instead of 'n' and 'p' for electrons) which stand for absorption and emission respectively. Under steady-state conditions these functions depend only on (t-t') which can be Fourier transformed to yield frequency domain correlation functions. At equilibrium (cf. Eq.(B.1.7))

$$g_{\alpha\beta}^{ab}(\hbar\omega) \;\; = \;\; a_{\alpha\beta}^{ph}(\hbar\omega) \, N(\hbar\omega)$$

$$g_{\alpha\beta}^{em}(\hbar\omega) \;\; = \;\; a_{\alpha\beta}^{ph}(\hbar\omega) \, (1 + N(\hbar\omega)) \qquad\qquad \text{(B.1.12)}$$

where the Fermi function has been replaced by the Bose function:

$$N(\hbar\omega) \;\; = \;\; (\exp(\hbar\omega/k_B T) - 1)^{-1} \qquad\qquad \text{(B.1.16)}$$

and the phonon spectral function in the eigenmode representation is given by

$$a_{\alpha\beta}^{ph}(\hbar\omega) \;=\; \delta_{\alpha\beta}\,\delta(\hbar\omega - \hbar\omega_\alpha) \qquad\qquad (B.1.17)$$

so that $g_{\alpha\beta}^{ab}(E) \;=\; \delta_{\alpha\beta}\,\delta(\hbar\omega - \hbar\omega_\alpha)N_\alpha$ , with $\quad N_\alpha \;\equiv\; N(\hbar\omega_\alpha)$

$$g_{\alpha\beta}^{ab}(t,t') \;=\; \delta_{\alpha\beta}\,N_\alpha\,\exp(-i\omega(t-t'))\;\exp(-\eta|t-t'|/\hbar)$$
$$(B.1.18)$$

and $\quad g_{\alpha\beta}^{em}(E) \;=\; \delta_{\alpha\beta}\,\delta(\hbar\omega - \hbar\omega_\alpha)(N_\alpha + 1)$

$$g_{\alpha\beta}^{em}(t,t') \;=\; \delta_{\alpha\beta}\,(N_\alpha + 1)\exp(-i\omega(t-t'))\exp(-\eta|t-t'|/\hbar)$$

***Energy-time relationship:*** The next point to note is that the wavefunction $\psi(t)$ or the field operator $c(t)$ are not observable quantities. What is observable are correlation functions like $\psi*(t')\,\psi(t)$ or $c^+(t')c(t)$ , somewhat the same way that the noise voltage $V(t)$ across a resistor is described by its correlation function $V(t')V(t)$. Under steady-state conditions, such two-time correlation functions depend only on time differences (t-t') and the corresponding Fourier transform variable is the energy E. For example, an electron with a waveunction $\psi(t) = \psi_0 \exp(-i\varepsilon t/\hbar)$ has a correlation function of the form

$$\psi*(t')\,\psi(t) = \psi_0 * \psi_0 \exp[-i\varepsilon\,(t-t')/\hbar]$$

and the Fourier transform with respect to (t-t') is proportional to the delta function $\delta(E-\varepsilon)$ at $E = \varepsilon$. Steady-state phenomena can be described in terms of such Fourier transformed quantities with 'E' as an independent variable as we have been doing following the introduction of broadening in Section 1.3. In each of the following Sections, I will first derive expressions in the time domain and then Fourier transform with respect to (t-t') to obtain energy domain expressions suitable for steady-state analysis.

## B.2. Non-equilibrium density matrix

***Partitioning:*** Operators obeying Eqs.(B.1.12) are generally referred to as Boson operators while those obeying Eqs.(B.1.2) are referred to as Fermion operators. At ***equilibrium***, the corresponding correlation functions are given by Eqs.(B.1.10) and (B.1.15) respectively. The problem is to calculate them in the channel driven away from equilibrium and our strategy will be to assume that the channel is driven by the contacts and by the phonon/photon baths which are maintained in their respective equilibrium states by the environment.

To implement this procedure, we partition the overall structure into a part that is of interest (labeled channel) and a reservoir (labeled the contact) having a large continuous density of states. We start from the equations decribing the composite system

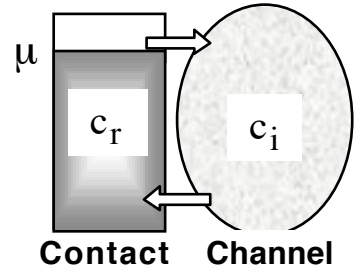

$$ i h \frac{d}{dt} \begin{Bmatrix} c_i \\ c_r \end{Bmatrix} = \begin{bmatrix} \varepsilon_i & \tau_{ir} \\ [\tau^+]_{ri} & \varepsilon_r \end{bmatrix} \begin{Bmatrix} c_i \\ c_r \end{Bmatrix} \qquad\qquad (B.2.1) $$

and obtain an effective equation for the channel of the form stated in Eq.(B.1.15), eliminating the reservoir variables 'r' ('i' and 'r' are assumed to represent eigenstates of the isolated channel and reservoir respectively) by assuming that they are maintained in equilibrium by the environment so that the corresponding fields are described by Eqs.(B.1.10). In this Section let me illustrate the approach assuming that the coupling with the contact involves purely elastic interactions, just as we assumed in our discussion of coherent transport in Chapters 8 and 9. As we will see the mathematics in this case will look just as if the quantities 'c' are ordinary complex numbers like the wavefunctions we have been using. We will not really need to make use of the fact that these are Fermion operators. But in the subsequent Sections we will discuss more general problems involving electron-phonon and electron-electron interactions, where we will make use of the properties of Fermion and Boson operators to obtain results that would be hard to rationalize from a one-electron picture.

***Eliminating the reservoir variables:*** The contact subset of Eqs.(B.2.1) yields

$$i\hbar \frac{d}{dt} c_r = (\varepsilon_r - i\eta) c_r + \sum_j [\tau^+]_{rj} c_j \qquad \text{(B.2.2)}$$

after adding an infinitesimal imaginary part $\eta$ to the energies that serves to broaden each level enough to make the density of states continuous. We can write the solution for the reservoir field in the form

$$c_r(t) = \underbrace{C_r(t)}_{\substack{\text{Isolated} \\ \text{Contact}}} + \underbrace{\sum_j \int_{-\infty}^{+\infty} dt_1 \, g_{rr}(t,t_1) [\tau^+]_{rj} c_j(t_1)}_{\text{Channel Induced}} \qquad \text{(B.2.3)}$$

where $\qquad g_{rr}(t,t') = (1/i\hbar)\, \vartheta(t-t')\, \exp{(-i\varepsilon_r - \eta)}\,(t-t') \qquad \text{(B.2.4)}$

represents the Green's function (or "impulse response") of the differential operator appearing in Eq.(B.2.2):

$$L_r \, g_{rr}(t,t') = \delta(t-t')$$

$$\text{with} \quad L_r \equiv i\hbar \frac{d}{dt} - (\varepsilon_r - i\eta) \qquad \text{(B.2.5)}$$

and $C_r(t)$ is the solution to the homogeneous equation: $\quad L_r \, C_r(t) = 0$

Physically, $C_r(t)$ represents the field operator in the ***isolated*** contact before connecting to the channel at $t = 0$. This allows us to use the law of equilibrium (see Eq.(B.1.10, 11)) to write down the corresponding correlation function ($f_r \equiv f_0(\varepsilon_r - \mu)$):

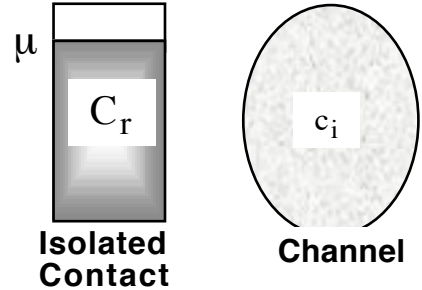

**Isolated Contact**    **Channel**

$$g_{rs}^n(E) = \delta_{rs}\, \delta(E - \varepsilon_r)\, f_r \qquad \text{(B.2.6a)}$$

$$g_{rs}^n(t,t') \equiv \langle C_s^+(t') C_r(t) \rangle$$

$$= \delta_{rs}\, f_r \, \exp{(-i\varepsilon_r(t-t'))}\, \exp{(-\eta\, |t-t'|)} \qquad \text{(B.2.6b)}$$

The channel itself is assumed to be completely empty prior to $t = 0$ and gets filled by the electrons "spilling over" from the contact after the connection is established at t = 0. This process is described by starting from the channel subset of Eq.(B.2.1):

$$ih \frac{d}{dt} c_i = \varepsilon_i c_i + \sum_j [\tau]_{ir} c_r$$

and substituting for $c_r(t)$ from Eq.(B.2.3):

$$ih \frac{d}{dt} c_i - \varepsilon_i c_i - \sum_j \int_{-\infty}^{+\infty} dt_1 \Sigma_{ij}(t,t_1) c_j(t_1) = S_i(t) \qquad (B.2.7)$$

$$\text{where} \quad \Sigma_{ij}(t,t') \equiv \sum_r [\tau]_{ir} g_{rr}(t,t') [\tau^+]_{rj} \qquad (B.2.8)$$

$$\text{and} \quad S_i(t) \equiv \sum_r [\tau]_{ir} C_r(t) \qquad (B.2.9)$$

***Equation relating correlation functions:*** Defining a Green's function for the integro-differential operator appearing in Eq.(B.2.7):

$$ih \frac{d}{dt} G_{ik}(t,t') - \varepsilon_i G_{ik}(t,t') - \sum_j \int_{-\infty}^{+\infty} dt_1 \Sigma_{ij}(t,t_1) G_{jk}(t_1,t')$$

$$= \delta_{ik} \delta(t-t')$$

we can write the solution to Eq.(B.2.7) in the form

$$c_i(t) = \sum_k \int_{-\infty}^{+\infty} dt_1 G_{ik}(t,t_1) S_k(t_1) \qquad (B.2.10)$$

so that the correlation function is given by

$$G_{ij}^n(t,t') \equiv < c_j^+(t') c_i(t) >$$

$$= \sum_{k,l} \int_{-\infty}^{+\infty} dt_1 \int_{-\infty}^{+\infty} dt_2 G_{ik}(t,t_1) G_{jl}^*(t',t_2) < S_l^+(t_2) S_k(t_1) >$$

Defining $\qquad \Sigma_{kl}{}^{in}(t_1, t_2) \;\; \equiv \;\; < S_l{}^+(t_2) S_k(t_1) > \qquad\qquad$ (B.2.11)

$\qquad$ and $\quad [G^+]_{lj}(t_2, t') \;\; = \;\; [G]_{jl}{}^*(t', t_2) \qquad\qquad$ (B.2.12)

we can write in matrix notation:

$$[G^n(t, t')] \;\; = \;\; \int_{-\infty}^{+\infty} dt_1 \int_{-\infty}^{+\infty} dt_2 \, [G(t, t_1)] \, [\Sigma^{in}(t_1, t_2)] \, [G^+(t_2, t')] \quad \text{(B.2.13)}$$

$\qquad$ where $\qquad L[G(t, t')] \;\; = \;\; [I] \, \delta(t - t')$

$$\qquad\qquad \text{with} \quad L \;\; = \;\; i\hbar \frac{d}{dt}[I] - [h] \; - \int_{-\infty}^{+\infty} dt_1 \, [\Sigma(t, t_1)] \qquad \text{(B.2.14)}$$

$$\qquad\qquad \text{and} \quad [\Sigma(t, t')] \;\; \equiv \;\; [\tau] \, [g(t, t')] \, [\tau^+] \qquad\qquad \text{(B.2.15)}$$

[h] being the Hamiltonian describing the isolated channel whose eigenstates are labeled by 'i'. Eq.(B.2.12) relates the correlation function in the channel to the inscattering function $\Sigma^{in}$ describing the correlation of the source term.

$\qquad$ Under steady-state conditions, all functions depend only on the difference between the two time coordinates so that the time integrals in Eq.(B.2.12) represent convolutions that turn into ordinary products if we Fourier transform with respect to the difference coordinate:

$$[G^n(E)] \;\; = \;\; [G(E)] \, [\Sigma^{in}(E)] \, [G^+(E)] \qquad\qquad\qquad \text{(B.2.16)}$$

$\qquad$ with $\quad L[G(E)] \;\; = \;\; [I],$

$$\qquad\qquad L \;\; = \;\; E[I] - [h] \; - \; [\Sigma(E)] \qquad\qquad\qquad \text{(B.2.17)}$$

$$\qquad\qquad\qquad \text{and} \quad [\Sigma(E)] \;\; \equiv \;\; [\tau] \, [g(E)] \, [\tau^+] \qquad\qquad \text{(B.2.18)}$$

***Inscattering function:*** To evaluate the inscattering function

$$\Sigma_{ij}^{in}(t, t') \;\; \equiv \;\; < S_j^+(t') S_i(t) >$$

we substitute for the source term from Eq.(B.2.9)

$$\Sigma_{ij}^{in}(t,t') \;=\; \sum_{s} \tau_{js}^{*} \sum_{r} \tau_{ir} <C_s^{+}(t')C_r(t)>$$

$$=\; \sum_{r,s} \tau_{ir}\, g_{rs}^{n}(t,t')\, [\tau^{+}]_{sj}$$

where $g_{rs}^{n}(t,t')$ is given by Eq.(B.2.6) since it represents the correlation function for the isolated contact in equilibrium if it were not connected to the channel. In matrix notation

$$[\Sigma^{in}(t,t')] \;=\; [\tau]\,[g^{n}(t,t')]\,[\tau^{+}] \tag{B.2.19}$$

Once again since this correlation function depends only on the difference between the two time arguments, we can Fourier transform with respect to the difference coordinate to write

$$[\Sigma^{in}(E)] \;=\; [\tau]\,[\,g^{n}(E)]\,[\tau^{+}] \tag{B.2.20}$$

***Inscattering vs. broadening:*** The anti-Hermitian component of the self-energy function (see Eq.(B.2.17)), also called the broadening function, is given by

$$[\Gamma(E)] \;=\; i\,[\Sigma(E) - \Sigma^{+}(E)] \;=\; [\tau]\,[\,a(E)]\,[\tau^{+}]$$

where $[a(E)]$ is the spectral function for the isolated contact: $[a] = i\,[\,g - g^{+}\,]$. If the contact is assumed to be in equilibrium with a Fermi function $f(E)$, then the inscattering function from Eq.(B.2.19) can be written as

$$[\Sigma^{in}(E)] \;=\; [\tau]\,[a(E)]\,[\tau^{+}]\,f(E)$$

so that the inscattering and the corresponding broadening are related :

$$[\Sigma^{in}(E)] \;=\; [\,\Gamma(E)]\;\; f(E) \tag{B.2.21}$$

### B.3. Inflow and outflow

We will now obtain an expression for the inflow and outflow, starting from the expression for the current (Note that $<c_i^+ c_i>$ tells us the number of electrons in state 'i')

$$
\begin{aligned}
I(t) &\equiv \sum_i \frac{d}{dt} < c_i^+(t) c_i(t) > \\
&= \frac{1}{i\hbar} \sum_i < c_i^+(t) \left( i\hbar \frac{d}{dt} c_i(t) \right) > \;\; - \;\; < \left( -i\hbar \frac{d}{dt} c_i^+(t) \right) c_i(t) >
\end{aligned}
$$

and substitute from Eq.(B.2.7) to obtain explicit expressions for the inflow and outflow. More generally we could define a two-time version I(t,t') as

$$
I(t, t') \;\equiv\; \sum_i \left( \frac{d}{dt} - \frac{d}{dt'} \right) \; < c_i^+(t') c_i(t) > \qquad\qquad (B.3.1)
$$

whose "diagonal" elements (t = t') give us the total current. The advantage of the two-time version is that for steady-state transport we can Fourier transform with respect to (t − t') to obtain the energy spectrum of the current. Substituting from Eq.(B.2.7) into Eq.(B.3.1) we obtain

$$
I(t, t') \;=\; I^{in}(t, t') - I^{out}(t, t')
$$

$$
\text{where} \quad I^{out}(t, t') = \;\; (-1/i\hbar) \sum_{i,j} \int_{-\infty}^{+\infty} dt_1 \, (\Sigma_{ij}(t, t_1) < c_i^+(t') c_j(t_1) >
$$

$$
- \;\; \Sigma_{ij}^*(t', t_1) < c_j^+(t_1) c_i(t) >)
$$

$$
= \;\; (-1/i\hbar) \sum_{i,j} \int_{-\infty}^{+\infty} dt_1 \, (\Sigma_{ij}(t, t_1) G_{ji}^n(t_1, t') \;\; - \;\; \Sigma_{ij}^*(t', t_1) G_{ij}^n(t, t_1))
$$

$$
\text{and} \quad I^{in}(t, t') \;\; = \;\; (1/i\hbar) \sum_i \, (c_i^+(t') S_i(t) - S_i^+(t') c_i(t))
$$

$$= \ (1/i\hbar) \sum_{i,j} \int_{-\infty}^{+\infty} dt_1 \ G_{ij}*(t',t_1) < S_j^+(t_1) S_i(t) > \ - \ G_{ij}(t,t_1) < S_i^+(t') S_j(t_1) >$$

(making use of Eq.(B.2.10))

$$= \ (1/i\hbar) \sum_{i,j} \int_{-\infty}^{+\infty} dt_1 \ G_{ij}*(t',t_1) \Sigma_{ij}^{in}(t,t_1) \ - \ G_{ij}(t,t_1) \Sigma_{ji}^{in}(t_1,t')$$

(making use of Eq.(B.2.11))

In matrix notation we can write [17]

$$I^{in}(t,t') = \ (1/i\hbar) \ \text{Trace} \int_{-\infty}^{+\infty} dt_1 \ [\Sigma^{in}(t,t_1)][G^+(t_1,t')]$$

$$- \ [G(t,t_1)][\Sigma^{in}(t_1,t')] \qquad (B.3.2)$$

$$I^{out}(t,t') = \ (1/i\hbar) \ \text{Trace} \int_{-\infty}^{+\infty} dt_1 \ [\Sigma(t,t_1)][G^n(t_1,t')]$$

$$- \ [G^n(t,t_1)][\Sigma^+(t_1,t')] \qquad (B.3.3)$$

Note that these are just the conduction currents to which one should add the displacement currents to get the net terminal current.

Once again, under steady-state conditions each of the quantities depends only on the difference between the two time arguments and on Fourier transforming with respect to the difference coordinate the convolution turns into a normal product:

$$I^{in}(E) \ = \ \text{Trace} \ [\Sigma^{in}(E) \ G^+(E) - G(E) \ \Sigma^{in}(E)]/i\hbar$$

$$= \ \text{Trace} \ [\Sigma^{in}(E) \ A(E)]/\hbar \qquad (B.3.4)$$

$$I^{out}(E) \ = \ \text{Trace} \ [G^n(E) \ \Sigma^+(E) - \Sigma(E) \ G^n(E)]/i\hbar$$

$$= \ \text{Trace} \ [\Gamma(E) \ G^n(E)]/\hbar \qquad (B.3.5)$$

***Multi-terminal devices:*** We have now obtained the basic expressions for inflow (Eq.(B.3.4)) and outflow (Eq.(B.3.5)) that we advertised in Fig.B.1.1. along with the kinetic equation (Eq.(B.2.15)). For simplicity we considered a channel connected to just one contact, but the results are readily extended to real devices with two or more contacts labeled by indices p,q. For example, Eqs.(B.2.12-14) are modified to read

$$[G^n(t,t')] \; = \; \sum_p \int_{-\infty}^{+\infty} dt_1 \int_{-\infty}^{+\infty} dt_2 \, [G(t,t_1)] \, [\Sigma^{in}(t_1,t_2)]^{(p)} \, [G^+(t_2,t')] \quad \text{(B.3.6)}$$

$$\text{where} \quad L\,[G(t,t')] \; = \; [I]\,\delta(t-t')$$

$$\text{with} \quad L \; \equiv \; i\hbar \frac{d}{dt}[I] - [h] \; - \; \sum_p \int_{-\infty}^{+\infty} dt_1 \, [\Sigma(t,t_1)]^{(p)} \quad \text{(B.3.7)}$$

$$\text{and} \quad\quad\quad [\Sigma(t,t')]^{(p)} \; \equiv \; [\tau]^{(p)} \, [g(t,t')]^{(p)} \, [\tau^+]^{(p)} \quad\quad \text{(B.3.8)}$$

while Eqs.(B.2.18), (B.3.1) and (B.3.2) become

$$[\Sigma^{in}(t,t')]^{(p)} \; = \; [\tau]^{(p)} \, [g^n(t,t')]^{(p)} \, [\tau^+]^{(p)} \quad\quad\quad \text{(B.3.9)}$$

$$I^{in}(t,t')^{(p)} = \; (1/i\hbar) \, \text{Trace} \int_{-\infty}^{+\infty} dt_1 \, [\Sigma^{in}(t,t_1)]^{(p)} \, [G^+(t_1,t')]$$

$$- \; [G(t,t_1)][\Sigma^{in}(t_1,t')]^{(p)}$$

$$\text{(B.3.10)}$$

$$I^{out}(t,t')^{(p)} = \; (-1/i\hbar) \, \text{Trace} \int_{-\infty}^{+\infty} dt_1 \, [\Sigma(t,t_1)]^{(p)} \, [G^n(t_1,t')]$$

$$- \; [G^n(t,t_1)][\Sigma^+(t_1,t')]^{(p)}$$

Under steady-state conditions the multiterminal versions take the form

$$[G^n(E)] \; = \; \sum_p \, [G(E)][\Sigma^{in}(E)]^{(p)} \, [G^+(E)] \quad\quad\quad \text{(B.3.11)}$$

$$L\,[G] \; = \; [I] \quad , \quad\quad L \; \equiv \; E\,[I] - [h] - \sum_p \, [\Sigma(E)]^{(p)} \quad \text{(B.3.12)}$$

$$[\Sigma(E)]^{(p)} \; \equiv \; [\tau]^{(p)} \, [g(E)]^{(p)} \, [\tau^+]^{(p)} \quad\quad\quad\quad \text{(B.3.13)}$$

$$[\Sigma^{in}(E)]^{(p)} \quad = \quad [\tau]^{(p)} \, [g^n(E)]^{(p)} \, [\tau^+]^{(p)} \qquad (B.3.14)$$

$$I^{in}(E)^{(p)} = \quad \text{Trace} \ [\Sigma^{in}(E)]^{(p)} \, [A(E)]/\hbar \qquad (B.3.15)$$

$$I^{out}(E)^{(p)} = \quad \text{Trace} \ [\Gamma(E)]^{(p)} \, [G^n(E)] \qquad (B.3.16)$$

The terminal current is given by the difference between the inflow and outflow:

$$I(E)^{(p)} = \quad (1/\hbar) \, \text{Trace} \ ([\Sigma^{in}(E)]^{(p)} \, [A(E)] - [\Gamma(E)]^{(p)} \, [G^n(E)])$$

Making use of Eq.(B.3.10) and the relation

$$A = \sum_q G\Gamma^{(q)}G^+ = \sum_q G^+\Gamma^{(q)}G$$

(which follows from Eq.(B.3.11)), we can write

$$I(E)^{(p)} = \quad (1/\hbar) \sum_q \text{Trace} \ (\Sigma^{in(p)} \, G\Gamma^{(q)}G^+ - \Gamma^{(p)} \, G\Sigma^{in(q)}G^+)$$

$$= \quad (1/\hbar) \sum_q \text{Trace} \ [\Gamma^{(p)} \, G\Gamma^{(q)}G^+] \, (f_p - f_q)$$

if $\Sigma^{in(p)} = \Gamma^{(p)} \, f_p$. We can then write the current as

$$I(E)^{(p)} = \quad (1/\hbar) \sum_q T^{(pq)} \, (f_p - f_q) \qquad (B.3.17)$$

in terms of the transmission function: $T^{(pq)} \equiv \ \text{Trace} \ [\Gamma^{(p)}G\Gamma^{(q)}G^+] \quad (B.3.18)$

  Linearizing Eq.(B.3.17) about the equilibrium electrochemical potential we obtain the standard Buttiker equations (using $\delta$ to denote the change from the equilibrium value)

$$\delta I(E)^{(p)} = \quad (1/\hbar) \sum_q T^{(pq)} \, (\delta\mu_p - \delta\mu_q) \qquad (B.3.19)$$

widely used in mesoscopic physics (see Ref.9e and Chapter 2 of Ref.9a).

*Supriyo Datta, Purdue University*

### B.4. Inelastic flow

Next let us consider a problem in which an electron can exit from the channel into the contact by emitting (creating) or absorbing (annihilating) a phonon. We start with the equation of motion for a channel field operator

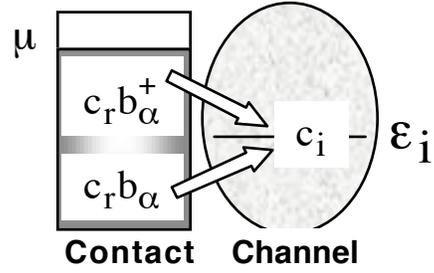

$$i\hbar \frac{d}{dt} c_i \; - \; \varepsilon_i \, c_i \; = \; s_i$$

$$\text{where} \quad s_i \; = \; \sum_{r,\alpha} \tau_{ir\,\alpha} \, c_r b_\alpha + \tau_{ri\alpha}{}^* \, c_r b_\alpha^+ \qquad (B.4.1)$$

Equations of motion of this type describing the time evolution of the field operators (in the Heisenberg picture) are obtained using the Heisenberg equations of motion for any operator 'P'

$$i\hbar \, dP/dt \; = \; PH - HP$$

where H is the second quantized Hamiltonian:

$$H \; = \; \sum_i \varepsilon_i \, c_i^{+} c_i \; + \; \sum_r \varepsilon_r \, c_r^{+} c_r \; + \; \sum_\alpha \hbar\omega_\alpha \, (b_\alpha^+ b_\alpha + 1/2)$$

$$+ \; \sum_{i,r,\alpha} (\tau_{ir\alpha} \, c_i^{+} c_r b_\alpha \; + \; \tau_{ir\alpha}^{*} \, c_r^{+} c_i b_\alpha^+ \; + \; \tau_{r\,i\alpha} \, c_r^{+} c_i b_\alpha \; + \; \tau_{r\,i\alpha}^{*} \, c_i^{+} c_r b_\alpha^+ )$$

$$(B.4.2)$$

The electron-phonon interaction term (last term in the above expression) is obtained by writing the potential U(x) felt by one electron in real space

$$U(x) \; = \; \sum_\alpha b_\alpha \, \tau_\alpha(x) \; + \; b_\alpha^{+} \, \tau_\alpha{}^*(x)$$

and then writing the second quantized version as we would for any other potential:

$$H_{el-ph} = \sum_{i,r} U_{ir} c_i^+ c_r + U_{ri} c_r^+ c_i$$

We will now evaluate the two source terms in Eq.(B.4.1) one by one.

***Evaluating*** $c_r b_\alpha$ ***:*** We start from the equations of motion for the individual electron and phonon operators

$$i h \frac{d}{dt} c_r = (\varepsilon_r - i\eta) c_r + \sum_{i,\alpha} \tau_{r i\alpha} c_i b_\alpha + \tau_{ir\alpha}{}^* c_i b_\alpha^+$$

$$i h \frac{d}{dt} b_\alpha = h\omega_\alpha b_\alpha + \sum_{i,r} \tau_{ri\alpha}{}^* c_i^+ c_r + \tau_{ir\alpha}{}^* c_r^+ c_i$$

and write an equation for the product

$$i h \frac{d}{dt} c_r b_\alpha = \left( i h \frac{d}{dt} c_r \right) b_\alpha + c_r \left( i h \frac{d}{dt} b_\alpha \right)$$

$$= \left( (\varepsilon_r - i\eta) c_r + \sum_{j,\beta} \tau_{r j\beta} c_j b_\beta + \tau_{jr\beta}{}^* c_j b_\beta^+ \right) b_\alpha$$

$$+ c_r \left( h\omega_\alpha b_\alpha + \sum_{j,s} (\tau_{s j\alpha}{}^* c_j^+ c_s + \tau_{js\alpha}{}^* c_s^+ c_j) \right)$$

which is simplified by replacing the products involving reservoir variables with their average values (which are time-independent numbers) to obtain

$$i h \frac{d}{dt} c_r b_\alpha - (\varepsilon_r - i\eta + h\omega_\alpha) c_r b_\alpha$$

$$= \sum_j \sum_{s,\beta} \tau_{js\beta}^* \left( < b_\beta^+ b_\alpha > \delta_{rs} + < c_r c_s^+ > \delta_{\alpha\beta} \right) c_j$$

$$= \sum_j \sum_{s,\beta} \tau_{js\beta}^* \left( < b_\beta^+ b_\alpha > < c_s^+ c_r > + < c_r c_s^+ > < b_\alpha b_\beta^+ > \right) c_j$$

For the last step we have made use of the relations $\delta_{rs} = c_r^+ c_s + c_s c_r^+$ (Eq.(B.1.3a)) and $\delta_{\alpha\beta} = b_\beta b_\alpha^+ - b_\alpha^+ b_\beta$ (Eq.(B.1.12a)).

As in the last Section we can write the solution as a sum of a solution to the homogeneous equation and a response induced by the source term:

$$c_r(t)b_\alpha(t) \;=\; C_r(t)B_\alpha(t) \;+\; \int_{-\infty}^{+\infty} dt_1 \sum_j c_j(t_1) \sum_{s,\beta} \tau_{js\beta}^* \, g_{r\alpha}(t,t_1)$$

$$\Big( <b_\beta^+(t_1)b_\alpha(t_1)> <c_s^+(t_1)c_r(t_1)> + <c_r(t_1)c_s^+(t_1)> <b_\alpha(t_1)b_\beta^+(t_1)> \Big)$$

where

$$i\hbar\frac{d}{dt}G_{r\alpha}(t,t_1) \;-\; (\varepsilon_r - i\eta + \hbar\omega_\alpha)\,G_{r\alpha}(t,t_1) \;=\; \delta(t-t_1)$$

To lowest order, we can write $c_r(t)b_\alpha(t) \;\approx\; G_{r\alpha}(t,t_1)\,c_r(t_1)b_\alpha(t_1)$, so that

$$c_r(t)b_\alpha(t) \;=\; C_r(t)B_\alpha(t) \;+\; \int_{-\infty}^{+\infty} dt_1 \sum_j c_j(t_1) \sum_{s,\beta} \tau_{js\beta}^*$$

$$\Big( <b_\beta^+(t_1)b_\alpha(t)> <c_s^+(t_1)c_r(t)> + <c_r(t_1)c_s^+(t_1)> <b_\alpha(t)b_\beta^+(t_1)> \Big)$$

$$=\; C_r(t)B_\alpha(t) \;+\; \int_{-\infty}^{+\infty} dt_1 \sum_j c_j(t_1) \sum_{s,\beta} \tau_{js\beta}^*$$

$$\Big( G_{\alpha\beta}^{ab}(t,t_1)\,G_{rs}^{n}(t,t_1) \;+\; G_{rs}^{p}(t,t_1)\,G_{\alpha\beta}^{em}(t,t_1) \Big) \qquad \text{(B.4.3a)}$$

***Evaluating*** $c_r b_\alpha^+$***:*** Now that we have evaluated $c_r b_\alpha$, the next step is to evaluate $c_r b_\alpha^+$. With this in mind, we start from

$$i\hbar\frac{d}{dt}c_r \;=\; (\varepsilon_r - i\eta)c_r \;+\; \sum_{i,\alpha} \tau_{ri\alpha} c_i b_\alpha + \tau_{ir\alpha}* \, c_i b_\alpha^+$$

and

$$-i\hbar\frac{d}{dt}b_\alpha^+ \;=\; \hbar\omega_\alpha\, b_\alpha^+ \;+\; \sum_{i,r} \tau_{ir\alpha} c_i^+ c_r \;+\; \tau_{ri\alpha} c_r^+ c_i$$

and write an equation for the product

$$ih\frac{d}{dt}c_r b_\alpha{}^+ = \left(ih\frac{d}{dt}c_r\right)b_\alpha{}^+ - c_r\left(-ih\frac{d}{dt}b_\alpha{}^+\right)$$

$$= \left((\varepsilon_r - i\eta)c_r + \sum_{j,\beta}\tau_{r\,j\beta}\,c_j b_\beta + \tau_{jr\beta}{}^*\,c_j b_\beta{}^+\right)b_\alpha{}^+$$

$$- c_r\left(\hbar\omega_\alpha\,b_\alpha{}^+ + \sum_{j,s}(\tau_{js\alpha}\,c_j{}^+c_s + \tau_{s\,j\alpha}\,c_s{}^+c_j)\right)$$

which is simplified by replacing the products involving reservoir variables with their average values as before:

$$ih\frac{d}{dt}c_r b_\alpha{}^+ - (\varepsilon_r - i\eta - \hbar\omega_\alpha)c_r b_\alpha{}^+$$

$$= \sum_j\sum_{s,\beta}\tau_{s\,j\beta}\left(<b_\beta b_\alpha{}^+>\delta_{rs} - <c_r c_s{}^+>\delta_{\alpha\beta}\right)c_j$$

$$= \sum_j\sum_{s,\beta}\tau_{s\,j\beta}\left(<b_\beta b_\alpha{}^+><c_s{}^+c_r> + <c_r c_s{}^+><b_\alpha{}^+b_\beta>\right)c_j$$

Once again we can write the solution as a sum of a solution to the homogeneous equation and a response induced by the source term:

$$c_r(t)b_\alpha{}^+(t) = C_r(t)B_\alpha{}^+(t) + \int_{-\infty}^{+\infty}dt_1\sum_j c_j(t_1)\sum_{s,\beta}\tau_{sj\beta}\,\overline{G}_{r\alpha}(t,t_1)$$

$$\left(<b_\beta(t_1)b_\alpha{}^+(t_1)><c_s{}^+(t_1)c_r(t_1)> + <c_r(t_1)c_s{}^+(t_1)><b_\alpha{}^+(t_1)b_\beta(t_1)>\right)$$

where

$$ih\frac{d}{dt}\overline{G}_{r\alpha}(t,t_1) - (\varepsilon_r - i\eta - \hbar\omega_\alpha)\overline{G}_{r\alpha}(t,t_1) = \delta(t - t_1)$$

Proceeding as before we write

$$c_r(t)b_\alpha{}^+(t) \approx C_r(t)B_\alpha{}^+(t) + \int_{-\infty}^{+\infty} dt_1 \sum_j c_j(t_1) \sum_{s,\beta} \tau_{sj\beta}$$

$$\left( <b_\beta(t_1)b_\alpha^+(t)> <c_s^+(t_1)c_r(t)> + <c_r(t)c_s^+(t_1)> <b_\alpha^+(t)b_\beta(t_1)> \right)$$

$$= C_r(t)B_\alpha{}^+(t) + \int_{-\infty}^{+\infty} dt_1 \sum_j c_j(t_1) \sum_{s,\beta} \tau_{sj\beta}$$

$$\left( G_{\alpha\beta}^{em}(t_1,t) G_{rs}^n(t,t_1) + G_{rs}^p(t,t_1) G_{\alpha\beta}^{ab}(t_1,t) \right) \qquad (B.4.3b)$$

Substituting Eqs.(B.4.3a,b) into Eq.(B.4.1) we obtain an equation for the channel field (cf. Eqs.(B.2.7)-(B.2.9)):

$$ih\frac{d}{dt}c_i - \varepsilon_i c_i - \sum_j \int_{-\infty}^{+\infty} dt_1 \Sigma_{ij}(t,t_1)c_j(t_1) = S_i(t) \qquad (B.4.4)$$

$$\text{where} \qquad S_i(t) \equiv \sum_{r,\alpha} \tau_{ir\alpha} C_r(t)B_\alpha(t) + \tau_{r\,i\alpha}^* C_r(t)B_\alpha^+(t) \qquad (B.4.5)$$

$$\Sigma_{ij}(t,t_1) \equiv \vartheta(t-t_1) \Gamma_{ij}(t-t_1) \qquad (B.4.6)$$

and

$$\Gamma_{ij}(t,t_1) = \sum_{r,s,\alpha,\beta} \tau_{ir\alpha} \tau_{js\beta}^* \left( G_{\alpha\beta}^{ab}(t,t_1) G_{rs}^n(t,t_1) + G_{rs}^p(t,t_1) G_{\alpha\beta}^{em}(t,t_1) \right)$$

$$+ \tau_{sj\beta} \tau_{r\,i\alpha}^* \left( G_{\alpha\beta}^{em}(t_1,t) G_{rs}^n(t,t_1) + G_{rs}^p(t,t_1) G_{\alpha\beta}^{ab}(t_1,t) \right) \qquad (B.4.7)$$

***Inscattering function:*** We evaluate the inscattering function

$$\Sigma_{ij}^{in}(t,t') \equiv <S_j^+(t')S_i(t)>$$

as before by substituting for the source term from Eq.(B.4.5)

$$\Sigma_{ij}^{in}(t,t') = \sum_{r,s,\alpha,\beta} \tau_{ir\alpha} \tau_{js\beta}^* <C_s^+(t')C_r(t)> <B_\beta^+(t')B_\alpha(t)>$$

$$+ \tau_{sj\beta} \tau_{ri\alpha}^* <C_s^+(t')C_r(t)> <B_\beta(t')B_\alpha^+(t)>$$

which yields

$$\Sigma_{ij}^{\ in}(t,t') \ = \ \sum_{r,s,\alpha,\beta} \tau_{ir\alpha} \, \tau_{js\beta}^{\ *} \, G_{rs}^n(t,t') G_{\alpha\beta}^{ab}(t,t') \ + \ \tau_{sj\beta} \, \tau_{ri\alpha}^{\ *} \, G_{rs}^n(t,t') \, G_{\beta\alpha}^{em}(t',t)$$

$$(B.4.8)$$

**Steady-state:** Under steady-state conditions we Fourier transform with respect to the difference time coordinate to obtain

$$\Gamma_{ij}(E) \ = \ \int_0^\infty d(\hbar\omega)/2\pi \sum_{r,s,\alpha,\beta}$$

$$\tau_{ir\alpha} \, \tau_{js\beta}^{\ *} \Big( \ G_{rs}^n(E-\hbar\omega) \, G_{\beta\alpha}^{ab}(\hbar\omega) \ + \ G_{rs}^p(E-\hbar\omega) \, G_{\beta\alpha}^{em}(\hbar\omega) \Big)$$

$$+ \ \tau_{sj\beta} \, \tau_{r\,i\alpha}^{\ *} \Big( \ G_{rs}^n(E+\hbar\omega) \, G_{\beta\alpha}^{em}(\hbar\omega) \ + \ G_{rs}^p(E+\hbar\omega) \, G_{\beta\alpha}^{ab}(\hbar\omega) \Big) \qquad (B.4.9)$$

from Eq.(B.4.7) and

$$\Sigma_{ij}^{\ in}(E) \ = \ \int_0^\infty d(\hbar\omega)/2\pi$$

$$\sum_{r,s,\alpha,\beta} \tau_{ir\alpha} \, \tau_{js\beta}^{\ *} \, G_{rs}^n(E-\hbar\omega) G_{\beta\alpha}^{ab}(\hbar\omega) \ + \ \tau_{sj\beta} \, \tau_{r\,i\alpha}^{\ *} \, G_{rs}^n(E+\hbar\omega) \, G_{\beta\alpha}^{em}(\hbar\omega)$$

$$(B.4.10)$$

from Eq.(B.4.8).

**Current:** Note that there is no need to re-derive the expressions for the inflow and the outflow (see Eqs.(B.3.2)-(B.3.5)) since these were derived making use of Eq.(B.2.7) which is still valid (see Eq.(B.4.4)). What has changed is the expression for the self-energy $[\Sigma]$ (cf. Eqs. (B.2.8) and (B.4.7)) and the inscattering $[\Sigma^{in}]$ (cf. Eqs.(B.2.19) and (B.4.8)). But the basic expressions for the current in terms of these quantities remains the same.

**B.5. Coulomb blockade / Kondo resonance**

Next we consider a problem involving purely coherent coupling to the contact, but we now take into account the Coulomb interaction between the two spin levels inside the channel denoted by 'c' and 'd'. For simplicity, we will only consider an equilibrium problem (with one contact) and include one level of each type in this discussion:

$$i\hbar \frac{d}{dt} c \;=\; \varepsilon \, c + \sum_r \tau_r \, c_r \;+\; U \, d^+ \, dc \qquad\qquad (B.5.1a)$$

$$i\hbar \frac{d}{dt} c_r \;=\; (\varepsilon_r - i\eta) \, c_r + \; \tau_r^* \, c \qquad\qquad (B.5.1b)$$

We can proceed as before to write Eq.(B.2.3) from Eq.(B.5.1b) and substitute back into Eq.(B.5.1a) to obtain the same equation (Eq.(B.2.7)) as before except for an additional term $U d^+ dc$ :

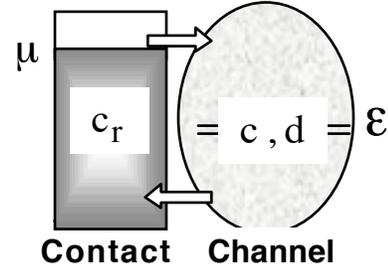

$$i\hbar \frac{d}{dt} c \;-\; \varepsilon c \;-\; \sum_j \int_{-\infty}^{+\infty} dt_1 \, \Sigma_0(t,t_1) c(t_1) \;=\; S(t) \;+\; U d^+(t) d(t) c(t) \qquad (B.5.2)$$

Fourier transforming we obtain

$$(E - \varepsilon - \Sigma_0) \, c \;=\; S^c \;+\; U\{d^+ dc\}(E)$$

$$\Sigma_0 \;=\; \sum_r \frac{|\tau_r|^2}{E - \varepsilon_r + i\eta} \quad , \quad S^c \;\equiv\; \sum_r \tau_r \, C_r \qquad (B.5.3)$$

where we have added the braces {} around $d^+ dc$ to indicate that we need the Fourier transform of the product (which is different from the product of the individual Fourier transforms).

Similarly we can work with the equations for the down-spin components:

$$i\hbar \frac{d}{dt} d = \varepsilon d + \sum_r \tau_r d_r + U c^+ c d \qquad \text{(B.5.4a)}$$

$$\text{and} \quad i\hbar \frac{d}{dt} d_r = (\varepsilon_r - i\eta) d_r + \tau_r^* d \qquad \text{(B.5.4b)}$$

$$\text{to obtain} \quad (E - \varepsilon - \Sigma_0) d = S^d + U\{c^+ c d\}(E)$$

$$\text{where} \quad S^d \equiv \sum_r \tau_r D_r \qquad \text{(B.5.5)}$$

What we propose to show in this Section is that if we treat the new terms $U d^+ d c$ and $U c^+ c d$ using a self-consistent field method, we will get a single peak for the spectral function as shown in Fig.A.1, but if we use a better approximation for these terms we will obtain two peaks as shown in Fig. B.5.1, which can be viewed as a combination of the two possibilities sketched in Fig.A.2. If we take our approximation a step further we will obtain the central peak around $E = \mu$ as shown in Fig.B.5.2. This peak is responsible for the increased resistivity in bulk metals with magnetic impurities at low temperatures as explained in the 1960s by Kondo. More recently experiments on single quantum dots and molecules in the 1990s have observed an enhanced conductance believed to arise from the same Kondo peak in the spectral function [16].

***Self-consistent field (scf) approximation:*** The simplest approximation is to write

$$\{c^+ c d\}(E) \approx <c^+ c> d(E)$$

$$\text{and} \quad \{d^+ d c\}(E) \approx <d^+ d> c(E) \qquad \text{(B.5.6)}$$

so that from Eqs.(B.5.3) and (B.5.5)

$$c = \underbrace{\frac{1}{E - \varepsilon - \Sigma_0 - U < d^+ d >}}_{\text{Gc(E)}} S^c$$

and
$$d = \underbrace{\frac{1}{E - \varepsilon - \Sigma_0 - U < c^+ c >}}_{G_d(E)} S^d$$

These equations can be solved self-consistently with the corresponding expressions for $< c^+ c >$ and $< d^+ d >$:

$$< c^+ c > = \int (dE/2\pi) \, |G_c(E)|^2 \, < S_c^+(E) S_c(E) >$$

$$= \int (dE/2\pi) \, |G_c(E)|^2 \, \Gamma(E) \, f_0(E - \mu)$$

Making use of Eq.(8.1.14), $\quad < c^+ c > = \int (dE/2\pi) \, A_c(E) \, f_0(E - \mu) \quad$ (B.5.7a)

Similarly, we can show that $\quad < d^+ d > = \int (dE/2\pi) \, A_d(E) \, f_0(E - \mu) \quad$ (B.5.7b)

This is essentially what is often called the unrestricted self-consistent field (scf) approximation (as compared to the restricted scf method described in Section 5). How can we go beyond this approximation?

**Coulomb blockade:** One approach is to replace Eq.(B.5.6) with a better approximation, starting from the time-dependent equation for '$d^+ dc$'

$$i h \frac{d}{dt} d^+ dc = (\varepsilon + U) d^+ dc + \sum_r (\tau_r^* d_r^+ c d + \tau_r d^+ d_s c + \tau_r d^+ dc_r) \quad (B.5.8)$$

$$\approx (\varepsilon + U) d^+ dc + \sum_r \tau_r d^+ d c_r \quad (B.5.9)$$

$$\rightarrow (E - \varepsilon - U) \{d^+ dc\} = \sum_r \tau_r d^+ d \, c_r$$

and making use of Eq.(B.5.2) to write

$$\{d^+ dc\}(E) = \sum_r \frac{\tau_r < d^+ d > C_r}{E - \varepsilon - U - \Sigma_0} = \frac{< d^+ d >}{E - \varepsilon - U - \Sigma_0} S^c \quad (B.5.10)$$

Substituting into Eq.(B.5.3) we obtain

$$(E - \varepsilon - \Sigma_0)\, c \;\; = \;\; \left(1 + \frac{U <d^+d>}{E - \varepsilon - U - \Sigma_0}\right) S^c \qquad (B.5.11)$$

from which we can write the Green's function, G(E)

$$c \;\; = \;\; \left(\frac{1}{E - \varepsilon - \Sigma_0} + \frac{U <d^+d>}{(E - \varepsilon - \Sigma_0)\,(E - \varepsilon - U - \Sigma_0)}\right) S^c$$

$$= \underbrace{\left(\frac{1 - <d^+d>}{E - \varepsilon - \Sigma_0} + \frac{<d^+d>}{E - \varepsilon - U - \Sigma_0}\right)}_{G(E)} S^c \quad (B.5.12)$$

The density of states, $D(E) = i\,[G - G^+]$ calculated using the above G(E) for a typical set of parameters is shown below. This is similar to the picture we expect under Coulomb blockade conditions (see Fig.A.2).

**Fig.B.5.1. DOS, $D(E) = i\,[G - G^+]$ calculated using G(E) from Eq.(B.5.12) with $\varepsilon = -0.1$ eV, U = +0.2 eV, $<d^+d> = 0.5$ and $\Sigma_0$ calculated from Eq.(B.4.3) assuming that the reservoir consists of a set of levels uniformly spaced by 0.4 meV with $\eta = 1$ meV and $\tau_r = 60$ meV.**

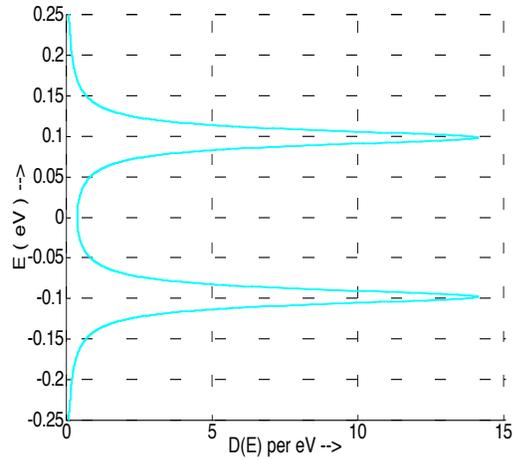

**Kondo resonance:** To go further we step back to Eq.(B.5.8)

$$i\hbar \frac{d}{dt} d^+ dc = (\varepsilon + U) d^+ dc + \sum_r (\tau_r^* d_r^+ c d + \tau_r d^+ d_s c + \tau_r d^+ dc_r)$$

$$\rightarrow \quad (E - \varepsilon - U)\{d^+ dc\} = \sum_r (\tau_r^* \{d_r^+ cd\} + \tau_r \{d^+ d_s c\} + \tau_r \{d^+ dc_r\}) \quad \text{(B.5.13)}$$

and try to do better than what we did in going to Eq.(B.4.9) by starting from the time-dependent equations for each the three quantities appearing in Eqs.(B.5.13):

$$i\hbar \frac{d}{dt} d^+ dc_r = (\varepsilon_r - i\eta) d^+ dc_r + \sum_r (\tau_r^* d^+ dc + \tau_r d^+ d_r c_r + \tau_r^* d_r^+ c_r d)$$

$$\approx (\varepsilon_r - i\eta) d^+ dc_r + \sum_r \tau_r^* d^+ dc$$

$$\{d^+ dc_r\} \approx <d^+ d> C_r + \sum_r \frac{\tau_r^*}{E - \varepsilon_r + i\eta} \{d^+ dc\} \quad \text{(B.5.14)}$$

$$i\hbar \frac{d}{dt} d^+ d_r c = \varepsilon_r d^+ d_r c + \sum_r (\tau_r^* d^+ dc + \tau_r d^+ d_r c_r - \tau_r^* d_r^+ d_r c)$$

$$\approx \varepsilon_r d^+ d_r c + \sum_r \tau_r^* d^+ dc - \tau_r^* f_r c$$

$$\{d^+ d_r c\} \approx \sum_r \frac{\tau_r^*}{E - \varepsilon_r + i\eta} \{d^+ dc\} - \frac{\tau_r^* f_r}{E - \varepsilon_r + i\eta} c \quad \text{(B.5.15)}$$

$$i\hbar \frac{d}{dt} d_r^+ cd = (\varepsilon_r - 2\varepsilon - U - i\eta) d_r^+ cd + \sum_r (\tau_r d^+ dc - \tau_r d_r^+ d_r c - \tau_r d_r^+ c_r c)$$

$$\approx (\varepsilon_r - 2\varepsilon - U - i\eta) d_r^+ cd + \sum_r (\tau_r d^+ dc - \tau_r f_r c)$$

$$\{d_r^+ cd\} \approx \sum_r \frac{\tau_r}{E + 2\varepsilon + U - \varepsilon_r + i\eta} \{d^+ dc\} - \frac{\tau_r f_r}{E + 2\varepsilon + U - \varepsilon_r + i\eta} c \quad \text{(B.5.16)}$$

Substituting Eqs.(B.5.14), (B.5.15) and (B.5.16) into Eq.(B.5.13) we obtain a better expression for $\{d^+dc\}$ than the last one (see Eq.(B.5.10))

$$(E - \varepsilon - U - 2\Sigma_0 - \Sigma_1)\{d^+dc\} \;=\; <d^+d> S^c - (\Sigma_2 + \Sigma_3)\,c$$

$$\rightarrow \quad \{d^+dc\} \;=\; \frac{<d^+d>}{E - \varepsilon - U - 2\Sigma_0 - \Sigma_1}\,S^c - \frac{\Sigma_2 + \Sigma_3}{E - \varepsilon - U - 2\Sigma_0 - \Sigma_1}\,c \quad \text{(B.5.17)}$$

$$\text{where} \quad \Sigma_1 \;\equiv\; \sum_r \frac{|\tau_r|^2}{E + 2\varepsilon + U - \varepsilon_r + i\eta}$$

$$\Sigma_2 \;\equiv\; \sum_r \frac{|\tau_r|^2 f_r}{E - \varepsilon_r + i\eta} \quad \text{and} \quad \Sigma_3 \;\equiv\; \sum_r \frac{|\tau_r|^2 f_r}{E + 2\varepsilon + U - \varepsilon_r + i\eta} \quad \text{(B.5.18)}$$

Substituting Eq.(B.5.17) back into Eq.(B.5.13) we obtain

$$\left(E - \varepsilon - \Sigma_0 + \frac{U(\Sigma_2 + \Sigma_3)}{E - \varepsilon - U - 2\Sigma_0 - \Sigma_1}\right)c \;=\; \left(1 + \frac{U <d^+d>}{E - \varepsilon - U - 2\Sigma_0 - \Sigma_1}\right)S^c \quad \text{(B.5.19)}$$

so that the Green's function G(E) (such that $c = G\,S^c$) is given by

$$G \;=\; \frac{1 - <d^+d>}{E - \varepsilon - \Sigma_0 + \dfrac{U(\Sigma_2 + \Sigma_3)}{E - \varepsilon - U - 2\Sigma_0 - \Sigma_1}}$$

$$+ \frac{<d^+d>}{E - \varepsilon - \Sigma_0 - U - \dfrac{U(\Sigma_0 + \Sigma_1 - \Sigma_2 - \Sigma_3)}{E - \varepsilon - 2\Sigma_0 - \Sigma_1}} \quad \text{(B.5.20)}$$

The density of states $D(E) = i[G - G^+]$ calculated using G(E) from Eq.(B.5.20) with the same parameters as in Fig.B.5.1 is shown in Fig.B.5.2 (also shown for reference is the result using G(E) from Eq.(B.5.12)). Note how the central "Kondo peak" grows in amplitude as the temperature is lowered. Indeed at low temperatures and also for stronger coupling to the reservoir, the treatment presented above may be inadequate making it necessary to go to higher orders (perhaps infinite!) in perturbation theory. For many years this was one of the frontier problems in many-electron theory [16].

**Fig.B.5.2. DOS, D(E) =**
$i[G - G^+]$ **calculated using**
**G(E) from Eq.(B.5.9) with the**
**same parameters as in**
**Fig.B.5.1 (also shown for**
**reference is the result using**
**G(E) from Eq.(B.5.12)). Note**
**how the central "Kondo peak"**
**grows in amplitude as the**
**temperature is lowered**
**(courtesy of A.W. Ghosh).**

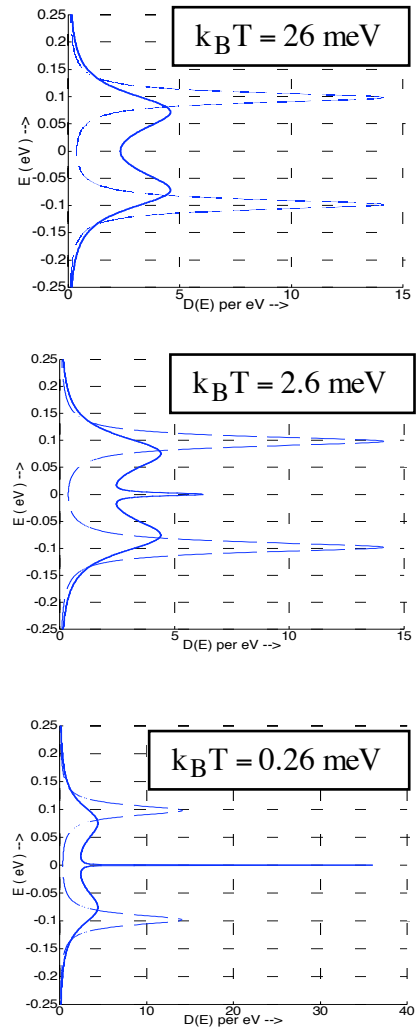

**Appendix C: MATLAB codes for numerical examples in**
              **Section 7, Appendix A, Appendix B**

Can be downloaded from www.nanohub.org,

where they can also be run without installation.

```
%Figs.7.1a,b
clear all

%Constants (all MKS, except energy which is in eV)
hbar=1.055e-34;q=1.602e-19;I0=q*q/hbar;

%Parameters
U0=0.25;kT=0.025;mu=0;ep=0.2;N0=0;
alphag=0;alphad=0.5;

%Energy grid
NE=501;E=linspace(-1,1,NE);dE=E(2)-E(1);
g2=0.005*ones(1,NE);%g1=g2;% use for Fig.7.2a
g1=0.005*(E+abs(E))./(E+E+1e-6);% use for Fig.7.2b
g=g1+g2;

%Bias
IV=101;VV=linspace(-1,1,IV);
for iV=1:IV
        Vg=0;Vd=VV(iV);
        %Vd=0;Vg=VV(iV);
                mu1=mu;mu2=mu1-Vd;UL=-(alphag*Vg)-(alphad*Vd);

U=0;%Self-consistent field
dU=1;
while dU>1e-6
        f1=1./(1+exp((E-mu1)./kT));
                f2=1./(1+exp((E-mu2)./kT));
                D=(g./(2*pi))./(((E-ep-UL-U).^2)+((g./2).^2));
                        D=D./(dE*sum(D));
        N(iV)=dE*2*sum(D.*((f1.*g1./g)+(f2.*g2./g)));
                Unew=U0*(N(iV)-N0);dU=abs(U-Unew);
                        U=U+0.1*(Unew-U);
end
I(iV)=dE*2*I0*(sum(D.*(f1-f2).*g1.*g2./g));
end

hold on
h=plot(VV,I,'b');
set(h,'linewidth',[2.0])
set(gca,'Fontsize',[25])
xlabel(' Voltage ( V ) --->')
ylabel(' Current ( A ) ---> ')
grid on

%Fig.7.3
clear all

%Constants (all MKS, except energy which is in eV)
hbar=1.055e-34;q=1.602e-19;I0=q*q/hbar;
```

```
%Parameters
kT1=0.026;kT2=0.025;ep=0.2;
alphag=1;alphad=0.5;

%Energy grid
NE=501;E=linspace(-1,1,NE);dE=E(2)-E(1);
g1=0.005*(E+abs(E))./(E+E+1e-6);% zero for negative E
g2=0.005*ones(1,NE);g1=g2;
g=g1+g2;

%Bias
IV=101;VV=linspace(-0.25,0.25,IV);
for iV=1:IV
        mu1=ep+VV(iV);mu2=mu1;
                f1=1./(1+exp((E-mu1)./kT1));
                f2=1./(1+exp((E-mu2)./kT2));
                D=(g./(2*pi))./(((E-ep).^2)+((g./2).^2));
                        D=D./(dE*sum(D));
I(iV)=dE*2*I0*(sum(D.*(f1-f2).*g1.*g2./g));
end

hold on
h=plot(VV,I,'b');
set(h,'linewidth',[2.0])
set(gca,'Fontsize',[25])
xlabel(' Voltage ( V ) --->')
ylabel(' Current ( A ) ---> ')
grid on
```

**%  Fig.7.5**
```
clear all

%Constants (all MKS, except energy which is in eV)
hbar=1.055e-34;q=1.602e-19;eps0=8.854E-12;epsr=4;
m=0.25*9.11e-31;%Effective mass
I0=q*q/hbar;

%Parameters
W=1e-6;L=10e-9;t=1.5e-9;%W=Width,L=Length of active region,t=oxide thickness
Cg=epsr*eps0*W*L/t;Cs=0.05*Cg;Cd=0.05*Cg;CE=Cg+Cs+Cd;U0=q/CE;
alphag=Cg/CE,alphad=Cd/CE
        %alphag=1;alphad=0.5;U0=0.25;

kT=0.025;mu=-0.2;
        v=1e5;%Escape velocity
                g1=hbar*v/(q*L);g2=g1;g=g1+g2;
                        %g1=0.005;g2=0.005;g=g1+g2;

%Energy grid
NE=501;E=linspace(-1,1,NE);dE=E(2)-E(1);
        D0=m*q*W*L/(pi*hbar*hbar);% Step Density of states per eV
        D=D0*[zeros(1,251) ones(1,250)];
        %D=(2*g/(2*pi))./((E.^2)+((g/2)^2));% Lorentzian Density of states per eV
                %D=D./(dE*sum(D));%Normalizing to one

%Reference number of electrons
f0=1./(1+exp((E-mu)./kT));N0=2*dE*sum(D.*f0);ns=N0/(L*W*1e4),%/cm^2
```

```
%Bias
IV=61;VV=linspace(0,0.6,IV);
for iV=1:IV
        Vg=0.5;Vd=VV(iV);
        %Vd=0.5;Vg=VV(iV);
                mu1=mu;mu2=mu1-Vd;UL=-(alphag*Vg)-(alphad*Vd);

U=0;%Self-consistent field
dU=1;
while dU>1e-6
        f1=1./(1+exp((E+UL+U-mu1)./kT));
                f2=1./(1+exp((E+UL+U-mu2)./kT));
        N(iV)=dE*sum(D.*((f1.*g1/g)+(f2.*g2/g)));
                Unew=U0*(N(iV)-N0);dU=abs(U-Unew);
                        U=U+0.1*(Unew-U);
end
I(iV)=dE*I0*(sum(D.*(f1-f2)))*g1*g2/g;
end

hold on
h=plot(VV,I,'b');
set(h,'linewidth',[2.0])
set(gca,'Fontsize',[25])
xlabel(' Voltage (V) --->')
ylabel(' Current (A) ---> ')
grid on
```

**%Fig.7.7**
```
clear all
%0-D with inelastic scattering

%Constants (all MKS, except energy which is in eV)
hbar=1.06e-34;q=1.6e-19;I0=q*q/(2*pi*hbar);

%Parameters
        H0=5;Ef=0;kT=0.0025;dE=0.0005;zplus=i*1e-12;gamma=0.1;
        D0=0;Dnu=0*[0.5 0.7];Nph=size(Dnu,2);
                hnu=[100 550];%Multiply by dE for actual hnu
                        Nhnu=1./((exp(dE*hnu./kT))-1);

%Bias
NV=203;VV=linspace(-0.51,0.5,NV);dV=VV(2)-VV(1);
for iV=1:NV
V=VV(iV);mu1=Ef;mu2=Ef-V;U1=(-0.5)*V;

        %Energy grid
        E=[mu2-(10*kT)-(10*dE):dE:mu1+(10*kT)+(10*dE)];
                if V<0
                        E=[mu1-(10*kT)-(10*dE):dE:mu2+(10*kT)+(10*dE)];
                end
                NE=size(E,2);[iV NE]
                        f1=1./(1+exp((E-mu1)./kT));
                        f2=1./(1+exp((E-mu2)./kT));

        %Initial guess
                n=zeros(1,NE);p=zeros(1,NE);
                sigin1=zeros(1,NE);sigout1=zeros(1,NE);
```

```
                sign2=zeros(1,NE);sigout2=zeros(1,NE);
                sign=0*ones(1,NE);sigout=0*ones(1,NE);
%Iterative solution of transport equation
change=1;it=1;
while change>1e-3
        for k=1:NE
        sig1=-i*gamma/2;gam1=i*(sig1-sig1');
        sig2=-i*gamma/2;gam2=i*(sig2-sig2');
                sign1(k)=f1(k)*gam1;sign2(k)=f2(k)*gam2;
                sigout1(k)=(1-f1(k))*gam1;sigout2(k)=(1-f2(k))*gam2;
                gamp=sign(k)+sigout(k);
        G=inv((E(k)+zplus)-H0-U1-sig1-sig2+(i*0.5*gamp));
                A=i*(G-G');
                        n(k)=real(G*((f1(k)*gam1)+(f2(k)*gam2)+sign(k))*G');
                        p(k)=A-n(k);
        end

        siginnew=D0*n;sigoutnew=D0*p;
        for iph=1:Nph
                inu=hnu(iph);
                if inu<NE
                ne=n([inu+1:NE]);ne=[ne zeros(1,inu)];
                na=n([1:NE-inu]);na=[zeros(1,inu) na];
                pe=p([inu+1:NE]);pe=[pe zeros(1,inu)];
                pa=p([1:NE-inu]);pa=[zeros(1,inu) pa];

        siginnew=siginnew+((Nhnu(iph)+1)*Dnu(iph)*ne)+(Nhnu(iph)*Dnu(iph)*na);
        sigoutnew=sigoutnew+(Nhnu(iph)*Dnu(iph)*pe)+((Nhnu(iph)+1)*Dnu(iph)*pa);
                end
        end

        change=sum(sum(abs(siginnew-sign)));
        change=change+sum(sum(abs(sigoutnew-sigout)));
                sign=((1-it)*sign)+(it*siginnew);
                sigout=((1-it)*sigout)+(it*sigoutnew);
end

        I1=real((sigout1.*n)-(sign1.*p));I1=sum(I1);
        I2=real((sigout2.*n)-(sign2.*p));I2=sum(I2);
        I3=real((sigout.*n)-(sign.*p));I3=sum(I3);

        I123=dE*[sum(I1) sum(I2) sum(I3)],%Normalized Conductance
        %IE=(dE/(V*V))*[sum(E.*I1) sum(E.*I2) sum(E.*I3)],%Normalized Power
        %kirchoff=[sum(I123) sum(IE)],%checking for conservation of current and energy current

II(iV)=sum(I2)*dE*I0;
end

G1=diff(II)./dV;VG=VV([2:NV]);
IETS=diff(G1)./dV;VETS=VV([3:NV]);

hold on
%h=plot(VV,II,'rx');
h=plot(VG,G1,'b--');
set(h,'linewidth',[2.0])
set(gca,'Fontsize',[24])
%xlabel(' Voltage ( V ) ---> ')
%ylabel(' d2I/dV2 ---> ')
```

```
%  Fig.A.4
clear all

%Constants (all MKS, except energy which is in eV)
hbar=1.055e-34;q=1.602e-19;I0=q*q/hbar;

%Parameters
U0=0.25;
kT=0.025;mu=0;ep=0.2;
g1=0.005;g2=0.005;g=g1+g2;
alphag=1;alphad=0.5;

%Bias
IV=101;VV=linspace(-.25,.75,IV);
for iV=1:IV
        Vg=VV(iV);UL=-(alphag*Vg);
f1=1/(1+exp((ep+UL-(U0/2)-mu)/kT));f2=1/(1+exp((ep+UL-(U0/2)-mu)/kT));
f1U=1/(1+exp((ep+UL+(U0/2)-mu)/kT));f2U=1/(1+exp((ep+UL+(U0/2)-mu)/kT));

P1=((g1*f1)+(g2*f2))/(1e-6+(g1*(1-f1))+(g2*(1-f2)));
P2=P1*((g1*f1U)+(g2*f2U))/(1e-6+(g1*(1-f1U))+(g2*(1-f2U)));
P0=1/(1+P1+P1+P2);P1=P1*P0;P2=P2*P0;
p0(iV)=P0;p1(iV)=P1;p2(iV)=P2;
end

hold on
h=plot(VV,p0,'bo');
h=plot(VV,p1,'b');
h=plot(VV,p2,'bx');
set(h,'linewidth',[2.0])
set(gca,'Fontsize',[25])
grid on
xlabel(' Gate voltage, VG ( volts ) --->')
ylabel(' Probability ---> ')
axis([-.2 .6 0 1])

%  Fig.A.5
clear all

%Constants (all MKS, except energy which is in eV)
hbar=1.055e-34;q=1.602e-19;I0=q*q/hbar;

%Parameters
U0=0.025;% U0 is 0.25 for part(a), 0.025 for part (b)
kT=0.025;mu=0;ep=0.2;N0=1;
g1=0.005;g2=0.005;g=g1+g2;
alphag=1;alphad=0.5;

%Bias
IV=101;VV=linspace(0,1,IV);
for iV=1:IV
        Vg=0;Vd=VV(iV);
        %Vd=0;Vg=VV(iV);
                mu1=mu;mu2=mu1-Vd;UL=-(alphag*Vg)-(alphad*Vd);
f1=1/(1+exp((ep+UL-(U0/2)-mu1)/kT));f2=1/(1+exp((ep+UL-(U0/2)-mu2)/kT));
f1U=1/(1+exp((ep+UL+(U0/2)-mu1)/kT));f2U=1/(1+exp((ep+UL+(U0/2)-mu2)/kT));
```

```
P1=((g1*f1)+(g2*f2))/(1e-6+(g1*(1-f1))+(g2*(1-f2)));
P2=P1*((g1*f1U)+(g2*f2U))/(1e-6+(g1*(1-f1U))+(g2*(1-f2U)));
P0=1/(1+P1+P1+P2);P1=P1*P0;P2=P2*P0;

I1(iV)=2*I0*((P0*g1*f1)-(P1*g1*(1-f1))+(P1*g1*f1U)-(P2*g1*(1-f1U)));
I2(iV)=2*I0*((P0*g2*f2)-(P1*g2*(1-f2))+(P1*g2*f2U)-(P2*g2*(1-f2U)));
end

%RSCF method (same as FP53)
%Energy grid
NE=501;E=linspace(-1,1,NE);dE=E(2)-E(1);
D=(g/(2*pi))./((E.^2)+((g/2)^2));% Lorentzian Density of states per eV
D=D./(dE*sum(D));%Normalizing to one

%Bias
for iV=1:IV
        Vg=0;Vd=VV(iV);
        %Vd=0;Vg=VV(iV);
                mu1=mu;mu2=mu1-Vd;UL=-(alphag*Vg)-(alphad*Vd);

U=0;%Self-consistent field
dU=1;
while dU>1e-6
        F1=1./(1+exp((E+ep+UL+U-mu1)./kT));
                F2=1./(1+exp((E+ep+UL+U-mu2)./kT));
        N(iV)=dE*2*sum(D.*((F1.*g1/g)+(F2.*g2/g)));
                Unew=U0*(N(iV)-N0);
                dU=abs(U-Unew);U=U+0.1*(Unew-U);
end
I(iV)=dE*2*I0*(sum(D.*(F1-F2)))*(g1*g2/g);
end

hold on
h=plot(VV,I1,'b');
h=plot(VV,I,'b--');
set(h,'linewidth',[2.0])
set(gca,'Fontsize',[25])
grid on
xlabel(' Drain Voltage, VD ( volts ) --->')
ylabel(' Current ( Amperes ) ---> ')
axis([0 1 0 1.4e-6])

%  Figs.B.5.1,B.5.2
clear all

NE=1001;E=linspace(-.25,.25,NE);zplus=i*1e-3;dE=E(2)-E(1);kT=.00026;
Nep=5001;ep=linspace(-1,1,Nep);tau=0.05;dep=ep(2)-ep(1);delta=3.117*tau*tau/2
ep0=-25*delta;U=50*delta;[U/pi abs(ep0)]/delta
D=ones(1,Nep);f=1./(1+exp(ep./kT));fK=1./(1+exp(E./kT));tau=0.06;

for kE=1:NE
s0(kE)=dep*tau*tau*sum(D./(E(kE)-ep+zplus));
s1(kE)=dep*tau*tau*sum(D./(E(kE)+ep0+ep0+U-ep+zplus));
s2(kE)=dep*tau*tau*sum(D.*f./(E(kE)-ep+zplus));
s3(kE)=dep*tau*tau*sum(D.*f./(E(kE)+ep0+ep0+U-ep+zplus));
end
```

```
g=U./(E-ep0-U-s0-s0-s1);
GK=(1+(0.5*g))./(E-ep0-s0+(g.*(s2+s3)));
G=(1+(0.5*U./(E-ep0-U-s0)))./(E-ep0-s0);
A=i*(G-conj(G))/(2*pi);dE*sum(A)
AK=i*(GK-conj(GK))/(2*pi);dE*sum(AK)
dE*sum(AK.*fK)
del=-dE*sum(imag(s0))

hold on
%h=plot(E,-imag(s0));
%h=plot(E,imag(s1));
%h=plot(E,imag(s2),'mx');
%h=plot(E,imag(s3),'m');
h=plot(A,E,'b--');
h=plot(AK,E,'b');
set(h,'linewidth',[2.0])
set(gca,'Fontsize',[24])
grid on
xlabel(' D(E) per eV --> ')
ylabel(' E ( eV ) --> ')
```